\documentclass[prb,longbibliography,twocolumn]{revtex4-1}

\usepackage{graphicx}
\usepackage{dcolumn}
\usepackage{bm}
\usepackage{hyperref}
\usepackage{amsmath, amssymb}
\usepackage{mathtools}
\usepackage{color}
\usepackage[none]{hyphenat}
\usepackage{ulem} 
\usepackage{xcolor}

\hypersetup{pdftex,
  breaklinks=true,
  colorlinks=true,
  urlcolor=blue,
  linkcolor=blue,
  citecolor=blue,
  }
\newcommand{\bpm}{B/M/B}

\begin{document}

\preprint{APS/123-QED}

\title{Understanding the Oxygen Reduction Reaction and Oxygen Evolution Reaction in Metal Intercalated Biphenylene Bilayers}

\author{Henri G. Mendon\c{c}a$^1$}
\email{irneh.gm@gmail.com}

\author{Pedro H. Souza$^2$}
\email{psouza8628@gmail.com}

\author{Walter Orellana$^2$}
\email{worellana@unab.cl}

\author{Roberto H. Miwa$^1$}
\email{hiroki@ufu.br}

\affiliation{$^1$Instituto de F\'isica, Universidade Federal de Uberl\^andia, 38400-902 Uberlandia, Minas Gerais, Brazil.}

\affiliation{$^2$Departamento de F\'isica y Astronom\'i­a,
Facultad de Ciencias Exactas, Universidad Andres Bello,
Santiago 8370136, Chile}

\date{\today}

\begin{abstract}

We conducted an {\it ab initio} study of the oxygen reduction reaction (ORR) and oxygen evolution reaction (OER) in metal-encapsulated biphenylene bilayers, B/M/B, with M = Ti, V, Cr, Mn, Fe, Co, Ni, Cu, Nb, Ru, W, Os and Pt. In most systems, the intercalated metal sits at the square carbon sites (C$^{468}$) of the biphenylene lattice. Using a computational 
hydrogen electrode approach, we evaluated the reaction energetics at these active sites. Several B/M/B systems show competitive ORR and OER performance. 
Among the investigated systems, Cu, Pt, Ru, and Mn exhibit the lowest ORR overpotentials of 0.42, 0.44, 0.50, and 0.56\,V, respectively, while Fe is identified as the most active catalyst for OER with an overpotential of 0.44\,V.
To understand the catalytic trends, we looked at the electronic structure through the metal $d-$band centers, the C$^{468}$ $p_z-$band centers, and the corresponding orbital charge populations. The band centers did not give a simple polynomial dependence on the overpotentials, though they did point to favorable electronic ranges for the best catalysts. The $d-$orbital charge population of the encapsulated metal, however, correlated most clearly with activity-especially for OER-yielding volcano-type plots. From these, B/Fe/B emerges as the best OER catalyst, while B/Mn/B lies closest to the ORR optimum. The $p-$orbital population at the active carbon site also captures the main trends, albeit less strongly. Overall, these results show that straightforward electronic descriptors can predict catalytic behavior in metal-encapsulated biphenylene bilayers and guide the search for efficient catalysts where the carbon framework itself drives the reactivity.
\end{abstract}

\maketitle

\section{\label{sec:Introduction}Introduction}

In recent years, two-dimensional (2D) catalysts have attracted significant attention as key components for sustainable development 
within a green economy \cite{RAZA2024}. Their applications span fuel cells \cite{Liu2022}, zinc-air batteries \cite{Li2025}, and water 
splitting \cite{Wang2025}. In particular, single-atom catalysts (SACs) have demonstrated remarkable performance in the hydrogen 
evolution reaction (HER) \cite{Attanayake2018,Sai2024}, oxygen reduction reaction (ORR) \cite{Liu2017,Xue2024,Tamtaji2024,Lucchetti2024},
and oxygen evolution reaction (OER) \cite{Sai2024,Li2025,Li2025-vac}. This performance originates from the ability to precisely tailor 
the electronic structure of isolated active sites, thereby enhancing catalytic activity \cite{walter2025,Xu2024,Zhou2023,Li2025-vac}.

Despite these advantages over commercial catalysts, SACs often face limitations in operational stability \cite{Zhao2025,Tsipoaka2025}. 
Under electrocatalytic conditions, changes in oxidation state can promote dissolution of the active site into the electrolyte
\cite{DiLiberto2024,Kumar2020,Liu2022,Bae2023}, resulting in progressive loss of catalytic efficiency. To mitigate these effects, several 
strategies have been proposed, including the deliberate introduction of vacancies to stabilize favorable oxidation states \cite{Li2025-vac-stability},
engineering of the coordination environment \cite{DiLiberto2024,walter2025}, and intercalation within 2D materials \cite{Jin2021}. Among
these, intercalation stands out as a particularly effective approach, as encapsulation shields the active sites from environmental degradation 
while preserving both catalytic activity and long-term stability \cite{Ma2024}.

The feasibility of such architectures is supported by experimental realization of layered materials with intercalated transition metals (TM), 
including graphene \cite{Yam2020,Hu2014,deng2015pene}, phosphorene \cite{Pang2018}, and MoS$_2$ \cite{Twitto2022,Zhao2018}, as 
well as metal nanoparticles encapsulated within carbon nanotubes \cite{deng2013iron,deng2013FeCo}. These systems demonstrate that 
encapsulation effectively suppresses corrosion \cite{Thomas2023,deng2017robust,Zhou2021,deng2013iron,deng2016catalysis}, 
thereby extending catalyst lifetime relative to conventional SACs. Beyond enhanced stability, intercalated structures, such as 
borophene--TM--borophene heterostructures \cite{Chang2023} and TM dichalcogenides encapsulated within h-BN layers \cite{Ahn2016},
exhibit tunable electronic properties \cite{deng2017robust,deng2016catalysis,He2018,Li2025-vac-stability,Zhou2021},
enabling rational optimization of reduction and evolution reactions \cite{Yang2024,Zhang2024,Song2025}. The electronic properties can be 
tuned such that catalytic activity is enhanced \cite{deng2017robust, He2018, Li2025-vac-stability, Zhou2021}. In intercalated systems, electronic tuning 
generally occurs through charge transfer \cite{uosaki2014boron, Lyalin2013}, modifications in the work function of an intercalated material
\cite{zhou2018heterostructures, chen2013}, and hybridization between $p$- and $d$-orbitals~\cite{deng2013iron,deng2013FeCo,deng2015pene}. 
The latter strategy offers the greatest potential for tuning catalytic activity due to the large number of elements with $d$-orbitals in their valence 
shell \cite{deng2017robust, Zhou2021}.

The biphenylene lattice crystallizes in the $Pmmm$ space group and remains dynamically and thermally stable up to 4000~K \cite{Luo2021}. 
Its electronic structure is metallic, with pronounced electron localization at the square-planar carbon sites \cite{Luo2021,ABDELSALAM2024,Ren2022}. 
This distinctive combination of properties has spurred extensive theoretical and experimental efforts exploring biphenylene across a broad range 
of applications, including charge transport \cite{Victor2025}, Li$^+$ and LiO$_2$ battery electrodes \cite{Guo2023, Chen2023}, and hydrogen 
storage \cite{MA2024ti-dec}. Beyond these applications, biphenylene stands out for its exceptional catalytic performance, particularly in 
electrocatalytic processes such as the hydrogen evolution reaction (HER) \cite{Mukesh2024, SAHOO2023, Wang2025, FengChen2025, Lebre2025}, 
CO$_2$ reduction reactions \cite{LI2024, Somaiya2024}, nitrogen reduction reaction (NRR) \cite{Pandiyan2024}, oxygen reduction reaction 
(ORR) \cite{Liu2021, Xu2023, FengChen2025}, and oxygen evolution reaction (OER) \cite{Wang2025, FengChen2025}.

In this work, we investigate the intercalation of transition metals in AA-stacked biphenylene bilayers (B/M/B, with M = Ti, V, Cr, Mn, Fe, Co, Ni, 
Cu, Nb, Ru, W, Os, and Pt) as a strategy to develop ORR/OER electrocatalysts in which the metallic centers are not directly exposed to the electrolyte. 
We propose that intercalating metal atoms between biphenylene layers enhances resistance to corrosion while promoting catalytic activity at the 
surface carbon atoms. We begin by examining the energetic and structural stability of the B/M/B systems, as discussed in Sec.~\ref{EandS}. 
In Sec.~\ref{catal}, we evaluate their catalytic activity, identify B/M/B with M $=$ Mn, Fe, Co, Ni, Cu, and Pt are particularly promising candidates, and provide a detailed analysis of 
the electronic properties underlying the superior performance, relative to other B/M/B systems, introducing descriptors to guide the rational 
design of high-performance catalysts.

\begin{figure*}
\centering
\includegraphics[width=0.7\linewidth]{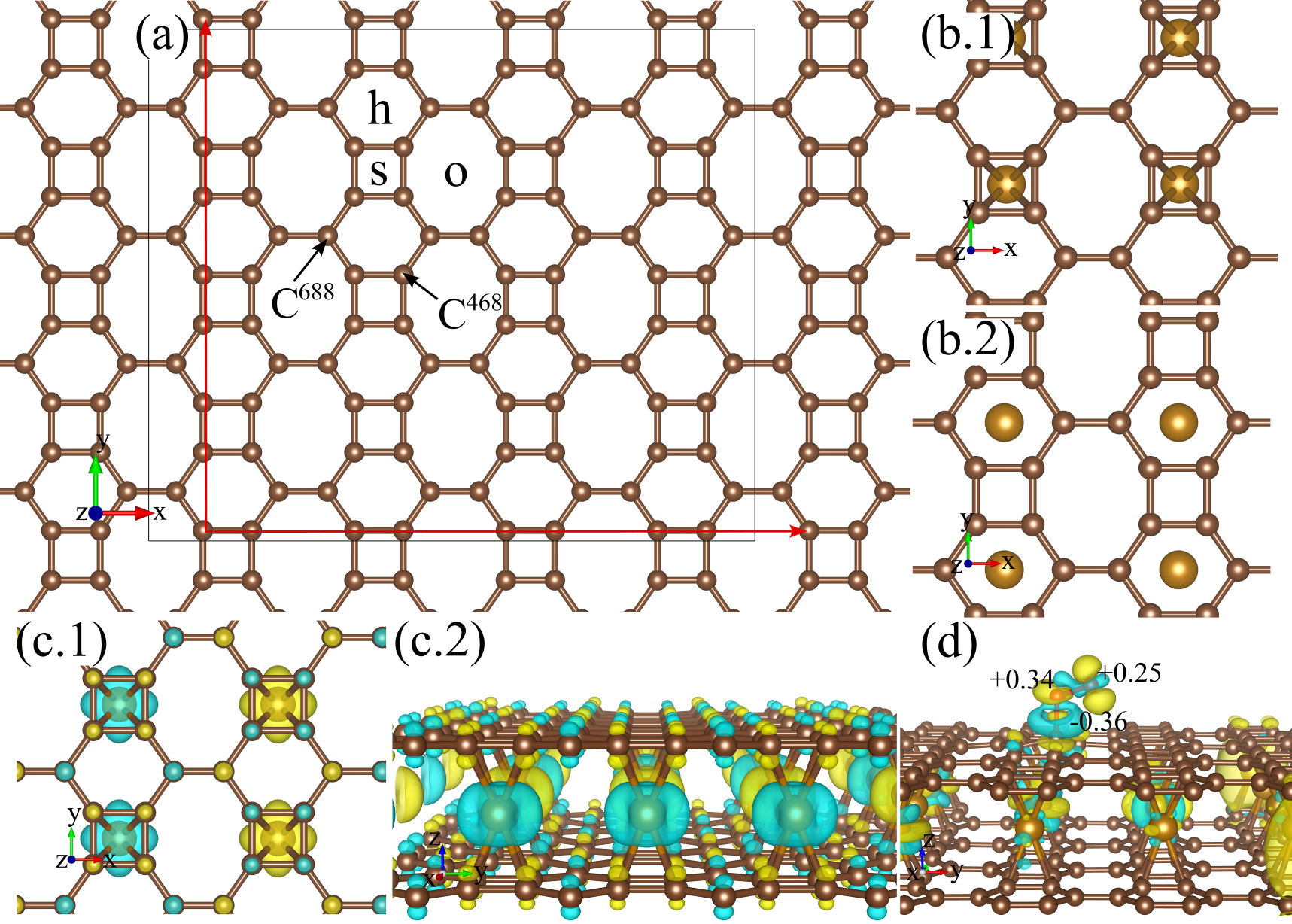}
\caption{(a) Top view of the equilibrium geometry of pristine biphenylene bilayer (B/B); h, s, and o represent intercalation sites for the 
metal atoms. C$^{468}$ indicates the carbon atom connecting the square, hexagonal, and octagonal sublattices, and C$^{688}$ 
indicates the carbon atom connecting hexagonal and octagonal sublattices; (b.1) B/Fe/B with Fe at the square-planar site, while (b.2) B/V/B at the 
hexagonal site; (c.1) shows the spin-density of B/Fe/B at the top view; (c.2) also shows the spin-density of B/Fe/B in the respective view. The yellow and blue densities represents the spin-up and spin-down densities, respectively. The magnetic configuration presented for B/Fe/B is AFx; (d) shows the charge density of B/Fe/B with O$_2$ adsorbed on C$^{468}$.
The isosurfaces of spin-density and charge density used was 0.01\,$e/$\AA$^3$ and 0.006\,$e/$\AA$^3$, respectively.}
\label{f1}
\end{figure*}

\section{\label{sec:Computational}Computational Details}

The calculations were performed using spin-polarized density functional theory (DFT) as implemented in the Vienna {\it Ab initio} Simulation 
Package (VASP)~\cite{Kresse1993}. Structural optimization and electronic structure calculations were carried out using the 
Perdew-Burke-Ernzerhof (PBE) exchange-correlation functional~\cite{paw1994,Perdew1996}. Dispersion interactions were accounted for 
using the vdW-DF2 functional~\cite{Lee2010}. The projector augmented-wave (PAW) method was employed to describe ion-electron 
interactions~\cite{paw1994}. Self-interaction effects associated with the $3d$ orbitals of transition metals were treated using the Hubbard $U$ correction. 
The $U$ parameters were determined for each transition metal via the linear response approach\cite{Cococcioni2005}, and the 
corresponding values are listed in Table\,T1 of the Supplemental Materials (SM)~\cite{SM}. A plane-wave energy cutoff of 500~eV was adopted 
based on previous studies~\cite{Mukesh2024,SAHOO2023}. The Brillouin zone was sampled using a $2\times2\times1$ Monkhorst-Pack 
$k$-point mesh~\cite{Monkhorst1976}, and ionic relaxations were performed with a force convergence criterion of 0.01~eV\AA$^{-1}$.

The biphenylene bilayer was modeled using a $4\times4$ surface unit cell in the AA stacking configuration, in which the interlayer region 
is intercalated with transition metal atoms, as shown in Fig.~\ref{f1}. The supercell has lattice parameters $\mathbf{a} = 18.07$~\AA\ and 
$\mathbf{b} = 15.26$~\AA, with a vacuum spacing of 15~\AA\ along the out-of-plane $\mathbf{c}$ direction. It contains 192 carbon atoms and 16 transition 
metal atoms. For the evaluation of electronic properties, a reduced $(1\times1)$ unit cell containing 48 carbon atoms and 4 metal atoms was 
employed. This choice represents the smallest system that accurately reproduces the magnetic phases under investigation. Further structural 
stability was examined by {\it ab initio} molecular dynamics (AIMD) simulations, performed in the NVT ensemble using a 
Nos\'e-Hoover thermostat approach at 400~K, with a total simulation time of 10~ps and a time step of 0.4~fs.

We calculated the C-$1s$ core-level binding energies ($BE$s) and the corresponding core-level shifts (CLSs) for the B/M/B systems using the VASP package~\cite{Kresse1993}. The calculations were performed within the $\Delta$ self-consistent field ($\Delta$SCF) approach~\cite{Deus2023}. Within this framework, the C-$1s$ core-level binding energy is defined as

\begin{equation}
BE = E^{(n-1)} - E^{(n)},
\end{equation}

where $E^{(n-1)}$ and $E^{(n)}$ are the total energies of the ionized system containing a C-$1s$ core hole and the ground-state system with $n$ electrons, respectively. The binding energies were computed for all crystallographically nonequivalent carbon atoms in the B/M/B structures. As a reference, we also calculated the C-$1s$ binding energies of pristine single-layer and AA-stacked biphenylene, denoted by $BE_0$. The corresponding core-level shifts were then obtained as

\begin{equation}
\Delta BE = BE - BE_0.
\end{equation}

To facilitate comparison with experiment, the calculated CLSs can be referenced to the measured C-$1s$ binding energy of pristine biphenylene, $BE_0^{\mathrm{exp}}\approx283.9$\,eV~\cite{Qitang2021}. Accordingly, the calculated absolute binding energies are given by

\begin{equation}
BE^{\mathrm{calc}} = \Delta BE + BE_0^{\mathrm{exp}}.
\end{equation}


With these values now comparable to experimental results, we also simulate the XPS spectra using the pseudo-Voigt approach~\cite{Ugolotti2017}, capturing the peaks corresponding to non-equivalent carbons in B/M/B. The parameters used in the convolution profiles are given by $\sigma = 0.38$ as Gaussian, $\gamma = 0.17$ as the Lorentzian width, and $\eta = 0.47$ as the mixing parameter of the two distributions~\cite{Ugolotti2017}.

The thermodynamic overpotentials for ORR and OER were evaluated within the computational hydrogen electrode (CHE) model proposed 
by N{\o}rskov and co-workers~\cite{Norskov2004}. In this approach, catalytic activity is described in terms of the adsorption strength of key 
intermediates (OH*, O*, and OOH*), quantified through their adsorption free energies on the active site, expressed as 
$\Delta G = \Delta E_{\text{DFT}} + \Delta E_{\text{ZPE}} - T\Delta S$. Here, $\Delta E_{\text{ZPE}}$ and $T\Delta S$ correspond to zero-point 
energy and entropic contributions, respectively, obtained from vibrational frequency calculations of the adsorbed species. Only the vibrational 
degrees of freedom of the intermediates were considered, while the catalyst surface was kept fixed. Thermodynamic and vibrational properties 
of gas-phase reference molecules were taken from the NIST database~\cite{nist}.

Solvation effects on adsorbate binding strength (OH* and OOH*), were included by accounting for hydrogen-bonding interactions. The 
corresponding corrections were computed for each structure using the continuum dielectric model implemented in 
VASPsol~\cite{VASPsol-Software, VASPsol2014-Dielectric, VASPsol2019-Electrolyte}. Post-processing and visualization were carried out 
using VASPKIT~\cite{WANG2021} and VESTA~\cite{Momma2008}. The biphenylene bilayer (B/B) has been theoretically predicted in 
AA\cite{Chowdhury2022}, AB, and AX stacking configurations~\cite{Lage2024}, all of which exhibit thermomechanical stability. 

\section{\label{sec:Results}Results}

\subsection{Energetic and Structural Properties}\label{EandS}

We begin determining the ground-state configurations of transition metals embedded in biphenylene bilayers by placing the metal atoms at the 
square (s), hexagonal (h), and octagonal (o) interlayer sites [Fig.~\ref{f1}(a)]. Owing to the partially filled $3d$ and $4d$ shells of most 
transition metals (M), four magnetic configurations 
($M_{\text{phase}}$) are considered at each site: ferromagnetic (FM), antiferromagnetic (AF), and stripe antiferromagnetic states AF-$x$ and AF-$y$. The latter 
are defined by antiferromagnetic (ferromagnetic) coupling along the $x$ ($y$) direction for AF-$x$, and along the $y$ ($x$) direction for AF-$y$.

The structural stability of the metal-intercalated biphenylene bilayers (B/M/B) is assessed via the intercalation energy $E^i_j$, defined as
\begin{equation}\label{eq:int}
E^i_j = \frac{1}{N_\mathrm{M}}\left(E_j[\mathrm{B/M}_j\mathrm{/B}] - E[\text{B/B}] - N_\text{M}E[\text{M}]\right),
\end{equation}
where $E_j[\mathrm{B/M}_j\mathrm{/B}]$ and $E[\mathrm{B/B}]$ are the total energies of the intercalated system (for $j =$ s, h, o) and the pristine bilayer, respectively. For each 
geometry, the lowest-energy magnetic phase is considered. Here, $E[\text{M}]$ is the energy of an isolated metal atom and $N_\text{M}$ is the number of intercalates. 
Negative $E^i_j$ indicates exothermic intercalation. We obtain that the most of intercalated metals preferentially occupy the square sites, B/M$_\mathrm{s}$/B [Fig.~\ref{f1}(b.1)], whereas Nb, V, and Ti favor hexagonal sites, B/M$_\mathrm{h}$/B [Fig.~\ref{f1}(b.2)]. The corresponding intercalation energies and ground-state magnetic phases ($M_{\text{phase}}$) are summarized in Table~\ref{t1}. Additional data for other geometries are provided in the SM in Tables T2--T5~\cite{SM}.

\begin{table}[ht]
\centering
\caption{Intercalation energy (in eV/M-atom) of the metal atoms between the square ($E^i_\text{s}$), octagonal ($E^i_\text{o}$), and hexagonal ($E^i_\text{h}$) 
sites of the biphenylene bilayer, and the (respective) magnetic phase ($M_\text{phase}$). 
The binding energies (in eV) of O$_2$ ($E_\mathrm{O_2}^b$) adsorbed on C$^{468}$ are listed in the rightmost column.}
\label{t1}
\begin{ruledtabular}
\begin{tabular}{lclccr}
M & $E^i_\text{s}$ & $M_\text{phase}$ & $E^i_\text{h}$& $E^i_\text{o}$ & $E_\mathrm{O_2}^b$ \\
\hline
Cr       & $-0.81$ & AF-$x$  & $-0.20$ &$-0.13$& $-0.30$\\
Mn       & $-2.45$ & AF & $-1.63$ &$-1.39$ & $-0.64$\\ 
Fe       & $-1.64$ & AF-$x$ & $-1.18$ &$-0.85$& $-0.57$\\
Co       & $-1.67$ &  AF & $-1.10$ &$-0.50$& $-0.35$\\
Ni       & $-1.70$ & AF  & $-0.74$ &$-0.69$ & $-0.48$\\
Cu       & $-0.87$ & FM & $-0.17$ &$-0.53$ & $-0.15$\\
Ru       & $-1.60$ & AF-$x$ & $-0.88$  &$-0.29$& $-0.48$\\
W        & $-3.92$ & AF & $-2.12$ &$-2.07$ & $+0.49$\\
Os       & $-4.19$ & FM & $-3.42$&$-2.88$ & $-0.04$\\
Pt       & $-3.04$ & FM & $-1.16$ &$-2.79$ & $-0.40$\\
\hline
M & $E^i_\text{h}$ & $M_\text{phase}$ & $E^i_\text{s}$& $E^i_\text{o}$& $E_\mathrm{O_2}^b$\\
\hline
Ti     & $-2.46$& AF & $-2.28$ &   $-2.26$ & $-0.76$\\
V      & $-2.47$& FM  & $-2.31$ &   $-1.77$ & $-1.42$\\
\end{tabular}
\end{ruledtabular}
\vspace{4pt}
\end{table}

\begin{figure}
\centering
\includegraphics[width=0.8\linewidth]{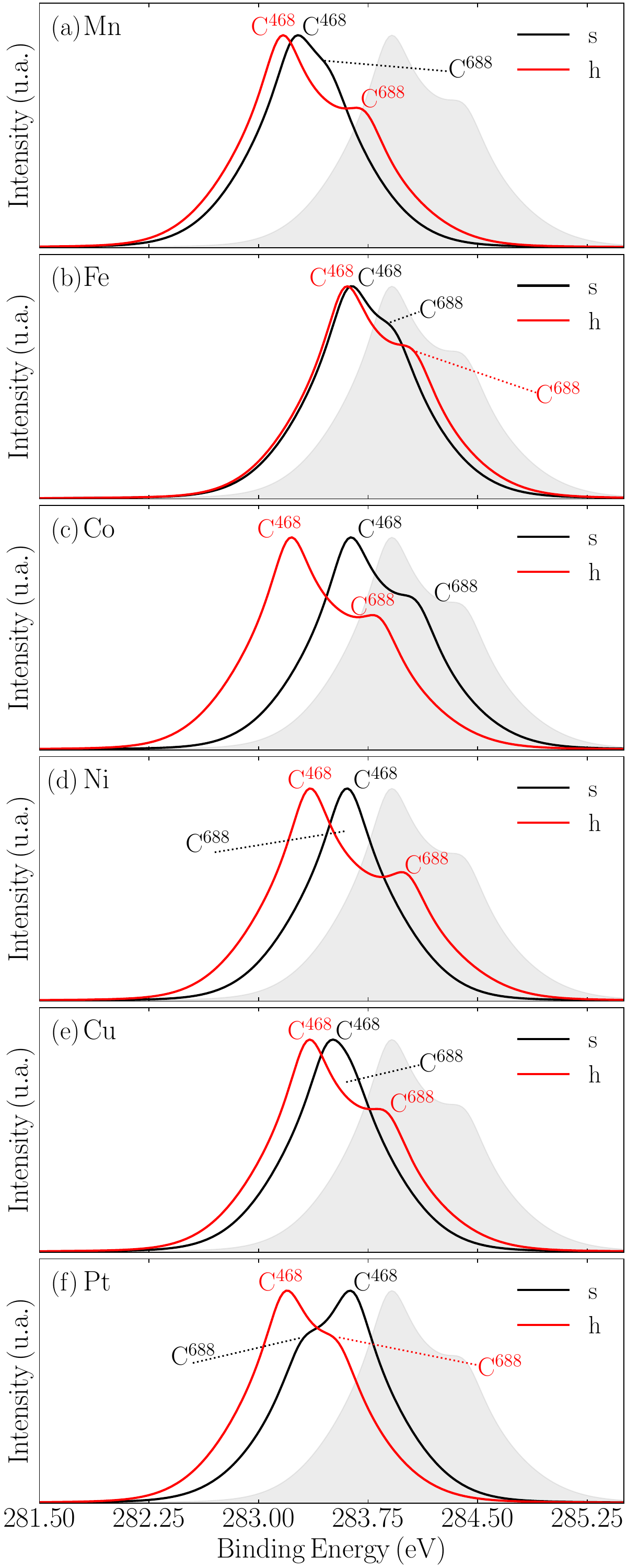}
\caption{Simulated XPS spectra for B/M/B with M = Mn, Fe, Co, Ni, Cu, and Pt, for metals intercalated in the s configuration (black curve) and h configuration (red curve). The region in gray indicates the experimental peak position~\cite{Qitang2021} and the convolution XPS spectra simulation for the biphenylene monolayer.}
\label{f_cls}
\end{figure}

The intercalated metal atoms form chemical bonds with the carbon atoms of the biphenylene bilayer, where the coordination environment depends on the intercalation site. At the square site, the intercalated metal atom ($\mathrm{M_s}$) exhibits fourfold coordination with the C$^{468}$ atoms. In contrast, at the hexagonal site ($\mathrm{M_h}$), the intercalated metal atom forms two bonds with the C$^{688}$ atoms and four bonds with the C$^{468}$ atoms. Bader charge analysis~\cite{bader} further reveals that the net charge transfer from the intercalated metal to the biphenylene bilayer also depends on the intercalation site. In particular, the amount of charge transferred to the C$^{468}$ and C$^{688}$ atoms differs between the $\mathrm{M_s}$ and $\mathrm{M_h}$ configurations. The corresponding equilibrium structural parameters and Bader charges are summarized in Table~T5 and T6 of the Supplemental Material.

Next, we examine the dependence of the C-$1s$ core-level binding energies on the metal intercalation site. As a reference, we first calculate the C-$1s$ binding energies of pristine biphenylene [Eq.~(3)]. The simulated XPS spectrum, shown as the shaded region in Fig.~\ref{f_cls}, exhibits two distinct binding-energy ($BE$) peaks. The lower-$BE$ peak is associated with the fourfold-coordinated C$^{468}$ atoms (283.90,eV), whereas the higher-$BE$ peak originates from the sixfold-coordinated C$^{688}$ atoms (284.42,eV), resulting in a binding-energy difference of $\Delta(\mathrm{C^{468}-C^{688}})=-0.52$\,eV.

Metal intercalation systematically shifts the C-$1s$ binding energies toward lower values relative to pristine biphenylene, resulting in negative core-level shifts (CLSs). This behavior is consistent with the net charge transfer from the intercalated metal atoms (M$_j$) to the biphenylene bilayer, as revealed by the Bader charge analysis. The calculated CLSs of the C$^{468}$ atoms (CLS$^{468}$) for the B/M$_\mathrm{s}$/B and B/M$_\mathrm{h}$/B structures, together with the corresponding binding-energy separations, $\Delta(\mathrm{C^{468}-C^{688}})$, are summarized in Table~\ref{t_cls}. Representative simulated C-$1s$ XPS spectra for B/M$_j$/B systems with M$_j$ = Mn, Fe, Co, Ni, Cu, and Pt are presented in Figs.~\ref{f_cls}(a)–(f). The spectra reveal that both the magnitude of the CLSs and the binding-energy separation between the C$^{468}$ and C$^{688}$ atoms depend on the chemical identity of the intercalated metal and its coordination site within the biphenylene bilayer, being systematically larger in B/M$_\mathrm{h}$/B. These findings demonstrate that the B/M$_j$/B systems exhibit distinct XPS fingerprints and provide valuable guidance for future experimental spectroscopic characterization.

\begin{table}[]
\caption{Calculated core-level shifts of C$^{468}$ with respect to the pristine biphenylene bilayer, CLS$^{468}$, and the C-$1s$ binding energy difference between the C$^{668}$ and C$^{468}$ carbon atoms of B/M/B with the metal intercalated at the square (M$_\mathrm{s}$) and hexagonal (M$_\mathrm{h}$) sites, $\Delta(\mathrm{C^{688}-C^{468}})$. The energies are in eV.}
\label{t_cls}
\begin{ruledtabular}
\begin{tabular}{lcccc}
\multicolumn{1}{l}{M$_j$} &
\multicolumn{2}{c}{B/M$_\mathrm{s}$/B} &
\multicolumn{2}{c}{B/M$_\mathrm{h}$/B} \\
\cline{2-3} \cline{4-5}
&  CLS$^{468}$ & $\Delta(\mathrm{C^{468}-C^{688}})$  
& CLS$^{468}$ & $\Delta(\mathrm{C^{468}-C^{688}})$  \\
\hline
Cr & $-0.63$ & $-0.19$ & $-0.77$ & $-0.52$\\
Mn & $-0.65$ & $-0.24$  & $-0.74$ & $-0.58$\\
Fe  & $-0.28$& $-0.32$ & $-0.31$ & $-0.48$\\
Co & $-0.28$ & $-0.47$  & $-0.68$ & $-0.62$\\
Ni & $-0.31$ & $-0.05$  & $-0.55$ & $-0.67$\\
Cu & $-0.42$ & $-0.16$ & $-0.56$ & $-0.55$\\
Ru & $-0.43$ & $-0.07$  & $-0.64$ & $-0.41$\\
W   & $-0.59$  & $-0.03$& $-0.63$ & $-0.64$\\
Os & $-0.91$ & $+0.04$  & $-0.68$ & $-0.26$\\
Pt & $-0.25$ &$+0.36$  & $-0.72$ & $-0.37$\\
Ti & $-0.80$ & $-0.40$ & $-0.63$ & $-0.63$\\
V  & $-0.64$ & $-0.28$ & $-0.63$ & $-0.60$\\
\end{tabular}
\end{ruledtabular}
\end{table}


\subsection{ORR and OER Activity of B/M/B}\label{catal}

Single-atom catalysts (SACs) have demonstrated remarkable catalytic activities for both the ORR and OER, often surpassing those of commercially available benchmark catalysts~\cite{Sai2024,walter2025}. However, their practical application is frequently limited by insufficient structural stability under operating conditions, where migration, aggregation, or detachment of the metal atom from the supporting material can lead to the progressive degradation of the catalytic performance~\cite{Zhao2025,DiLiberto2024,Tsipoaka2025,Kumar2020,Liu2022,Bae2023}.

In contrast, the metal atoms in {\bpm} are fully encapsulated between the two biphenylene layers, providing an effective confinement that hinders metal migration and detachment and is therefore expected to enhance the long-term structural stability of the catalyst. Moreover, charge analysis reveals electron transfer from the intercalated metal atoms to the biphenylene framework, leading to a more electron-rich carbon network~\cite{Li2023}. Such charge redistribution is expected to modify the C-$2p$ electronic states and, consequently, tune the adsorption strength of ORR/OER intermediates, thereby enhancing the catalytic activity~\cite{Zhou2021}.

\subsubsection{Reaction Free Energies of ORR/OER intermediates}


We begin by examining the O$_2$ adsorption on the biphenylene bilayer encapsulated by transition metals,

\begin{equation}
    \mathrm{O}_2 + \mathrm{B/M/B} \rightarrow \mathrm{O}_2^\ast, 
\end{equation}

The carbon atoms forming the square sublattice [C$^{468}$ in Fig.~\ref{f1}(a)] are identified as the active sites because they exhibit stronger O$_2$ adsorption than the C$^{688}$ atoms and possess a higher local electron density (see Figs.~S3 and S4 of the Supplemental Material~\cite{SM}), favoring the adsorption of ORR/OER reaction intermediates on the biphenylene framework~\cite{Luo2021}. For example, in B/Fe$_\mathrm{s}$/B, the calculated O$_2$ binding energies are $-0.57$ and $-0.06$,eV for adsorption at the C$^{468}$ and C$^{688}$ sites, respectively. The optimized adsorption geometry of O$_2$ on the C$^{468}$ site, together with the corresponding charge transfer upon adsorption, is shown in Fig.~\ref{f1}(d).

The calculated O$2$ binding energies ($E_{\mathrm{O}_2}^{b}$) at the C$^{468}$ active sites for all B/M/B systems are summarized in Table~\ref{t1}. With the exception of B/Nb/B and B/W/B\,\footnote{The positive $E{\mathrm{O}_2}^{b}$ values obtained for B/Nb/B and B/W/B originate from the pronounced displacement of the intercalated Nb and W atoms from their equilibrium square-site positions upon O$_2$ adsorption. This structural distortion significantly increases the total energy of the adsorbed system, as illustrated in Fig.~S2 of the Supplemental Material~\cite{SM}.} all systems exhibit negative O$_2$ binding energies, indicating that O$_2$ adsorption on the C$^{468}$ active sites is thermodynamically favorable.

\begin{figure*}
    \centering
    \includegraphics[width=1\linewidth]{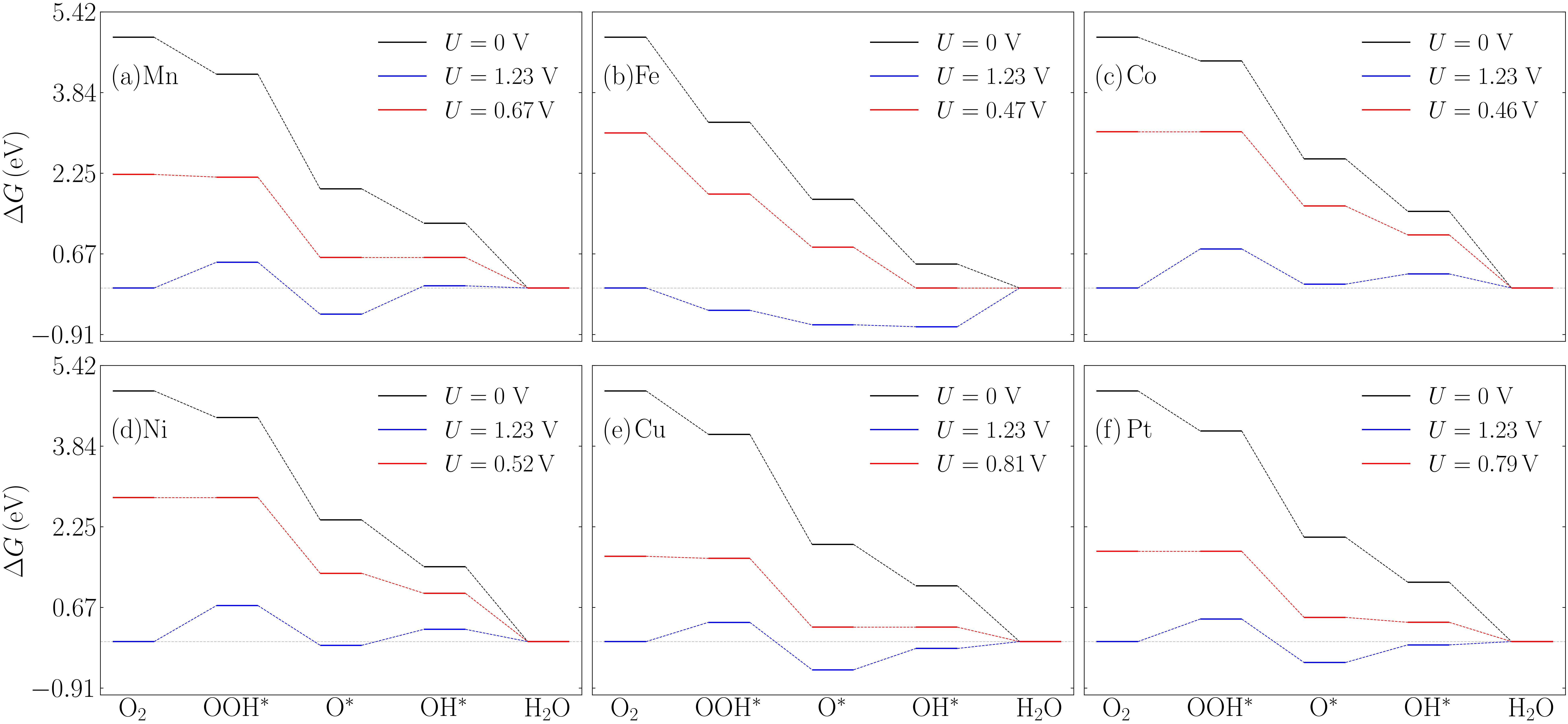}
    \caption{Reaction free energy diagrams for the elementary steps of the ORR on the B/M/B, for M $=$ Mn, Fe, Co, Ni, Cu and Pt, respectively in panels (a)-(f), at different applied potentials: $U = 0$\,V (black), $U = 1.23$\,V (blue), and the onset potential $U = U_{onset}$\,(red). All energies are referenced to standard conditions.}
    \label{f2}
\end{figure*}
Focusing on the ORR and OER processes, in an alkaline solution, the former can be described as the following steps,
\begin{eqnarray}
    \mathrm{O_2^\ast+H_2O}+e^-&\leftrightarrows&\mathrm{OOH^\ast+OH^-} \\
    \mathrm{OOH^\ast}+e^-&\leftrightarrows&\mathrm{O^\ast+OH^-}\\
    \mathrm{O^\ast+H_2O}+e^-&\leftrightarrows&\mathrm{OH^\ast+OH^-} \\
    \mathrm{OH^\ast}+e^-&\leftrightarrows&\mathrm{OH^-}+\ast.
\end{eqnarray}
Likewise, in an alkaline solution, the OER steps proceed in a reverse order.

We evaluate the ORR and OER activities of all {\bpm} systems using the computational hydrogen electrode (CHE) model developed by N{\o}rskov {\it et al}.~\cite{Norskov2004}. Within this framework, the adsorption free energies of the key reaction intermediates, namely $\mathrm{OOH}^\ast$, $\mathrm{O}^\ast$, and $\mathrm{OH}^\ast$, are calculated following Ref.~\cite{Xu2024} as

\begin{equation}
\Delta G = \Delta E_\mathrm{DFT} + E_\mathrm{ZPE} - T\Delta S + \Delta G_\mathrm{solv},
\end{equation}

where $\Delta E_\mathrm{DFT}$ is the adsorption energy obtained from density functional theory, $E_\mathrm{ZPE}$ is the zero-point energy correction, $T\Delta S$ is the entropic contribution derived from vibrational frequency calculations of the adsorbed species (with the catalyst surface kept fixed), and $\Delta G_\mathrm{solv}$ is the solvation correction. Further details of the free-energy calculations and the ORR/OER mechanisms are provided in Sec.~III of the Supplemental Material~\cite{SM}.

Through the Gibbs free-energy profiles of the elementary ORR and OER reaction steps, we determine the onset potentials, i.e., the minimum (maximum) electrode potential at which all ORR (OER) elementary steps become thermodynamically downhill with respect to the Gibbs free energy, $\Delta G$~\cite{kulkarni2018understanding}. The onset potentials for ORR and OER, denoted by $U_\mathrm{onset}(\mathrm{ORR})$ and $U_\mathrm{onset}(\mathrm{OER})$, respectively, are used to calculate the corresponding overpotentials ($\eta$), 
\begin{eqnarray}
    \eta_\mathrm{ORR} &=& U_0 - U_\mathrm{onset}(\mathrm{ORR}) \\
    \eta_\mathrm{OER} &=& U_\mathrm{onset}(\mathrm{OER}) - U_0,
\end{eqnarray}
where $U_0=1.23 \mathrm{V}$ is the standard electrode equilibrium potential.

\begin{figure*}
    \centering
    \includegraphics[width=1\linewidth]{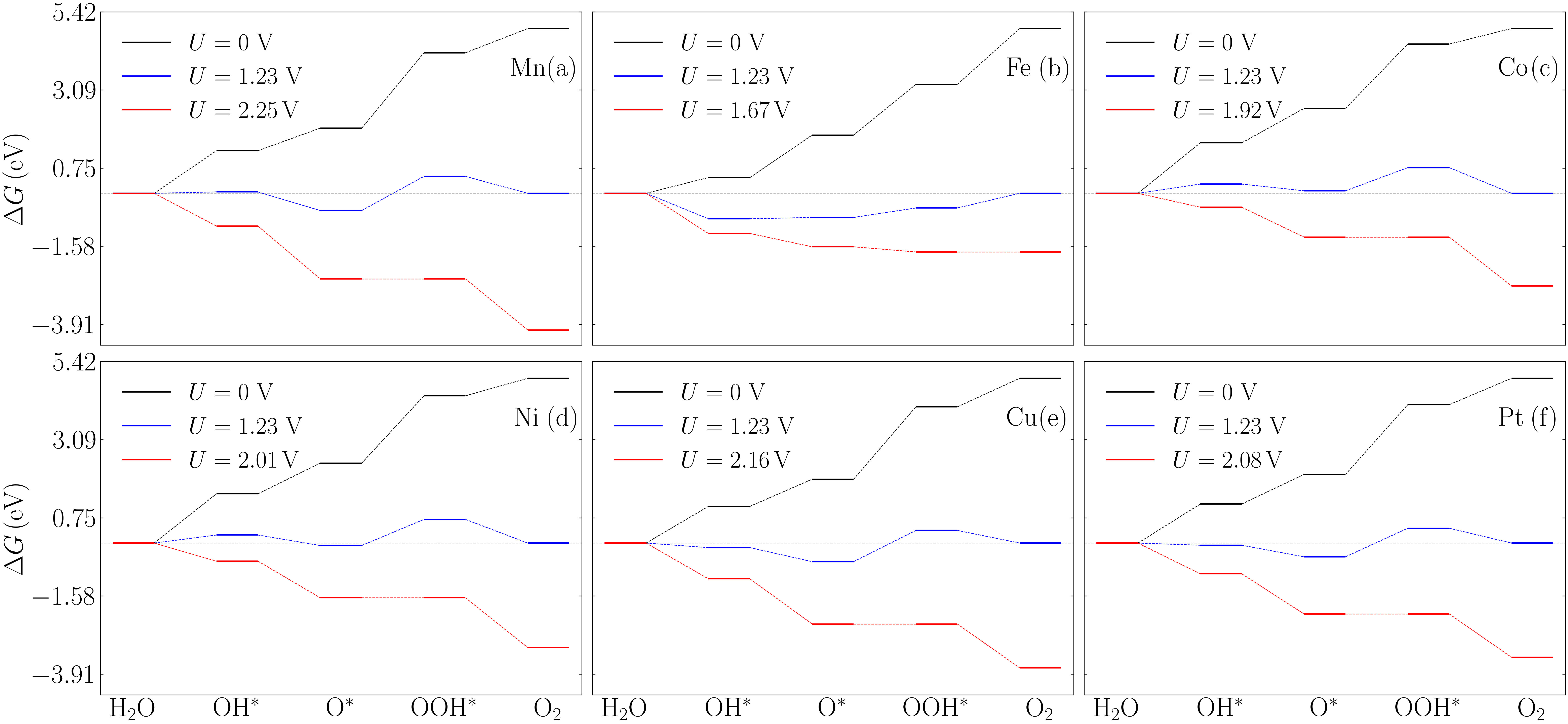}
    \caption{Reaction free energy diagrams for the elementary steps of the OER on the B/M/B, for M $=$ Mn, Fe, Co, Ni, Cu and Pt, respectively in panels (a)-(f), at different applied potentials: $U = 0$\,V (black), $U = 1.23$\,V (blue), and the onset potential $U = U_{onset}$\,(red). All energies are referenced to standard conditions.}
    \label{f3}
\end{figure*}

The Gibbs free-energy diagrams for the ORR and OER on the B/M$_\mathrm{s}$/B systems with M $=$ Mn, Fe, Co, Ni, Cu, and Pt (Figs.~\ref{f2} and \ref{f3}) show that (i) at zero electrode potential relative to the reversible hydrogen electrode ($U=0$), the Gibbs free-energy profiles exhibit a cascade-like behavior, characteristic of efficient ORR/OER catalysts ~\cite{Norskov2004, Xu2024}, whereas (ii) at the equilibrium potential ($U=U_0$), the rate-determining step depends on the intercalated metal. Specifically, the potential-limiting step is $\mathrm{O^\ast}\rightarrow\mathrm{OH^\ast}$ for Mn and Cu, $\mathrm{OH^\ast}\rightarrow\mathrm{H_2O}$ for Fe, and $\mathrm{O_2}\rightarrow\mathrm{OOH^\ast}$ for Co, Ni, and Pt. The Gibbs free-energy diagrams for the remaining B/M/B systems are provided in Figs.~S5--S10 of the Supplemental Material (SM)~\cite{SM}.


The calculated thermodynamic onset (limiting) potentials and corresponding overpotentials for the ORR ($\eta_{\mathrm{ORR}}$) and OER ($\eta_{\mathrm{OER}}$), summarized in Table~\ref{t2}, reveal that the B/M$_\mathrm{s}$/B systems with M $=$ Mn, Fe, Co, Ni, Cu, Ru, and Pt exhibit ORR catalytic activities comparable to that of the theoretical Pt(111) benchmark. Notably, B/Cu/B and B/Pt/B outperform Pt(111), exhibiting $\eta_{\mathrm{ORR}}$ values of 0.42 and 0.44,V, respectively. Likewise, the OER activities of the B/M$_\mathrm{s}$/B systems with M $=$ Fe, Co, Ni, and Pt are comparable to that of the theoretical IrO$2$ benchmark. Among them, B/Fe$_\mathrm{s}$/B stands out with an overpotential of $\eta_{\mathrm{OER}} = 0.44$\,V, which is substantially lower than that of IrO$_2$. The Gibbs free-energy profiles at the equilibrium potential ($U = 1.23$\,V, blue curves) and at the corresponding onset potential ($U = U_\mathrm{onset}$, red curves) are also shown for both reactions in Figs.~\ref{f2} and \ref{f3}. These results indicate that, beyond protecting against structural degradation through the shielding effect of the bilayer architecture~\cite{Kumar2023}, M-intercalated biphenylene bilayers (B/M/B) constitute a promising class of  electrocatalysts for both the ORR and OER~\cite{Wang2018,Li2025}.

\begin{table}[]
\caption{Calculated ORR and OER onset potentials ($U_\text{onset}$, in V) and overpotentials ($\eta$, in V) in {\bpm} for all M with $E_\mathrm{O_2}^b<0$. The benchmark Pt(111)\cite{Hansen2008} and IrO$_2$\cite{Xu2024} are included for comparison.}
\label{t2}
\begin{ruledtabular}    
\begin{tabular}{lrccc}
\multicolumn{1}{l}{M} &
\multicolumn{2}{c}{ORR} &
\multicolumn{2}{c}{OER} \\
  \cline{2-3} \cline{4-5} 
 & $U_\text{onset}$ & $\eta $ & $U_\text{onset}$ & $\eta$  \\
 \hline
Ti         & $ -1.96$ & $3.19$  & $2.94$&$1.71$\\
V         & $ -1.00$ & $2.23$  & $2.69$&$1.46$\\
Cr         & $ -0.32$ & $1.55$  & $2.81$&$1.58$\\
Mn         & $ 0.67$ & $0.56$  & $2.25$&$1.02$\\
Fe         & $ 0.47$ & $0.76$  & $1.67$&$0.44$\\
Co         & $ 0.46$ & $0.77$  & $1.92$&$0.69$\\
Ni         & $ 0.52$ & $0.71$  & $2.01$&$0.78$\\
Cu         & $ 0.81$ & $0.42$  & $2.16$&$0.93$\\
Ru         & $ 0.73$ & $0.50$  & $2.18$&$0.95$\\
Os         & $ -3.11$ & $4.34$  & $5.93$&$4.70$\\
Pt         & $ 0.79$ & $0.44$  & $2.08$&$0.85$\\
\hline
Pt($111$) & $0.75$ & $0.48$ & - & -\\
IrO$_2$($110$) & - & - & $1.88$ & $0.65$ \\
\end{tabular}
\end{ruledtabular}
\end{table}

\begin{figure}
    \centering
    \includegraphics[width=1\linewidth]{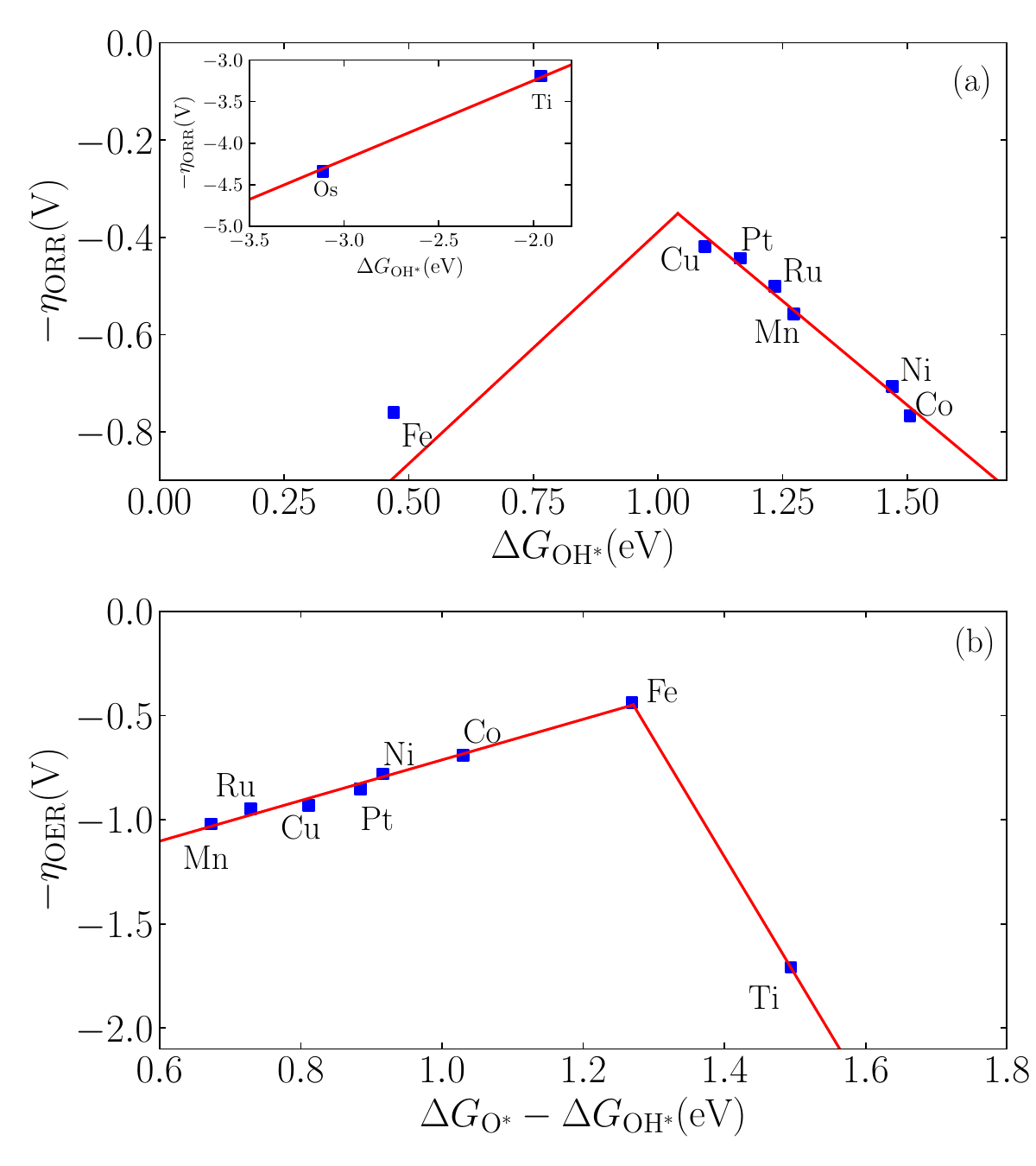}
    \caption{Volcano plots of the overpotentials for ORR and OER as a function of their respective universal descriptors, where panel (a) shows $-\eta (\mathrm{ORR})$ against $\Delta G_\mathrm{OH^\ast}$, while panel (b) shows $-\eta(\mathrm{OER})$ against $\Delta G_\mathrm{O^\ast} - \Delta G_\mathrm{OH^\ast}$.}
    \label{f4}
\end{figure}

To further assess the catalytic performance of the B/M/B systems, we compare our calculated overpotentials with predictions based on established adsorption free-energy descriptors. Figure~\ref{f4}(a) presents the ORR overpotential as a function of $\Delta G_{\mathrm{OH}^\ast}$, whereas Fig.~\ref{f4}(b) shows the OER overpotential as a function of $\Delta G_{\mathrm{O}^\ast}-\Delta G_{\mathrm{OH}^\ast}$, the universal activity descriptor for the OER~\cite{walter2025,deng2017robust}. The resulting volcano plots indicate that the minimum overpotentials are obtained for $\Delta G_{\mathrm{OH}^\ast}\approx1.04$\,eV in the ORR and $\Delta G_{\mathrm{O}^\ast}-\Delta G_{\mathrm{OH}^\ast}\approx1.27$\,eV in the OER. These optimal descriptor values are consistent with previous theoretical studies~\cite{Qiao2024,walter2025}.

Regarding structural stability, the calculated intercalation energies, $E^{i}_{j}$, indicate that the insertion of metal atoms between the square lattices of biphenylene is an exothermic process. The intercalation energies range from $-0.81$ to $-3.04$,eV per metal atom for B/Cr/B and B/Pt/B, respectively. Focusing on the B/M/B systems identified as promising ORR/OER electrocatalysts, the calculated intercalation energies range from $-0.87$\,eV per metal atom for B/Cu/B to $-1.70$,eV per metal atom for B/Ni/B.

\begin{figure}
\centering
\includegraphics[width=1\linewidth]{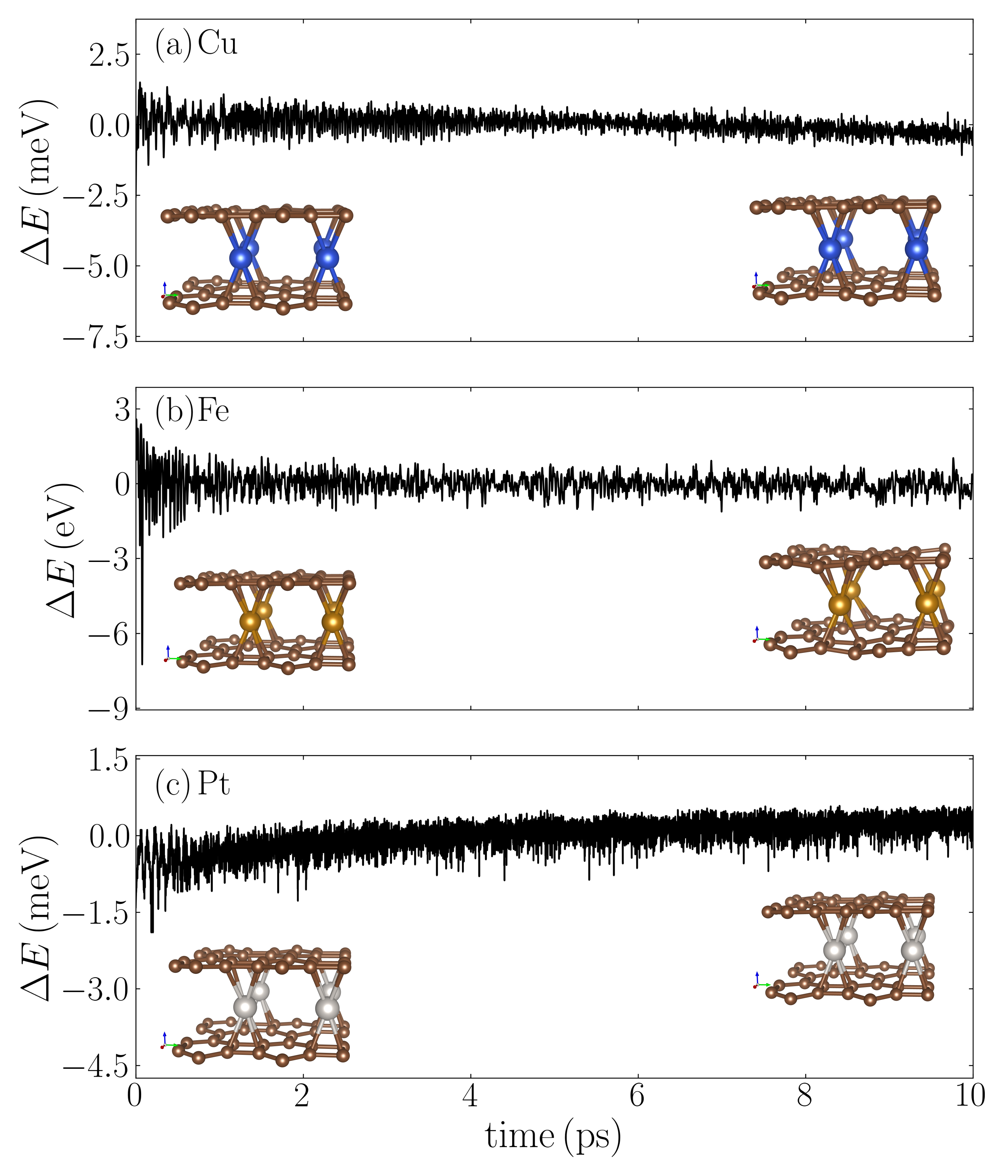}
\caption{Energy variation as a function of time of the AIMD simulation for the B/Cu/B (painel (a)), B/Fe/B (painel (b)) and B/Pt/B (painel (c)) systems at 400~K. Insets show ball-and-stick 
representations of the atomic configurations at the initial stage (left) and after 10~ps (right), and time steps of 0.4~fs.}
\label{f5}
\end{figure}

The thermal stability of the B/M/B systems was further assessed by \textit{ab initio} molecular dynamics (AIMD) simulations performed at 400~K for 10~ps. This temperature was chosen to approximate the operating conditions of proton exchange membrane fuel cells, which typically operate in the 350--400~K temperature range and involve both ORR and OER processes~\cite{OTHMAN2012}. As shown in Fig. ~\ref{f5}, the intercalated metal atoms (M = Cu, Fe, and Pt) remain stably confined within the biphenylene bilayer throughout the AIMD simulations, with no evidence of metal diffusion or clustering, confirming the structural stability of the B/M/B systems. The corresponding AIMD results for the other B/M/B structures, presented in Fig. S12 of the Supplemental Material~\cite{SM}, exhibit similar behavior, demonstrating that the metal atoms remain mostly at the squared site between the biphenylene layers during the entire simulation.

Additionally, we performed an AIMD simulation for the B/Fe/B system in the presence of water at its experimental density (1\,g\,cm$^{-3}$). After water stabilization, the simulation was carried out at 400\,K for 5\,ps using a time step of 1\,fs. As shown in Fig. S13 of the Supplemental Material~\cite{SM}, no reactions between the water molecules and the biphenylene surface were observed throughout the simulation. These results indicate that the biphenylene bilayer effectively acts as a protective barrier, shielding the intercalated metal centers from direct interaction with the aqueous environment where ORR/OER occur.

\subsubsection{Electronic Properties and catalytic-activity descriptors}
Next, we analyze the electronic structure of the B/M/B systems to establish correlations between their catalytic activity and their electronic properties. We show in detail the band structures, density of states (DOS), and projected density of states (PDOS) for all B/M/B systems in Section IV in SM~\cite{SM}.

We first consider the relative positions of the $d$ orbitals of M$_\mathrm{s}$ and the $p_z$ orbitals of C$^{468}$. These positions are quantified by the $d-$ and $p_z-$band centers, respectively, as defined in Eqs.~\ref{eq.d-band} and \ref{eq.p-band}.
The $d-$band center~\cite{LIU2026, zhou2023_bi} is given by 
\begin{equation}\label{eq.d-band}
\varepsilon_d = \frac{\int_{-\infty}^{E_{\text{vac}}} \varepsilon g_d(\varepsilon) d\varepsilon}{\int_{-\infty}^{E_{\text{vac}}} g_d(\varepsilon) d\varepsilon},
\end{equation}
where $g_d(\varepsilon)$ is the spin-dependent PDOS projected onto the M-$d$ orbitals. The Kohn-Sham eigenvalues ($\varepsilon$) are chosen in relation to the Fermi level.
This relation can be used as a descriptor for ORR~\cite{LIU2026} and OER~\cite{zhou2023_bi}, where the range of $\varepsilon_d$ should be $-3 \,\mathrm{eV} \leq \varepsilon_d \leq -1$\,eV, in relation to the Fermi level, to minimize the overpotentials in Metal-Organic Framework systems as verified in Ref.\cite{walter2025}.

For metal-free catalysts~\cite{zhou2018heterostructures,Zhou2021, xie2023}, the energy position of the $p-$orbitals has also been used as a descriptor for ORR and OER overpotentials~\cite{Zhou2021, zhou2018heterostructures,xie2023,GUO2024_pband}. In this context, the energy position of the $p_z$ band 
center ($\varepsilon_{p_z}$), defined as
\begin{equation}\label{eq.p-band}
\varepsilon_{p_z} = \frac{\int_{-\infty}^{E_{\text{vac}}} \varepsilon g_{p_z}(\varepsilon) d\varepsilon}{\int_{-\infty}^{E_{\text{vac}}} g_{p_z}(\varepsilon) d\varepsilon}.
\end{equation}
Here, $g_{p_z}(\varepsilon)$ is the PDOS projected onto the C$^{468}$-$2p_z$ orbitals for each spin channel. Therefore, we calculated $\varepsilon_d$ and $\varepsilon_{p_z}$ for all B/M/B systems and summarized these results in Table~\ref{t3}.


\begin{figure}
\centering
\includegraphics[width=1.\linewidth]{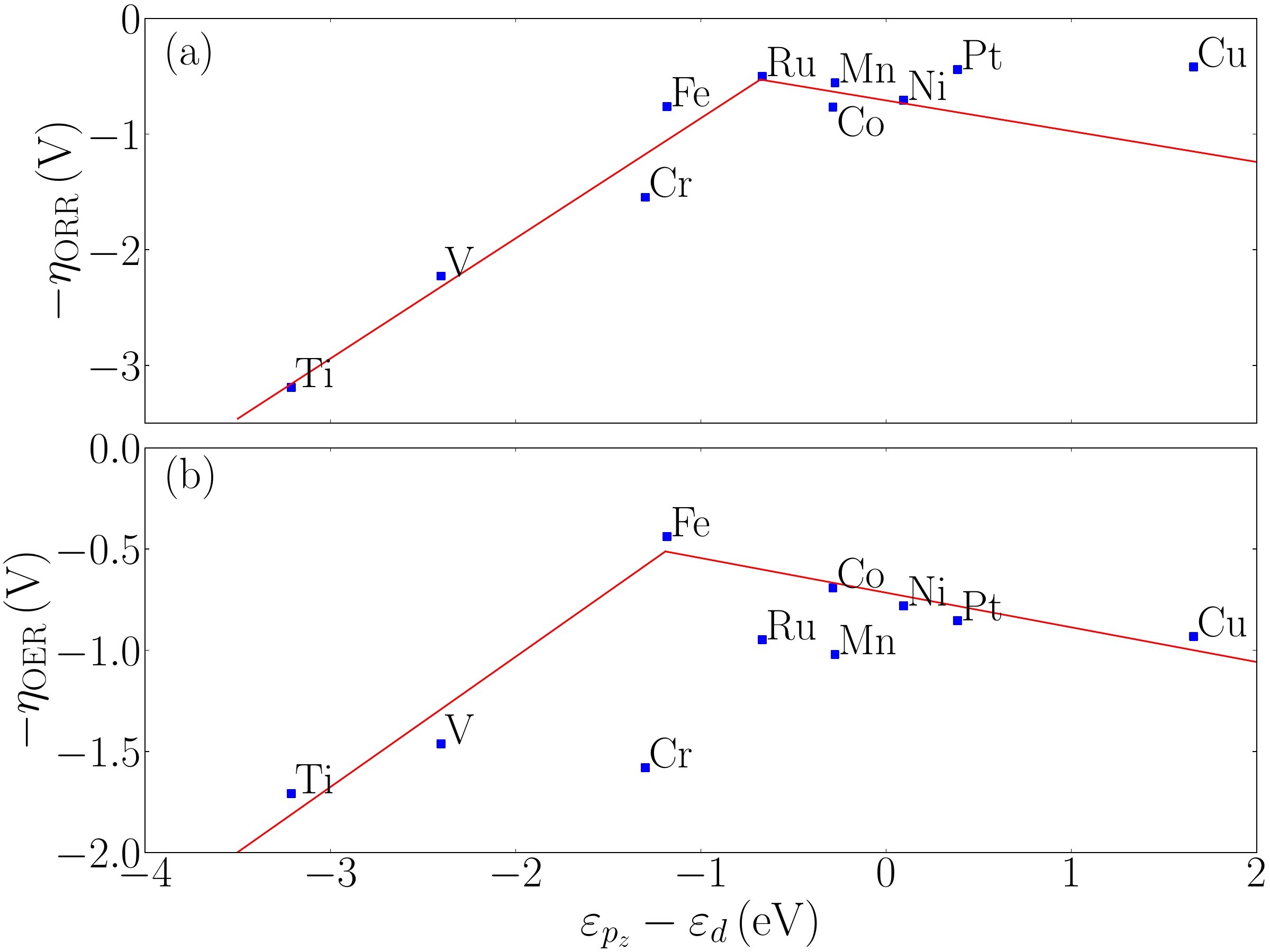}
\caption{ORR and OER overpotentials, respectively in panels (a) and (b), as a function of the difference between the $p_z-$band center and the $d-$band center for each B/M/B with $\eta_\mathrm{ORR} < 3.5$\,eV and $\eta_\mathrm{OER} < 2.0$\,eV are indicated by blue squares. The $p_z-$band center and the $d-$band center are calculated relative to the Fermi level.}
\label{f6}
\end{figure}

Neither $\eta_\mathrm{ORR}$ nor $\eta_\mathrm{OER}$ exhibits a clear linear dependence on the $p_z-$ or $d-$band center. The B/M/B systems span a narrow $\varepsilon_{p_z}$ range of $[-2.8,-2.4]$\,eV, indicating that the C$^{468}$ $p_z-$orbitals remain largely localized despite changes in the intercalated metal. In contrast, the $d$-band center exhibits a more discernible relation with catalytic activity: the M = Mn, Fe, Co, Ni, Ru, and Pt systems fall within the optimal range reported for oxygen electrocatalysis, $[-3,-1]$\,eV~\cite{walter2025}. B/Cu/B is a notable exception, with $\varepsilon_d\approx-4.5$\,eV while exhibiting ORR and OER overpotentials comparable to those of Pt(111)~\cite{Hansen2008} and IrO$_2$~\cite{Xu2024}, respectively.

\begin{table}[]
\caption{Calculated $p_z-$ and $d-$band centers for each B/M/B systems, in relation to the Fermi level, the difference between $p_z-$ and $d-$band centers (both in eV), and charge population of $d-$orbital from intercalated metal and $p-$orbital from C$^{468}$ (in $e^-$) for all B/M/B.}
\label{t3}
\begin{ruledtabular}    
\begin{tabular}{lrcccc}
\multicolumn{1}{l}{M} &
\multicolumn{3}{c}{Band center} &
\multicolumn{2}{c}{Charge population} \\
  \cline{2-4} \cline{5-6} 
 & $\varepsilon_{p_z}$ & $\varepsilon_d$ & $\varepsilon_{p_z}-\varepsilon_d$ & $p-$orbital & $d-$orbital  \\
 \hline
Ti         & $ -2.89$ & $+0.32$ &$-3.21$  & $1.760$&$2.581$\\
V         & $-2.85$ & $-0.45$ & $-2.40$ & $1.758$&$3.534$\\
Cr         & $-2.96$ & $-1.66$& $-1.30$ & $1.752$&$4.498$\\
Mn         & $-2.63$ & $-2.35$& $-0.28$ & $1.752$&$5.062$\\
Fe         & $-2.95$ & $-1.76$& $-1.18$ & $1.753$&$6.571$\\
Co         & $-2.73$ & $-2.75$& $-0.29$ & $1.748$&$7.457$\\
Ni         & $-2.68$ & $-2.78$& $+0.09$ & $1.744$&$8.835$\\
Cu         & $-2.75$ & $-4.41$& $+1.66$ & $1.740$&$9.415$\\
Nb         & $-2.90$ & $+0.09$& $-2.99$ & $1.752$&$5.062$\\
Ru         & $-2.70$ & $-2.03$& $-0.67$ & $1.747$&$6.579$\\
W         & $-3.32$ & $-1.12$ & $-2.19$ & $1.758$&$4.174$\\
Os         & $-3.45$ & $-2.08$& $-1.37$ & $1.753$&$6.300$\\
Pt         & $-2.53$ & $-2.91$& $+0.39$ & $1.734$&$8.080$\\
\end{tabular}
\end{ruledtabular}
\end{table}

Although the $d$-band center is a useful descriptor for transition-metal active sites~\cite{Li2025,LIU2026}, its applicability to B/M/B systems is less direct because the active site is C$^{468}$. We therefore examine the relative positions of the $d-$ and $p_z-$band centers, as variations in their relative energy have been shown to correlate with enhanced catalytic activity~\cite{Liu2023,Bai2026,Wang2025_delta}. Figure~\ref{f6} shows $\eta_\mathrm{ORR}$ and $\eta_\mathrm{OER}$ as functions of ($\varepsilon_{p_z}-\varepsilon_d$) in panels (a) and (b), respectively.

For both reactions, the overpotentials exhibit a piecewise-linear dependence on $(\varepsilon_d-\varepsilon_{p_z})$, resulting in a volcano-like relationship consistent with the Sabatier principle~\cite{Norskov2004}. For ORR, shown in panel (a), B/Cu/B and B/Pt/B were excluded from the linear fits because they deviate markedly from the trend followed by the remaining B/M/B systems. For OER [panel (b)], the M = Cr, Mn, and Ru systems were similarly excluded. After these exclusions, the corresponding $R^2$ values~\cite{DiBucchianico2008} for the left and right branches are $0.95$ and $0.45$ for ORR and $0.95$ and $0.83$ for OER, respectively. The resulting volcano peaks occur at B/Ru/B for ORR, with $(\varepsilon_d-\varepsilon_{p_z})\approx-0.68$\,eV, and at B/Fe/B for OER, with $(\varepsilon_d-\varepsilon_{p_z})\approx-1.19$\,eV. These results suggest that the relative position of the $d-$ and $p_z-$band centers may serve as a descriptor of catalytic activity in B/M/B systems.

\begin{figure*}
\centering
\includegraphics[width=1\linewidth]{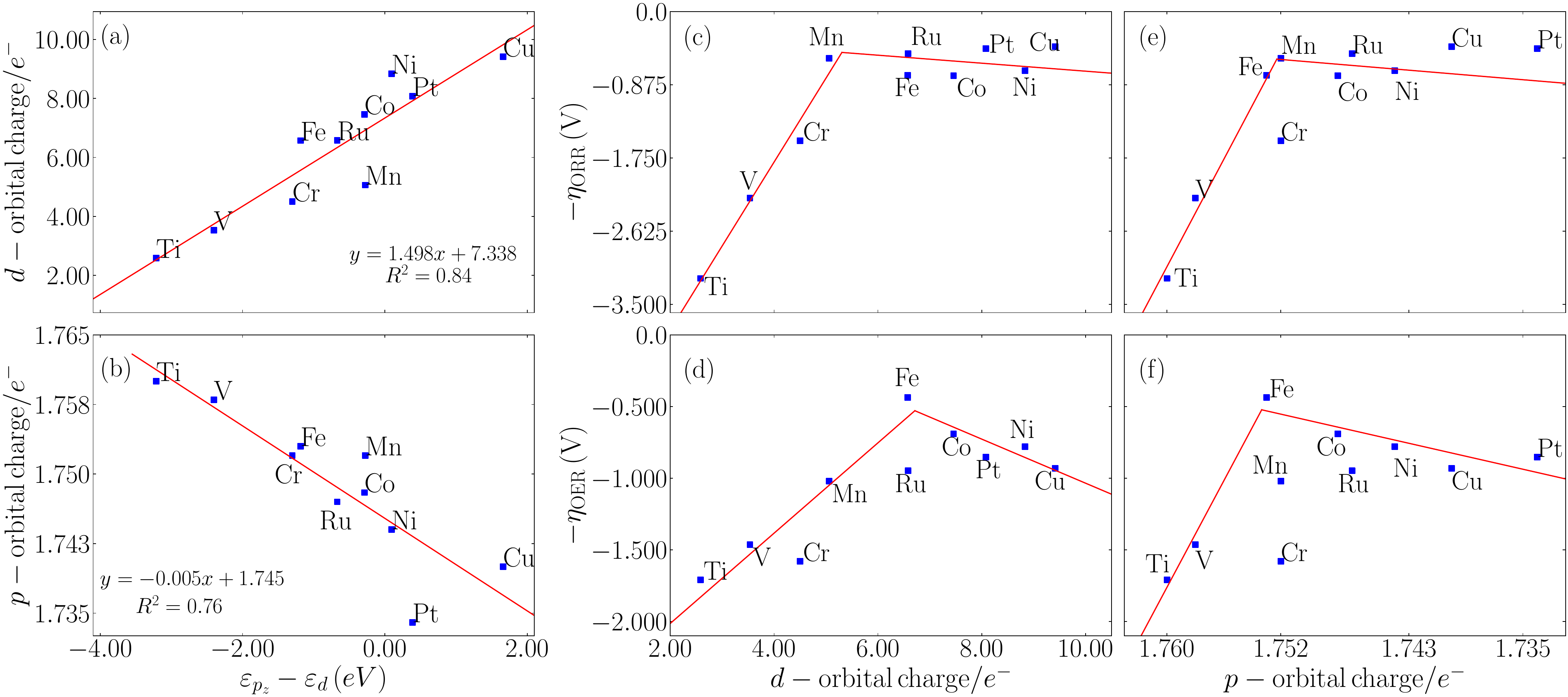}
\caption{Charge populations of the $d$ and $p$ orbitals as a function of $(\varepsilon_d-\varepsilon_{p_z})$ are shown in panels (a) and (b), respectively. The B/M/B systems are the same as those considered in Figs.~\ref{f5} and \ref{f6}. Each system is represented by a blue square, while the red lines denote linear fits to the corresponding charge populations. The ORR and OER overpotentials as a function of the $d-$orbital charge population of the intercalated metal are presented in panels (c) and (d), respectively. B/M/B systems with $-\eta_\mathrm{ORR}<-3.5$,eV and $-\eta_\mathrm{OER}<-2.5$\,eV are highlighted by blue squares. The corresponding ORR and OER overpotentials as a function of the $p-$orbital charge population are shown in panels (e) and (f), respectively, with the same systems highlighted. The red lines denote linear fits of $\eta_\mathrm{ORR}$ and $\eta_\mathrm{OER}$ as functions of the orbital charge population, revealing volcano-like relationships for both reactions.
}
\label{f_rel}
\end{figure*}

Considering the volcano-like dependence on $(\varepsilon_d-\varepsilon_{p_z})$ shown in Fig.~\ref{f6}, we next examine how this electronic descriptor is related to the orbital charge populations. As shown in Figs.~\ref{f_rel}(a) and \ref{f_rel}(b), the charge populations of the metal $d-$orbitals and the C$^{468}$ $p-$orbitals exhibit a nearly linear dependence on $(\varepsilon_d-\varepsilon_{p_z})$. The charge populations were obtained by projecting the Kohn-Sham wave functions onto atom-centered orbitals~\cite{vaspwiki_lorbit}. This linear relationship establishes a direct connection between the relative alignment of the metal $d$ states and the C$^{468}$ $p_Z$ states and their corresponding orbital occupations, thereby motivating the use of orbital charge populations as alternative electronic descriptors for catalytic activity.

Notably, in Fig.~\ref{f_rel}, both the ORR and OER overpotentials exhibit volcano-like dependences on the charge populations of the $d$ orbitals [panels (c) and (d)] and $p$ orbitals [panels (e) and (f)]. Table~\ref{t3} summarizes the charge populations of the $d-$orbitals of the intercalated metals and the $p-$orbitals of C$^{468}$. For the $d-$orbital charge population, the ORR volcano peak occurs at $\approx 5.31\,e^-$, with $R^2=0.97$ and $0.10$ for the left and right branches, respectively. For OER, the peak is located at $\approx 6.51\,e^-$, with $R^2=0.85$ and $0.82$ for the left and right branches, respectively.

In panel (c), B/Mn/B lies closest to the volcano summit. Although B/Cu/B exhibits a lower overpotential than B/Mn/B, it deviates from the volcano-like trend and was therefore excluded from the corresponding linear fits. In panel (d), B/Fe/B lies at the volcano summit, corresponding to the most favorable catalytic performance for OER. The B/Ru/B system was excluded from the linear fits because its $d-$orbital charge population is close to that of B/Fe/B but exhibits a significant deviation from the corresponding trend.

Similarly, in panel (e), the ORR volcano summit occurs at a $p-$orbital charge population of $\approx 1.752\,e^-$, with B/Mn/B located at the optimum. For OER [panel (f)], the summit occurs at $\approx 1.753\,e^-$, with B/Fe/B at the optimum. To maximize the $R^2$ values of the linear fits, three systems were excluded in each case. For ORR, B/Cr/B, B/Cu/B, and B/Pt/B were excluded, yielding $R^2\approx0.99$ and $0.15$ for the left and right branches, respectively. For OER, B/Cr/B, B/Mn/B, and B/Ru/B were excluded, resulting in $R^2\approx0.99$ and $0.75$ for the left and right branches, respectively. The origin of these deviations from the trends observed for the remaining B/M/B systems warrants further investigation.

These results establish a direct relationship between the relative positions of the $p_z-$ and $d-$band centers and the charge populations of the corresponding orbitals. They further suggest that the $d-$ and $p-$orbital charge populations may serve as qualitative descriptors of catalytic activity in B/M/B systems, as both exhibit volcano-like relationships with the ORR and OER overpotentials, consistent with the Sabatier principle.

\section{\label{sec:Conclusion}Conclusions}

We performed an {\it ab initio} investigation of the catalytic activity (ORR and OER) in metal-encapsulated biphenylene bilayers, B/M/B, with M = Os, W, Pt, Fe, Ni, Co, Ru, Cu, Cr, Mn, Nb, V, and Ti. Except for the last three elements, we find that metal intercalation occurs between the carbon atoms forming the square sites in biphenylene, C$^{468}$.

The evaluation of the ORR and OER processes at the surface C$^{468}$ sites, based on the theoretical framework developed by N{\o}rskov {\it et al.} \cite{Norskov2004}, reveals that the M-intercalated biphenylene bilayer (B/M/B) exhibits catalytic performance competitive for ORR and OER. 
The catalytic activity was quantified through overpotentials ($\eta$), and for B/M/B systems, the $\eta$ follows the usual linear relationships with $\Delta G_\mathrm{OH^\ast}$ for ORR and $\Delta G_\mathrm{O^\ast}-\Delta G_\mathrm{OH^\ast}$ for OER.

The analysis of the electronic structure of the B/M/B systems showed that the relative position of the $d-$band center with respect to the $p-$band center exhibits a tendency to describe the Sabatier principle, particularly for OER. The occurrence of a volcano-plot-like behavior of the overpotentials and $(\varepsilon_{p_z} - \varepsilon_d)$ indicates an intrinsic correlation between the catalytic activity and the electronic structure, modulated by the chemical environment, of the B/M/B systems. Additionally, we show that $(\varepsilon_{p_z} - \varepsilon_d)$ is directly related to the charge populations of the $p-$ and $d-$orbitals.

Furthermore, the metal $d-$orbital charge population exhibits a more robust correlation with the overpotentials, particularly for OER, enabling the construction of volcano-like relationships. These trends identify B/Fe/B as the optimal system for OER and B/Mn/B as the system closest to the ORR optimum, further supporting the use of simple electronic descriptors to rationalize catalytic activity in B/M/B systems.

The $p-$orbital charge population of the active C$^{468}$ site also exhibits volcano-like relationships with the overpotentials, yielding the same optimal systems for ORR and OER. However, compared with the metal $d-$orbital charge population, the $p-$orbital charge exhibits weaker correlations, requiring the exclusion of more B/M/B systems and yielding lower $R^2$ values. Thus, while the active-site $p-$orbital charge captures general trends in catalytic activity, the metal $d-$orbital charge population provides a more robust electronic descriptor for rationalizing and predicting the catalytic performance of B/M/B systems.

These findings suggest that such electronic features constitute key design principles for nonmetallic catalysts mediated by the encapsulation of metallic elements.

\begin{acknowledgments}
The authors acknowledge the Laborat\'orio Nacional de Computa\c{c}ao Cient\'ifica (LNCC-SCAFMat2), Centro Nacional de Processamento de Alto Desempenho (CENAPAD-SP) for computer time. HGM and RHM acknowledge financial support from CNPq, INCT-Nanocarbono, INCT-Materials Informatics, Coordena\c{c}ao de Aperfei\c{c}oamento de Pessoal de Ensino Superior (CAPES). WO and PHS acknowledge financial support from the National Agency for Research and Development of Chile (ANID) through project FONDECYT 1230138 and 
FONDECYT POSTDOC 3240440.
\end{acknowledgments}

\bibliography{apssamp}
\end{document}


\title{Supplementary Material: Understanding the Oxygen Reduction Reaction and Oxygen Evolution Reaction in Metal Intercalated Biphenylene Bilayers}

\author{Henri G. Mendonça}
\email{irneh.gm@gmail.com}
\affiliation{Instituto de Física, Universidade Federal de Uberlândia, 38400-902 Uberlândia, Minas Gerais, Brazil}

\author{Pedro H. Souza}
\email{psouza8628@gmail.com}
\affiliation{Departamento de Física y Astronomía,
Facultad de Ciencias Exactas, Universidad Andres Bello,
Santiago 8370136, Chile}

\author{Walter Orellana}
\email{worellana@unab.cl}
\affiliation{Departamento de Física y Astronomía,
Facultad de Ciencias Exactas, Universidad Andres Bello,
Santiago 8370136, Chile}

\author{Roberto H. Miwa}
\email{hiroki@ufu.br}
\affiliation{Instituto de Física, Universidade Federal de Uberlândia, 38400-902 Uberlândia, Minas Gerais, Brazil}

\maketitle

\section{Structural Details}
In the table below, we show the $U$ values in eV for all B/M/B, calculated by linear response according to Ref.\,\cite{Cococcioni2005}. For this calculation, we use the ($1\times 1$) supercell, with 48 carbon atoms and 4 M atoms and the most stable magnectic phase, where the geometry used is with M on C$^{468}$ biphenylene.

\begin{table}[H]
\centering
\caption{The $U$ values calculated by linear response \cite{Cococcioni2005} for transitions metals.}
\label{tab:u_two_rows}
\begin{ruledtabular}
\begin{tabular}{cccccccccccccc}
TM & Cr & V & Mn & Os & W & Nb & Ti & Fe & Pt & Co & Ni & Cu & Ru \\
\hline
$U$ &3.59 & 3.17 & 4.85 & 1.01 & 0.82 & 1.15 & 2.97 & 3.29 & 1.65 & 3.53 & 5.50 & 4.20 & 3.74 \\
\end{tabular}
\end{ruledtabular}
\end{table}

\begin{figure}[H]
    \centering
    \includegraphics[width=0.99\linewidth]{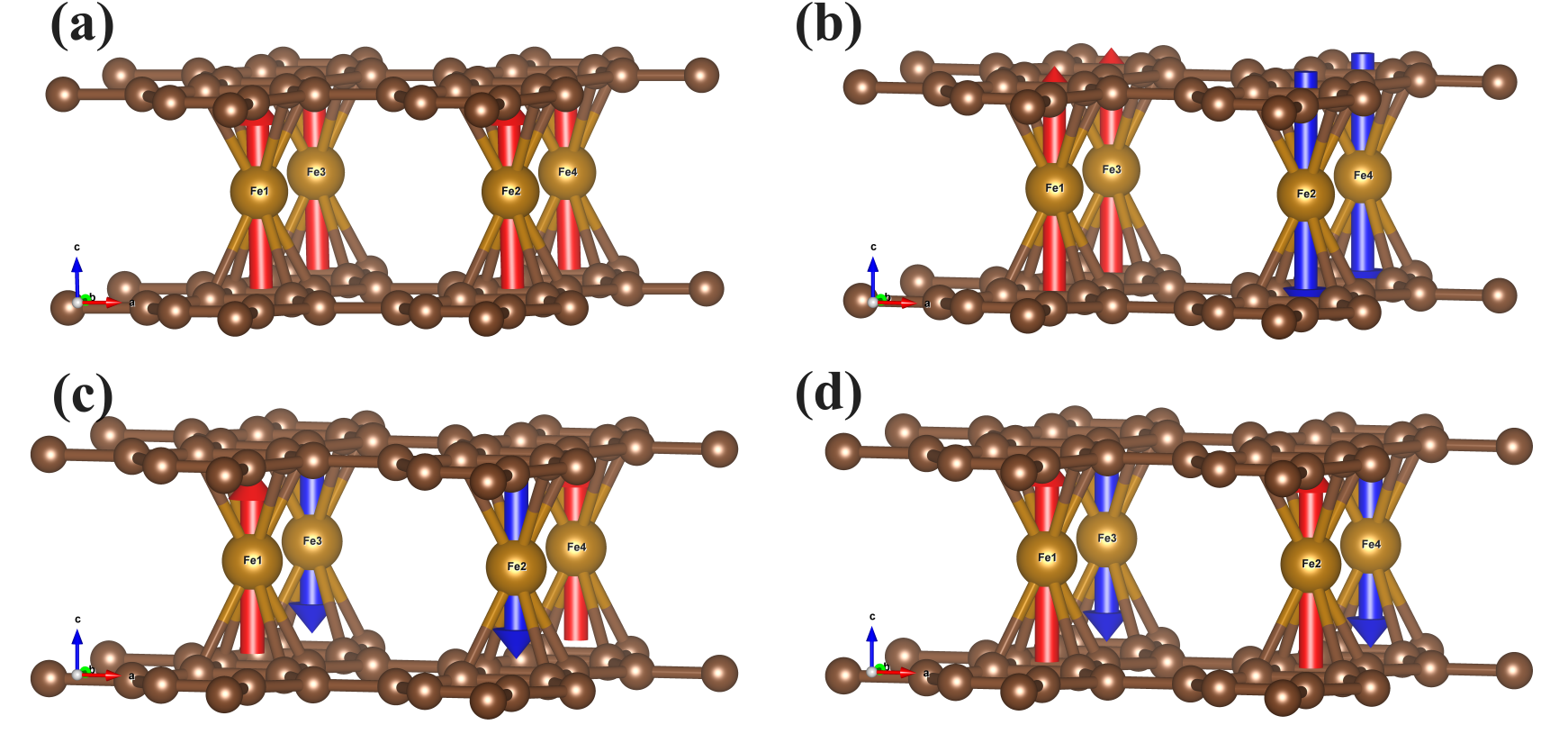}
    \caption{We present in panels (a)-(d) the respective magnetic phases FM, AF-y, AF, and AF-x, after convergence, for the B/Fe/B system. For B/Fe/B, the most stable magnetic phase at the C$^{468}$ site is AF-x.}
    \label{fig:fig_mag}
\end{figure}

We calculated the most stable magnetic phase for each of the B/M/B systems by considering four possible magnetic configurations: FM, AF-x, AF, and AF-y, as represented in panels (a)-(d) of Fig.\,\ref{fig:fig_mag} for the B/Fe/B system.

TABLEs T.2-T.4 compile, for the four considered magnetic orders FM, AF-y, AF, and AF-x, the energetic and magnetic results for each of the three structural geometries (s, h, and o). For each case, we list the energy difference relative to the lowest-energy configuration, defined as $\Delta E^j = E^j - E^j_{stable}$, in eV per cell; positive values indicate configurations that are less stable relative to the minimum. Additionally, we report the local magnetic moments, in $\mu B$, of the four transition metals atoms in the cell. Reference configurations ($\Delta E = 0$) have been highlighted in red in the tables.

\begin{table}[H]
\centering
\begin{ruledtabular}
\begin{tabular}{l|rrr|rrrr|rrrr|rrrr}
\textbf{Co} & s & h & o & \multicolumn{4}{c}{$m_\text{s}$} & \multicolumn{4}{c}{$m_\text{h}$} & \multicolumn{4}{c}{$m_\text{o}$} \\ \hline
FM    & 0.666 & 0.196 & \hgm{0.000} & 1.7& 1.7& 1.7& 1.7& 2.0& 2.0& 2.0& 2.0& 1.9& 1.9& 1.9& 1.9\\ 
AF-y & 0.254 & 0.601 & 0.060 & 1.6& 1.6& -1.6& -1.6& 2.0& 2.0& -2.0& -2.0& 2.0& 2.0& -2.0& -2.0\\ 
AF    & \hgm{0.000} & 0.542 & 0.081 & 1.7& -1.7& -1.7& 1.7& 2.0& -2.0& -2.0& 2.0& 2.0& -2.0& -2.0& 2.0\\ 
AF-x & 0.154 & \hgm{0.000} & 0.059 & 1.7& -1.7& 1.7& -1.7& 1.9& -1.9& 1.9& -1.9& 1.9& -1.9& 1.9& -1.9\\ 
\hline
\textbf{Cr} & s & h & o & \multicolumn{4}{c}{$m_\text{s}$} & \multicolumn{4}{c}{$m_\text{h}$} & \multicolumn{4}{c}{$m_\text{o}$} \\ \hline
FM    & 0.093 & 0.024 & 0.107 & 4.1& 4.1& 4.1& 4.1& 4.2& 4.2& 4.2& 4.2& 4.4& 4.4& 4.4& 4.4\\ 
AF-y & 0.259 & \hgm{0.000} & 0.089 & 3.8& 3.8& -3.8& -3.8& 4.1& 4.1& -4.1& -4.1& 4.4& 4.4& -4.4& -4.4\\ 
AF    & 0.687 & 0.529 & 0.083 & 3.7& -3.7& -3.7& 3.7& 4.1& -4.1& -4.1& 4.1& 4.4& -4.4& -4.4& 4.4\\ 
AF-x & \hgm{0.000} & 0.015 & \hgm{0.000} & 3.9& -3.9& 3.9& -3.9& 4.2& -4.2& 4.2& -4.2& 4.4& -4.4& 4.4& -4.4\\ 
\hline
\textbf{Cu} & s & h & o & \multicolumn{4}{c}{$m_\text{s}$} & \multicolumn{4}{c}{$m_\text{h}$} & \multicolumn{4}{c}{$m_\text{o}$} \\ \hline
FM    & \hgm{0.000} & 7$\times$10$^{-5}$ & \hgm{0.000} & 0.0& 0.0& 0.0& 0.0& 0.0& 0.0& 0.0& 0.0& 0.0& 0.0& 0.0& 0.0\\ 
AF-y & 1$\times$10$^{-4}$ & 8$\times$10$^{-5}$ & 5$\times$10$^{-4}$ & 0.0& 0.0& 0.0& 0.0& 0.0& 0.0& 0.0& 0.0& 0.0& 0.0& 0.0& 0.0\\ 
AF    & 2$\times$10$^{-5}$ & 8$\times$10$^{-5}$ & 5$\times$10$^{-5}$ & 0.0& 0.0& 0.0& 0.0& 0.0& 0.0& 0.0& 0.0& 0.0& 0.0& 0.0& 0.0\\ 
AF-x & 5$\times$10$^{-5}$ & \hgm{0.000} & 3$\times$10$^{-4}$ & 0.0& 0.0& 0.0& 0.0& 0.0& 0.0& 0.0& 0.0& 0.0& 0.0&0.0 &0.0 \\ 
\hline
\textbf{Fe} & s & h & o & \multicolumn{4}{c}{$m_\text{s}$} & \multicolumn{4}{c}{$m_\text{h}$} & \multicolumn{4}{c}{$m_\text{o}$} \\ \hline
FM    & 0.588 & 0.190 & \hgm{0.000} & 1.7& 1.7& 1.7& 1.7& 3.0& 3.0& 3.0& 3.0& 3.3& 3.3& 3.3& 3.3\\ 
AF-y & 0.555 & 0.911 & 0.241 & 1.7& 1.7& -1.7& -1.7& 2.7& 2.7& -2.7& -2.7& 3.1& 3.1& -3.1& -3.1\\ 
AF    & 0.683 & 0.895 & 0.278 & 1.5& -1.5& -1.5& 1.5& 2.7& -2.7& -2.7& 2.7& 3.2& -3.2& -3.2& 3.2\\ 
AF-x & \hgm{0.000} & \hgm{0.000} & 0.236 & 2.1& -2.1& 2.1& -2.1& 2.9& -2.9& 2.9& -2.9& 3.3& -3.3& 3.3& -3.3\\ 
\hline
\textbf{Mn} & s & h & o & \multicolumn{4}{c}{$m_\text{s}$} & \multicolumn{4}{c}{$m_\text{h}$} & \multicolumn{4}{c}{$m_\text{o}$} \\ \hline
FM    & 0.041 & \hgm{0.000} & 0.018 & 4.2& 4.2& 4.2& 4.2& 4.5& 4.5& 4.5& 4.5& 4.5& 4.5& 4.5& 4.5\\ 
AF-y & 0.123 & 0.015 & 0.106 & 4.3& 4.3& -4.3& -4.3& 4.5& 4.5& -4.5& -4.5& 4.6& 4.6& -4.6& -4.6\\ 
AF    & \hgm{0.000} & 0.356 & 0.047 & 4.2& -4.2& -4.2& 4.2& 4.4& -4.4& -4.4& 4.4& 4.6&-4.6 &-4.6 &4.6 \\ 
AF-x & 0.196 & 0.359 & \hgm{0.000} & 4.2& -4.2& 4.2& -4.2& 4.4& -4.4& 4.4& -4.4& 4.6& -4.6& 4.6& -4.6\\ 
\end{tabular}

\caption{Relative energy difference ($\Delta E$, eV per cell) between the FM, AF-y, AF, and AF-x magnetic configurations for the s, h, and o geometries of M $=$ Co, Cr, Cu, Fe and Mn, along with the local magnetic moments of the four M atoms ($m_\text{s}$, $m_\text{h}$, and $m_\text{o}$, in $\mu B$). The $\Delta E = 0$ indicates the most stable configuration; values in red highlight the most stable geometry for each magnetic phase.}
\end{ruledtabular}
\end{table}

\begin{table}[H]
\centering
\begin{ruledtabular}
\begin{tabular}{l|rrr|rrrr|rrrr|rrrr}
\textbf{Nb} & s & h & o & \multicolumn{4}{c}{$m_\text{s}$} & \multicolumn{4}{c}{$m_\text{h}$} & \multicolumn{4}{c}{$m_\text{o}$} \\ \hline
FM    & 0.209 & 0.361 & 0.117 & 1.1& 1.1& 1.1& 1.1& 1.1& 1.1& 1.1& 1.1& 0.7& 0.7& 0.7& 0.7\\ 
AF-y & 0.080 & 0.189 & 0.074 & 1.3& 1.3& -1.3& -1.3& 1.1& 1.1& -1.1& -1.1& 1.0& 1.0& -1.0& -1.0\\ 
AF    & 0.855 & \hgm{0.000} & \hgm{0.000} & 0.02& -0.02& -0.02& 0.02& 1.1& -1.1& -1.1& 1.1& 1.1& -1.1& -1.1& 1.1\\ 
AF-x & \hgm{0.000} & 0.329 & 0.085 & 1.2& -1.2& 1.2& -1.2& 1.1& -1.1& 1.1& -1.1& 0.7& -0.7& 0.7& -0.7\\ 
\hline
\textbf{Ni} & s & h & o & \multicolumn{4}{c}{$m_\text{s}$} & \multicolumn{4}{c}{$m_\text{h}$} & \multicolumn{4}{c}{$m_\text{o}$} \\ \hline
FM    & 0.009 & 0.019 & 0.268 & 0.0& 0.0& 0.0& 0.0& 0.9& 0.9& 0.9& 0.9& 0.9& 0.9& 0.9& 0.9\\ 
AF-y & 0.008 & 0.247 & \hgm{0.000} & 0.1& 0.1& -0.1& -0.1& 0.9& 0.9& -0.9& -0.9& 1.0& 1.0& -1.0& -1.0\\ 
AF    & \hgm{0.000} & 0.234 & 0.002 & 0.2& -0.2& -0.2& 0.2& 0.9& -0.9& -0.9& 0.9& 1.0& -1.0& -1.0& 1.0\\ 
AF-x & 0.009 & \hgm{0.000} & 0.015 & 0.0& 0.0& 0.0& 0.0& 0.9& -0.9& 0.9& -0.9& 1.0& -1.0& 1.0& -1.0\\ 
\hline
\textbf{Os} & s & h & o & \multicolumn{4}{c}{$m_\text{s}$} & \multicolumn{4}{c}{$m_\text{h}$} & \multicolumn{4}{c}{$m_\text{o}$} \\ \hline
FM    & \hgm{0.000} & \hgm{0.000} & \hgm{0.000} & 0.8& 0.8& 0.8& 0.8& 0.3& 0.3& 0.3& 0.3& 1.1& 1.1& 1.1& 1.1\\ 
AF-y & 0.075 & 0.001 & 4.653 & 0.7& 0.7& -0.7& -0.7& 0.3& 0.3& -0.3& -0.3& 0.0& 0.0& 0.0& 0.0\\ 
AF    & 0.169 & 0.098 & 4.654 & 0.8& -0.8& -0.8& 0.8& 0.0& 0.0& 0.0& 0.0& 0.0& 0.0& 0.0& 0.0\\ 
AF-x & 0.035 & 0.098 & 0.001 & 0.7& -0.7& 0.7& -0.7& 0.0& 0.0& 0.0& 0.0& 1.1& -1.1& 1.1& -1.1\\ 
\hline
\textbf{Pt} & s & h & o & \multicolumn{4}{c}{$m_\text{s}$} & \multicolumn{4}{c}{$m_\text{h}$} & \multicolumn{4}{c}{$m_\text{o}$} \\ \hline
FM    & \hgm{0.000} & \hgm{0.000} & 3$\times$10$^{-4}$ & 0.0& 0.0& 0.0& 0.0& 0.0& 0.0& 0.0& 0.0& 0.0& 0.0& 0.0& 0.0\\ 
AF-y & 1$\times$10$^{-5}$ & 0.001 &3$\times$10$^{-4}$ & 0.0& 0.0& 0.0& 0.0& 0.0& 0.0& 0.0& 0.0& 0.0& 0.0& 0.0& 0.0\\ 
AF    & 2$\times$10$^{-5}$ & 4$\times$10$^{-5}$ & 9$\times$10$^{-5}$ & 0.0& 0.0& 0.0& 0.0& 0.0& 0.0& 0.0& 0.0& 0.0& 0.0& 0.0& 0.0\\ 
AF-x & 0.035 & 3$\times$10$^{-4}$ & \hgm{0.000} & 0.0& 0.0& 0.0& 0.0& 0.0& 0.0& 0.0& 0.0& 0.0& 0.0& 0.0& 0.0\\ 
\hline
\textbf{Ru} & s & h & o & \multicolumn{4}{c}{$m_\text{s}$} & \multicolumn{4}{c}{$m_\text{h}$} & \multicolumn{4}{c}{$m_\text{o}$} \\ \hline
FM    & 0.142 & 0.049 & 0.223 & 1.2& 1.2& 1.2& 1.2& 0.8& 0.8& 0.8& 0.8& 2.8& 2.8& 2.8& 2.8\\ 
AF-y & 0.126 & \hgm{0.000} & 0.129 & 1.2& 1.2& -1.2& -1.2& 0.9& 0.9& -0.9& -0.9& 2.6& 2.6& -2.6& -2.6\\ 
AF    & 0.138 & 0.076 & 0.201 & 1.1& -1.1& -1.1& 1.1& 0.9& -0.9& -0.9& 0.9& 2.6& -2.6& -2.6& 2.6\\ 
AF-x & \hgm{0.000} & 0.135 & \hgm{0.000} & 1.4& -1.4& 1.4& -1.4& 0.8& -0.8& 0.8& -0.8& 2.6& -2.6& 2.6& -2.6\\ 
\end{tabular}

\caption{Relative energy difference ($\Delta E$, eV per cell) between the FM, AF-y, AF, and AF-x magnetic configurations for the s, h, and o geometries of M $=$ Nb, Ni, Os, Pt and Ru, along with the local magnetic moments of the four M atoms ($m_\text{s}$, $m_\text{h}$, and $m_\text{o}$, in $\mu B$). The $\Delta E = 0$ indicates the most stable configuration; values in red highlight the most stable geometry for each magnetic phase.}
\end{ruledtabular}
\end{table}

\begin{table}[H]
\centering
\begin{ruledtabular}
\begin{tabular}{l|rrr|rrrr|rrrr|rrrr}
\textbf{Ti} & s & h & o & \multicolumn{4}{c}{$m_\text{s}$} & \multicolumn{4}{c}{$m_\text{h}$} & \multicolumn{4}{c}{$m_\text{o}$} \\ \hline
FM    & \hgm{0.000} & 0.085 & \hgm{0.000} & 1.3& 1.3& 1.3& 1.3& 1.5& 1.5& 1.5& 1.5& 1.3& 1.3& 1.3& 1.3\\ 
AF-y & 0.127 & 0.607 & 0.975 & 1.2& 1.2& -1.2& -1.2& 1.5& 1.5& -1.5& -1.5& 1.6& 1.6& -1.6& -1.6\\ 
AF    & 0.164 & \hgm{0.000} & 0.975 & 1.1& -1.1& -1.1& 1.1& 1.7& -1.7& -1.7& 1.7& 1.6& -1.6& -1.6& 1.6\\ 
AF-x & 0.077 & 0.030 & 1.288 & 1.2& -1.2& 1.2& -1.2& 1.5& -1.5& 1.5& -1.5& 1.4& -1.4& 1.4& -1.4\\ 
\hline
\textbf{V} & s & h & o & \multicolumn{4}{c}{$m_\text{s}$} & \multicolumn{4}{c}{$m_\text{h}$} & \multicolumn{4}{c}{$m_\text{o}$} \\ \hline
FM    & 0.366 & \hgm{0.000} & 0.238 & 2.1& 2.1& 2.1& 2.1& 2.4& 2.4& 2.4& 2.4& 2.9& 2.9& 2.9& 2.9\\ 
AF-y & \hgm{0.000} & 0.248 & 0.096 & 2.3& 2.3& -2.3& -2.3& 2.6& 2.6& -2.6& -2.6& 3.0& 3.0& -3.0& -3.0\\ 
AF    & 0.127 & 0.234 & 0.096 & 2.1& -2.1& -2.1& 2.1& 2.5& -2.5& -2.5& 2.5& 3.0& -3.0& -3.0& 3.0\\ 
AF-x & 0.134 & 0.045 & \hgm{0.000} & 2.3& -2.3& 2.3& -2.3& 2.5& -2.5& 2.5& -2.5& 2.9& -2.9& 2.9& -2.9\\ 
\hline
\textbf{W} & s & h & o & \multicolumn{4}{c}{$m_\text{s}$} & \multicolumn{4}{c}{$m_\text{h}$} & \multicolumn{4}{c}{$m_\text{o}$} \\ \hline
FM    & 0.721 & \hgm{0.000} & 1$\times$10$^{-5}$ & 1.7& 1.7& 1.7& 1.7& 1.0& 1.0& 1.0& 1.0& 0.0& 0.0& 0.0& 0.0\\ 
AF-y & 0.261 & 0.025 &\hgm{0.000} & 2.2& 2.2& -2.2& -2.2& 0.8& 0.8& -0.8& -0.8& 0.0& 0.0& 0.0& 0.0\\ 
AF    & \hgm{0.000} & 0.005 & 3$\times$10$^{-5}$ & 2.0& -2.0& -2.0& 2.0& 0.7& -0.7& -0.7& 0.7& 0.0& 0.0& 0.0& 0.0\\ 
AF-x & 0.658 & 0.108 & 2$\times$10$^{-6}$ & 1.8& -1.8& 1.8& -1.8& 0.6& -0.6& 0.6& -0.6& 0.0& 0.0& 0.0& 0.0\\ 
\end{tabular}

\caption{Relative energy difference ($\Delta E$, eV per cell) between the FM, AF-y, AF, and AF-x magnetic configurations for the s, h, and o geometries of M $=$ Ti, V and W, along with the local magnetic moments of the four M atoms ($m_\text{s}$, $m_\text{h}$, and $m_\text{o}$, in $\mu B$). The $\Delta E = 0$ indicates the most stable configuration; values in red highlight the most stable geometry for each magnetic phase.}
\end{ruledtabular}
\end{table}

In the table below, we show the bond length between the metal transition intercalated in the B/B and the closest C$^{468}$ in the biphenylene bilayer, for the M intercalated in s and h configuration. Also, we show the distance ($d$) between the B/B in the same configurations. The average bilayer distance is $3.93$\,\AA and $3.78$\,\AA for the s and h configurations, respectively. For the B/B without intercalated metals the $d=3.39$\,\AA
\begin{table}[H]
\caption{the bond length between the metal transition intercalated in the B/B and the closest C$^{468}$, and the distance ($d$) between the B/B.}
\label{t_distance}
\begin{ruledtabular}    
\begin{tabular}{lcccccccc}
\multicolumn{1}{l}{M} &
\multicolumn{2}{c}{s configuration} &
\multicolumn{2}{c}{h configuration} \\
  \cline{2-3} \cline{4-5} 
 & Bond length M-C$^{468}$ (\AA) & $d$ (\AA) & Bond length M-C$^{468}$ (\AA) & $d$ (\AA)  \\
 \hline
Ti         & $ 2.31$ & $4.13$  & $2.38$&$3.88$\\
V         & $ 2.25$ & $3.98$  & $2.32$&$3.71$\\
Cr         & $ 2.28$ & $4.06$  & $2.45$&$4.05$\\
Mn         & $ 2.31$ & $4.12$  & $2.42$&$3.95$\\
Fe         & $ 2.11$ & $3.68$  & $2.30$&$3.68$\\
Co         & $ 2.11$ & $3.67$  & $2.25$&$3.55$\\
Ni         & $ 2.06$ & $3.59$  & $2.30$&$3.67$\\
Cu         & $ 2.16$ & $3.78$  & $2.35$&$3.79$\\
Nb         & $ 2.38$ & $4.27$  & $2.42$&$3.96$\\
Ru         & $ 2.20$ & $3.92$  & $2.28$&$3.64$\\
W         & $ 2.29$ & $4.07$  & $2.30$&$3.68$\\
Os         & $ 2.20$ & $3.86$  & $2.25$&$3.55$\\
Pt         & $ 2.25$ & $3.99$  & $2.46$&$4.06$\\
\hline
\end{tabular}
\end{ruledtabular}
\end{table}

For the analysis of binding energy of O$_2$, we can define the binding energy as
\begin{equation}
    E^b_{\mathrm{O_2}} = E_{\mathrm{O}_2^\ast} - (E_\ast + E_\mathrm{O_2}),
\end{equation}
where $E_{\mathrm{O}_2^\ast}$ is the total energy of the systems B/M/B when the O$_2$ is adsorbed on C$^{468}$, $E_\ast$ is the total energy of B/M/B and $E_\mathrm{O_2}$ is the total energy of O$_2$ molecule in the triplet configuration.

\begin{figure}[H]
    \centering
    \includegraphics[width=1\linewidth]{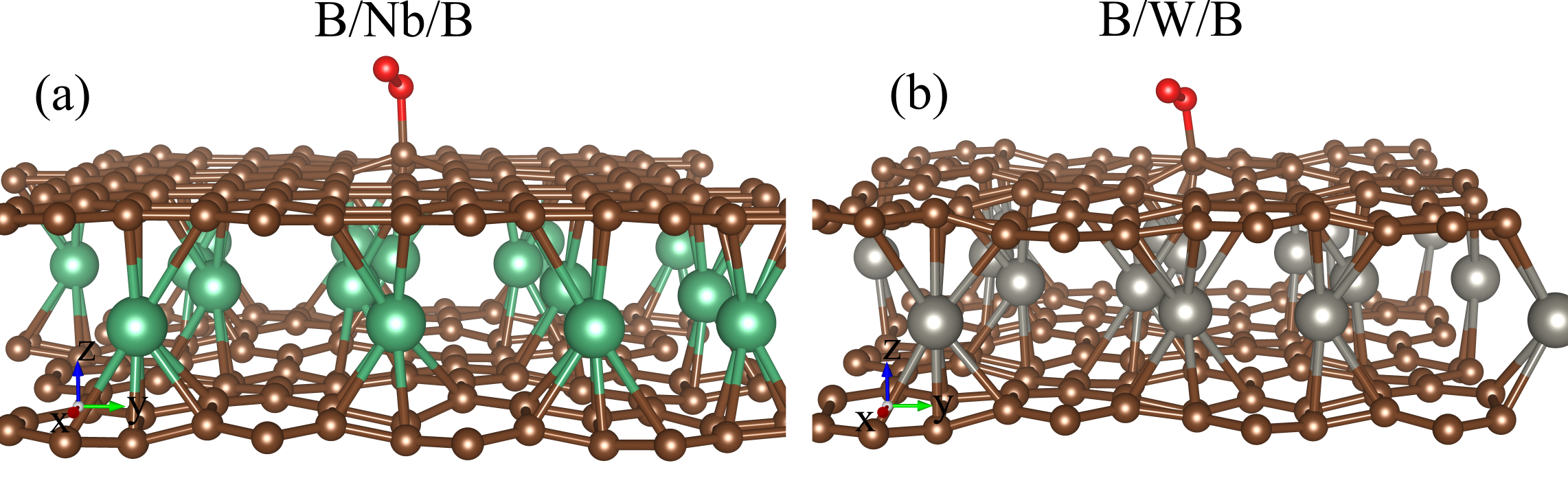}
    \caption{Perspective views of the optimized geometries for the B/Nb/B (a) and B/W/B (b) systems following the adsorption of the O$2$ at the square-planar sites (C$^{468}$).}
    \label{fig:nb_w_o2}
\end{figure}

At the point of view of O$_2$ adsorption, only B/Nb/B and B/W/B present the $E^b_{\mathrm{O_2}}>0$, indicating the unfavorable adsorption in this systems. This results are due to the fact that when O$_2$ is adsorbed onto B/M/B, the M atoms throughout the structure are displaced from the square sites of the B/B, causing the system energy to increase considerably, as show in the Figure~\ref{fig:nb_w_o2}.

\section{Bader Charge}
In this section, we present the results of charge transfer, via Bader charge analysis\,\cite{HENKELMAN2006} and net charge density, from the M atoms to the C$^{468}$ sites, as summarized in Table T.5. For each B/M/B system, we show the average charge lost by M and gained by the C$^{468}$ atoms.

\begin{table}[ht]
\centering
\caption{Mean value of Bader charge of transition metals atoms ($\Delta q_M$) and mean value of Bader charge of carbon atoms in Square ($\Delta q_{C^{468}}$). While other carbon atoms also accept electron from the M, the electron transfer to the carbon atoms at the square-planar sites is substantially more pronounced. The mean values of Bader charge are in $[\times 10^{13}e^-/\text{cm}^2]$ unit.}
\begin{tabular}{l*{8}{c}}
\hline\hline
 & Cr & V & Mn & Os & W & Nb & Ti  \\
\hline
$\Delta q_M$ &
$-2.554$ & $-2.932$ & $-2.671$ & $-1.568$ & $-2.903$ & $-3.629$ &
$-3.222$  \\
$\Delta q_{C^{468}}$ &
$+1.578$ & $+1.832$ & $+1.669$ & $+0.998$ & $+1.796$ & $+2.232$ &
$+2.014$   \\
\hline\hline
 &Fe & Pt & Co & Ni & Cu & Ru & \\
\hline
$\Delta q_M$ & $-1.858$ & $-0.929$ & $-1.858$ & $-1.081$ & $-1.684$ & $-1.713$ \\
$\Delta q_{C^{468}}$& $+1.161$& $+0.581$ & $+1.143$ & $+0.708$ & $+1.016$ & $+1.070$ \\
\hline\hline
\end{tabular}
\label{tab:bader}
\end{table}

\begin{figure}[H]
    \centering
    \includegraphics[width=0.8\linewidth]{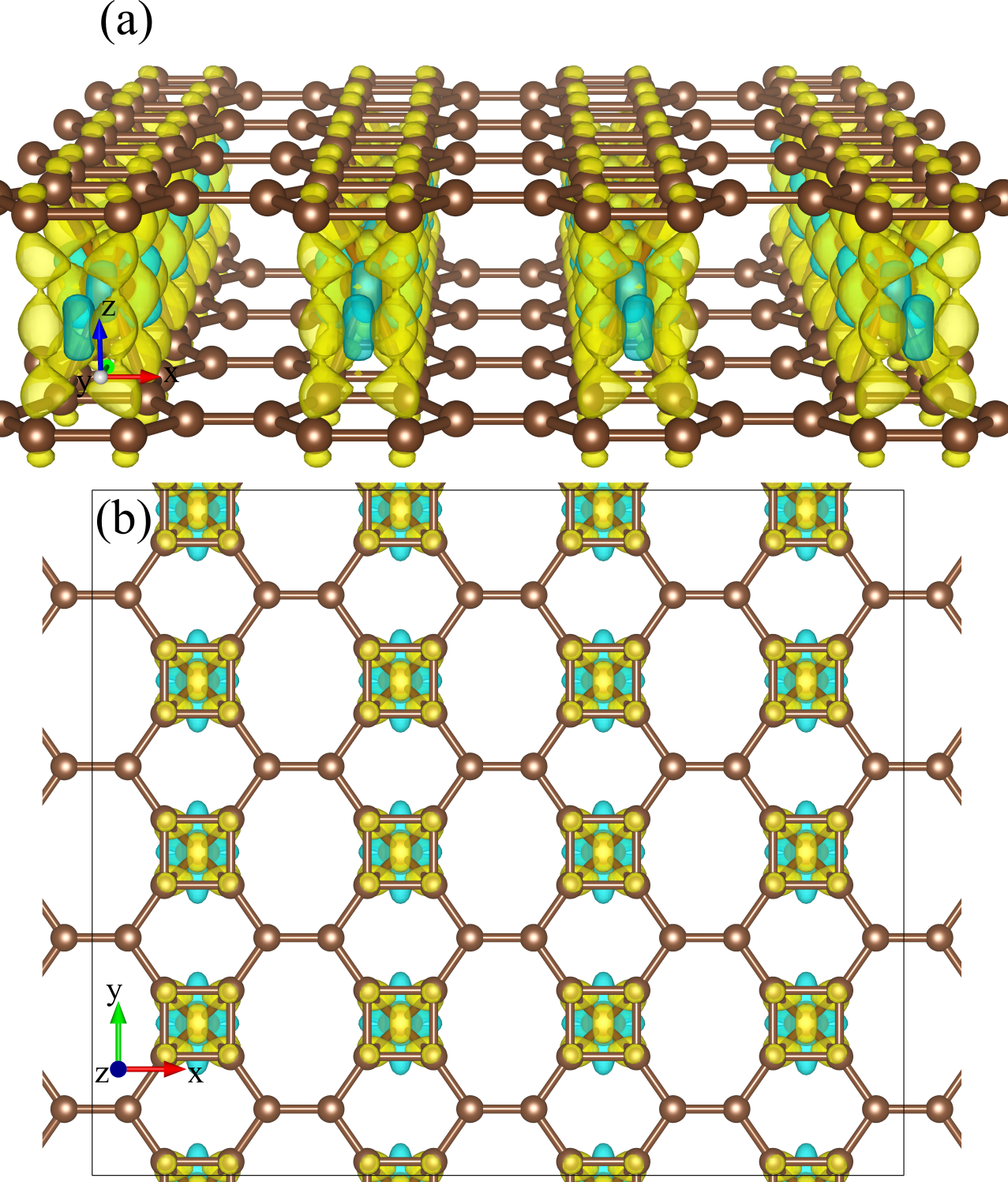}
    \caption{The $\Delta \rho$ for B/Fe/B. Panel (a) is the b-view of B/Fe/B, and panel (b) is the c-view of B/Fe/B. Yellow color indicated the region to gain electrons and blue color indicated the regions to lose electrons. Isosurface: $0.01\text{ }e^-/$\AA.}
    \label{fig:rho}
\end{figure}

We performed the net charge density difference of B/M/B for M$=$Fe, given by
\begin{equation}
    \Delta \rho = \rho^\text{B/M/B} - (\rho^\text{M} + \rho^\text{B/B}),
\end{equation}
where $\rho^\text{B/M/B}$, $\rho^\text{M}$ and $\rho^\text{B/B}$ are the charge density of B/M/B, M e B, respectively. Considering, B/Fe/B the $\Delta \rho$ are represented in Fig. \ref{fig:rho}. The carbons in the square gain electrons, as indicated by \ref{tab:bader}, whereas the Fe atoms lose electrons.

\begin{figure}[H]
    \centering
    \includegraphics[width=1\linewidth]{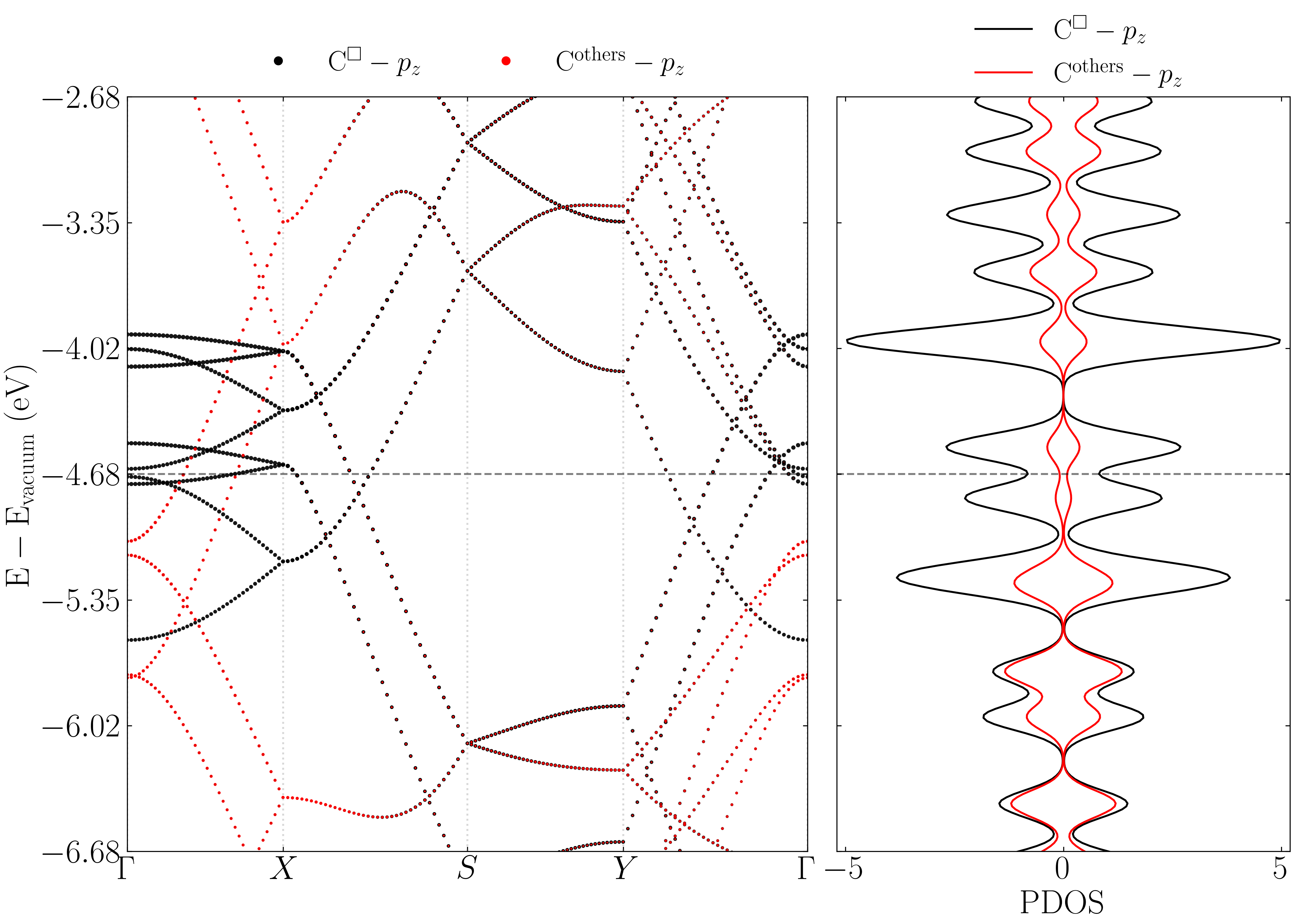}
    \caption{The left panel shows the band structure projected onto the $p_z$ orbitals of the square-site carbon atoms (black dots - C$^{468}$) and the non-square carbon atoms (red dots - C$^\text{688}$). The right panel displays the PDOS of the $p_z$ orbital for C$^{468}$ (black curve) and C$^\text{688}$ (red curve). The dashed line in both panels indicates the Fermi level. It is notable that near $E_F$, the majority contribution to the $p_z$-orbital comes from the C$^{468}$ sites. This suggests that the greater density of states at C$^{468}$ compared to $^\text{688}$ renders C$^{468}$ more reactive\,\cite{Luo2021}, and indicates that C$^{468}$ should be the preferential adsorption sites for O$^\ast$, OH$^\ast$, and OOH$^\ast$.}
    \label{fig:band_pdos_bi}
\end{figure}

In the Fig.\,\ref{fig:band_pdos_bi}, we present the project bands (left panel) and PDOS (right panel) of the $p_z-$orbital of the square-site carbon atoms (C$^{468}$) and the non-square carbon atoms (C$^\text{688}$) for B/B. It is notable that near $E_F$, the majority contribution to the $p_z$-orbital comes from the C$^{468}$ sites. This suggests that the greater density of states at C$^{468}$ compared to $^\text{688}$ renders C$^{468}$ more reactive\,\cite{Luo2021}, and indicates that C$^{468}$ should be the preferential adsorption sites for O$^\ast$, OH$^\ast$, and OOH$^\ast$.

\newpage
\section{Gibbs Free Energy}
We estimated the reaction energies for the ORR and OER on TB/M/B by calculating the adsorption free energies of three key intermediates ($\text{OOH}^\ast$, $\text{O}^\ast$, and $\text{OH}^\ast$)~\cite{Xu2024} at the square site of the biphenylene framework ~\cite{Luo2021}. For standardization, these energies were referenced to the free energies of equivalent stoichiometric amounts of gaseous $\text{H}_2\text{O}$ and $\text{H}_2$, utilizing the formalisms described in the following equations. The asterisks denote species adsorbed on the square B site.
\begin{equation}\label{eq:deltaG}
\begin{aligned}
\Delta G_{\text{OOH}^\ast} &= G(\text{OOH}^\ast) - G(\ast) + 1.5\,G(\text{H}_2) - 2\,G(\text{H}_2\text{O}),\\[6pt]
\Delta G_{\text{O}^\ast}   &= G(\text{O}^\ast)   - G(\ast) + G(\text{H}_2) - G(\text{H}_2\text{O}),\\[6pt]
\Delta G_{\text{OH}^\ast}  &= G(\text{OH}^\ast)  - G(\ast) + 0.5\,G(\text{H}_2) - G(\text{H}_2\text{O}).
\end{aligned}
\end{equation}

The reaction of the ORR can be written as\,\cite{Xu2024},
\begin{equation} \label{eq:1}
    \text{O}_2 + \text{H}_2\text{O} + e^- + \ast \to \text{OOH}^\ast + \text{OH}^-,
\end{equation}
\begin{equation}\label{eq:2}
    \text{OOH}^\ast + e^- \to \text{O}^\ast + \text{OH}^-,
\end{equation}
\begin{equation}\label{eq:3}
    \text{O}^\ast + \text{H}_2\text{O} + e^- \to \text{OH}^\ast + \text{OH}^-,
\end{equation}
\begin{equation}\label{eq:4}
    \text{OH}^\ast + e^- \to \text{OH}^- + \ast,
\end{equation}
where the Gibbs free energy at zero potential, relative to the reversible hydrogen electrode (RHE), for each chemical reaction steps are given by,
\begin{equation}\label{eq:deltaG_orr}
\begin{aligned}
    \Delta G_1 &= \Delta G_{\text{OOH}^\ast} -4.92, \\
    &  & \\
    \Delta G_2 &=\Delta G_{\text{O}^\ast} - \Delta G_{\text{OOH}^\ast}, \\
    & & \\
    \Delta G_3 &= \Delta G_{\text{OH}^\ast}- \Delta G_{\text{O}^\ast}, \\   
    & & \\
    \Delta G_4 &= -\Delta G_{\text{OH}^\ast}.
\end{aligned}
\end{equation}

The reaction of the OER can be written as\,\cite{Xu2024},
\begin{equation} \label{eq:5}
     \text{OH}^- + \ast\to \text{OH}^\ast + e^- ,
\end{equation}
\begin{equation}\label{eq:6}
    \text{OH}^\ast +\text{OH}^- \to \text{O}^\ast + \text{H}_2\text{O} + e^-,
\end{equation}
\begin{equation}\label{eq:7}
    \text{O}^\ast + \text{OH}^- \to \text{OOH}^\ast +e^-,
\end{equation}
\begin{equation}\label{eq:8}
     \text{OOH}^\ast + \text{OH}^-\to  \text{O}_2 + \text{H}_2\text{O} + e^- + \ast,
\end{equation}

The Gibbs free energy at zero potential, relative to RHE, for OER are given by:
\begin{equation}\label{eq:deltaG_oer}
\begin{aligned}
    \Delta G_5 &= \Delta G_{\text{OH}^\ast}, \\
    &  & \\
    \Delta G_6 &= \Delta G_{\text{O}^\ast} - \Delta G_{\text{OH}^\ast}, \\
    & & \\
    \Delta G_7 &= \Delta G_{\text{OOH}^\ast}- \Delta G_{\text{O}^\ast}, \\   
    & & \\
    \Delta G_8 &= 4.92 -\Delta G_{\text{OOH}^\ast}.
\end{aligned}
\end{equation}

The onset potential ($U_{onset}$) for ORR/OER is determined by,
\begin{equation}
    U_{onset}(\text{ORR}) = - \text{max} \{\Delta G_1, \Delta G_2, \Delta G_3, \Delta G_4 \}/e^-,
\end{equation}
\begin{equation}
    U_{onset}(\text{OER}) = \text{max} \{\Delta G_5, \Delta G_6, \Delta G_7, \Delta G_8 \}/e^-,
\end{equation}
and the overpotential ($\eta$) for each reaction is defined by
\begin{equation}
    \eta_\text{ORR} \approx U_0 - U_{onset}(\text{ORR}),
\end{equation}
\begin{equation}
    \eta_\text{OER} \approx U_{onset}(\text{OER})- U_0,
\end{equation}
where $U_0 = 1.23 \text{ V}$. 
\newpage
\subsection{Numerical Results for Gibbs free energies}

In this section, we present the numerical results obtained from the catalysis calculations using DFT. In the TABLE T.6 contains the calculated values for $E_{ZPE}$ and $TS$ at $T = 300$ K. TABLE T.7 presents the values for $\Delta G_{\text{OOH}^\ast}$, $\Delta G_{\text{O}^\ast}$, and $\Delta G_{\text{OH}^\ast}$, as per Eqs.\,\ref{eq:deltaG}, for each B/M/B system, including solvation corrections. TABLE T.8 provides the values for the ORR steps $\Delta G_1$, $\Delta G_2$, $\Delta G_3$, and $\Delta G_4$, given by Eqs.\,\ref{eq:deltaG_orr}, respectively, also including solvation corrections. Also in TABLE T.8 shows the values for the OER steps $\Delta G_5$, $\Delta G_6$, $\Delta G_7$, and $\Delta G_8$, given by Eqs.\,\ref{eq:deltaG_oer}, with solvation corrections. Finally, TABLE T.9 presents the solvation correction values calculated via VASPsol\,\cite{VASPsol-Software, VASPsol2014-Dielectric, VASPsol2019-Electrolyte} for each B/M/B system upon adsorption of OH$^\ast$ and OOH$^\ast$.

\begin{table}[H]
\centering
\caption{Zero-point energies ($E_{ZPE}$) and entropic contributions ($TS$) at $T = 300$ K for each B/M/B system with adsorbed O$^\ast$, OH$^\ast$, and OOH$^\ast$ species on the catalyst. For these calculations, the B/M/B substrate was kept fixed while the O$^\ast$, OH$^\ast$, and OOH$^\ast$ adsorbates were allowed to relax freely.}
\label{tab:zpe-entropy}
\begin{tabular}{lcccccc}
\hline
\hline
Metals & $E_{ZPE}(\text{O}^\ast)$  & $TS(\text{O}^\ast)$& $E_{ZPE}(\text{OH}^\ast)$ & $TS(\text{OH}^\ast)$ &$E_{ZPE}(\text{OOH}^\ast)$ & $TS(\text{OOH}^\ast)$ \\
\hline
Cr       & $0.081$ &$0.048$& $0.386$  & $0.071$& $0.454$ & $0.096$ \\
V        & $0.082$ &$0.047$& $0.382$  & $0.077$ &$0.464$ & $0.131$\\
Mn       & $0.081$ &$0.050$& $0.387$  & $0.068$& $0.447$ & $0.104$\\
Os       & $0.084$ &$0.046$& $0.384$  & $0.075$& $0.467$ & $0.130$\\
W        & $0.103$ &$0.039$& $0.383$  & $0.075$& $0.465$ & $0.131$\\
Nb       & $0.092$ &$0.041$& $0.390$  & $0.065$& $0.444$ & $0.112$\\
Ti       & $0.081$ &$0.053$& $0.389$  & $0.067$& $0.460$ & $0.132$\\
Fe       & $0.082$ &$0.045$& $0.387$  & $0.068$ & $0.456$ & $0.142$\\
Pt       & $0.079$ &$0.050$& $0.382$  & $0.076$ & $0.461$ & $0.136$\\
Co       & $0.082$ &$0.046$& $0.386$  & $0.070$ & $0.457$ & $0.143$\\
Ni       & $0.087$ &$0.033$& $0.388$  & $0.069$ & $0.460$ & $0.138$\\
Cu       & $0.081$ &$0.048$& $0.390$  & $0.066$& $0.463$ & $0.130$\\
Ru       & $0.082$ &$0.047$& $0.390$  & $0.067$ &$0.451$ & $0.165$\\
\hline
\hline
\end{tabular}
\vspace{4pt}
\end{table}

\begin{table}[H]
\centering
\caption{Gibbs free energy adsorption for each M considered the species OOH$^\ast$, O$^\ast$ and OH$^\ast$, and with the solvation effect includes.}
\label{tab:g_ads}
\begin{tabular}{lccccc}
\hline
\hline
Metals & $\Delta G_{\text{OOH}^\ast}$ (eV) & $\Delta G_{\text{O}^\ast}$ (eV)& $\Delta G_{\text{OH}^\ast}$ (eV)  \\
\hline
Ti       & $1.98$ &$-0.47$& $-1.96$   \\
V        & $2.23$ &$0.75$& $1.75$   \\
Cr       & $5.24$ &$5.23$& $2.42$   \\
Mn       & $4.20$ &$1.95$& $1.27$   \\
Fe        & $3.25$ &$1.74$& $0.47$   \\
Co       & $4.46$ &$2.54$& $1.51$   \\
Ni       & $4.40$ &$2.39$& $1.47$   \\
Cu       & $4.07$ &$1.91$& $1.09$   \\
Nb       & $-7.10$ &$-8.04$& $-8.97$   \\
Ru       & $4.14$ &$1.96$& $1.23$   \\
W       & $-8.10$ &$-15.22$& $-11.17$   \\
Os       & $-1.01$ &$-1.60$& $-3.11$   \\
Pt       & $4.13$ &$2.05$& $1.17$   \\
\hline
\hline
\end{tabular}
\vspace{4pt}
\end{table}

\begin{table}[H]
\centering
\caption{Gibbs free energy steps for each M in the ORR with $\Delta G_1$-$\Delta G_4$ and to the OER with $\Delta G_5$-$\Delta G_8$.}
\label{tab:orr}
\begin{tabular}{lcccccccc}
\hline
\hline
Metals & $\Delta G_{1}$ (eV) & $\Delta G_{2}$ (eV)& $\Delta G_{3}$ (eV) & $\Delta G_4$ & $\Delta G_{5}$ (eV) & $\Delta G_{6}$ (eV)& $\Delta G_{7}$ (eV) & $\Delta G_8$ \\
\hline
Ti       & $-2.94$ &$-2.45$& $-1.49$ & $1.96$ & $-1.96$ &$1.49$& $2.45$ & $2.94$  \\
V        & $-2.69$ &$-1.48$& $1.00$ & $-1.75$ & $1.75$ &$-1.00$& $1.48$ & $2.69$  \\
Cr       & $0.32$ &$-0.004$& $-2.81$ & $-2.42$ & $2.42$ &$2.81$& $0.004$ & $-0.32$ \\
Mn       & $-0.72$ &$-2.25$& $-0.67$ & $-1.27$ & $1.27$ &$0.67$& $2.25$ & $0.72$ \\
Fe        & $-1.67$ &$-1.51$& $-1.27$ & $-0.47$ & $0.47$ &$1.27$& $1.51$ & $1.67$ \\
Co       & $-0.46$ &$-1.92$& $-1.03$ & $-1.51$ & $1.51$ &$1.03$& $1.92$ & $0.46$  \\
Ni       & $-0.52$ &$-2.01$& $-0.92$ & $-1.47$ & $1.47$ &$0.92$& $2.01$ & $0.52$ \\
Cu       & $-0.85$ &$-2.16$& $-0.81$ & $-1.09$ & $1.09$ &$0.81$& $2.16$ & $0.85$ \\
Nb       & $-12.02$ &$-0.93$& $-0.94$ & $8.97$ & $-8.97$ &$0.94$& $0.93$ & $12.02$ \\
Ru       & $-0.78$ &$-2.18$& $-0.73$ & $-1.23$ & $1.23$ &$0.73$& $2.18$ & $0.78$ \\
W       & $-13.02$ &$-7.12$& $4.05$ & $11.17$ & $-11.17$ &$-4.05$& $7.12$ & $13.02$ \\
Os       & $-5.93$ &$-0.58$& $-1.51$ & $3.11$ & $-3.11$ &$1.51$& $0.58$ & $5.93$ \\
Pt       & $-0.79$ &$-2.08$& $-0.88$ & $-1.17$ & $1.17$ &$0.88$& $2.08$ & $0.79$ \\
\hline
\hline
\end{tabular}
\vspace{4pt}
\end{table}

\begin{table}[H]
\centering
\caption{Binding energy energy for solvation correction for each metals.}
\label{tab:oer}
\begin{tabular}{lccccc}
\hline
\hline
Metals & $E^{solv}_b{(\text{OH}^\ast)}$ (eV) & $E^{solv}_b{(\text{OOH}^\ast)}$ (eV) \\
\hline
Ti       & $-1.08$ & $0.004$ \\
V        &$ 1.97$ & $-0.86$ \\
Cr       & $-0.62$ & $-0.55$ \\
Mn       & $-1.29$ & $0.14$ \\
Fe       & $0.13$ & $0.07$ \\
Co       & $-0.08$ & $0.11$ \\
Ni       & $0.14$ & $0.14$ \\
Cu       & $0.06$ & $0.13$ \\
Nb       & $0.32$ & $0.37$ \\
Ru       & $-0.38$ & $-0.20$ \\
W        & $0.05$ & $0.17$ \\
Os       & $-0.57$ & $-0.52$ \\
Pt       & $0.06$ & $0.13$ \\
\hline
\hline
\end{tabular}
\vspace{4pt}
\end{table}

\newpage
We present the onset potential ($U_{onset}$) and overpotentials ($\eta$) for B/Nb/B and B/W/B, where for ORR we obtain $U_{onset}= -8.97$\,V and $\eta=10.20$\,V for B/Nb/B, and $U_{onset}= -11.17$\,V and $\eta=12.40$\,V for B/W/B. Addictionally, for OER we obtain $U_{onset}=13.02$\,V and $\eta=11.79$\,V for B/Nb/B, and $U_{onset}=12.02$\,V and $\eta=10.79$\,V for B/W/B.

\begin{figure}[H]
  \centering
  \begin{subfigure}[b]{0.5\textwidth}
    \centering
    \includegraphics[width=\linewidth,height=0.4\textheight,keepaspectratio]{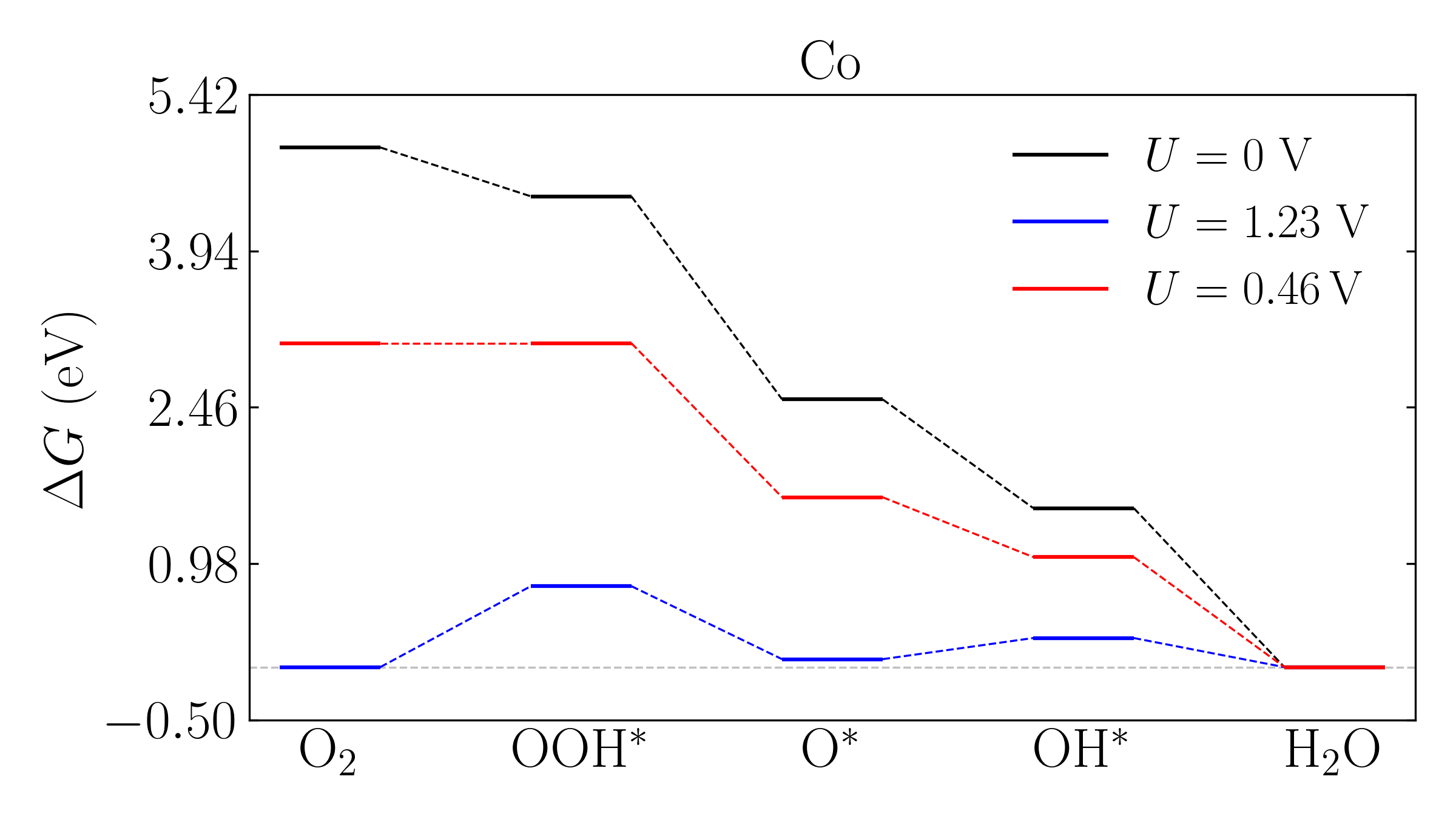}
  \end{subfigure}\hfill%
  \begin{subfigure}[b]{0.5\textwidth}
    \centering
    \includegraphics[width=\linewidth,height=0.4\textheight,keepaspectratio]{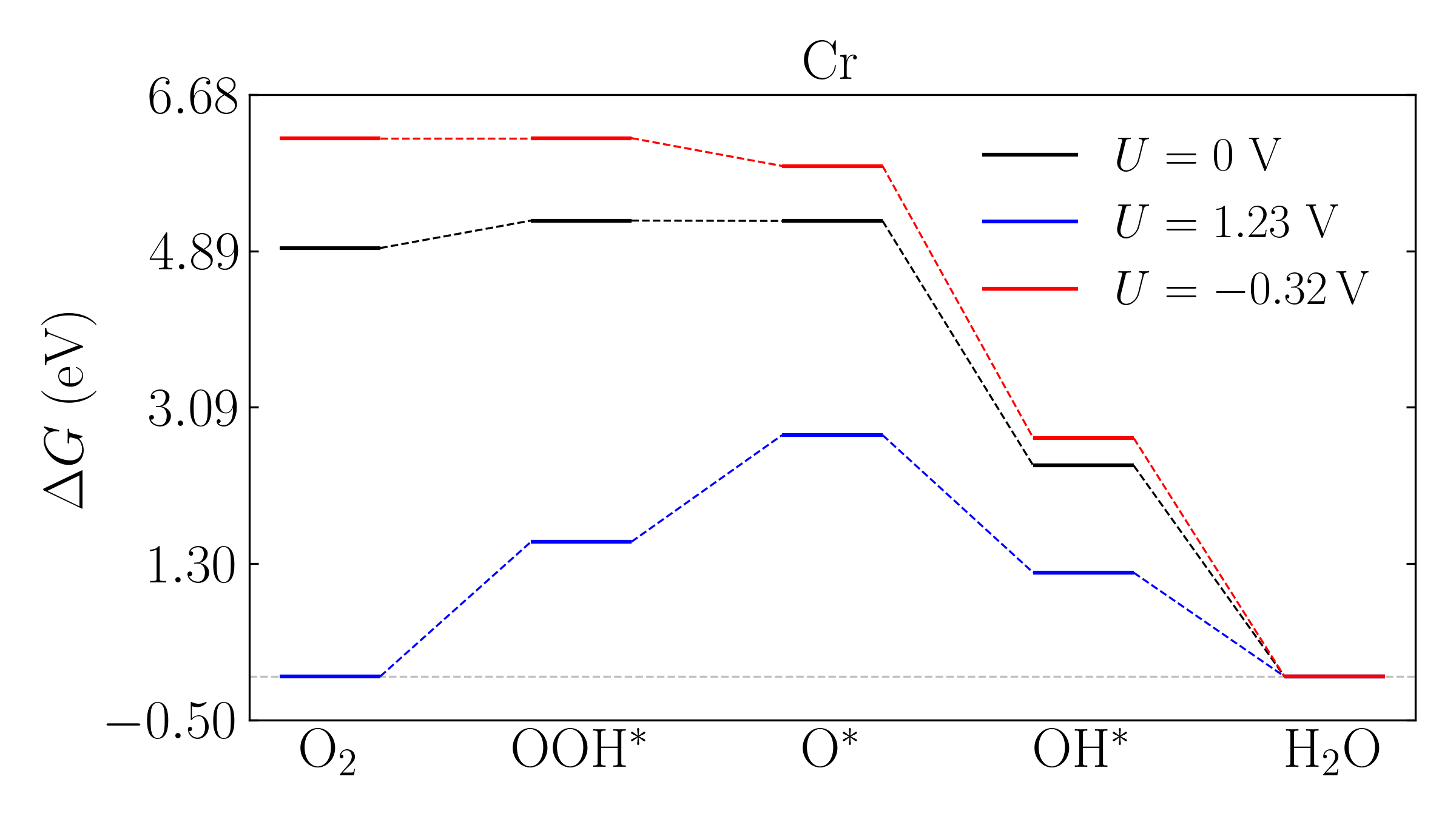}
  \end{subfigure}
  
  \par\vspace{-8pt}
  \begin{subfigure}[b]{0.5\textwidth}
    \centering
    \includegraphics[width=\linewidth,height=0.4\textheight,keepaspectratio]{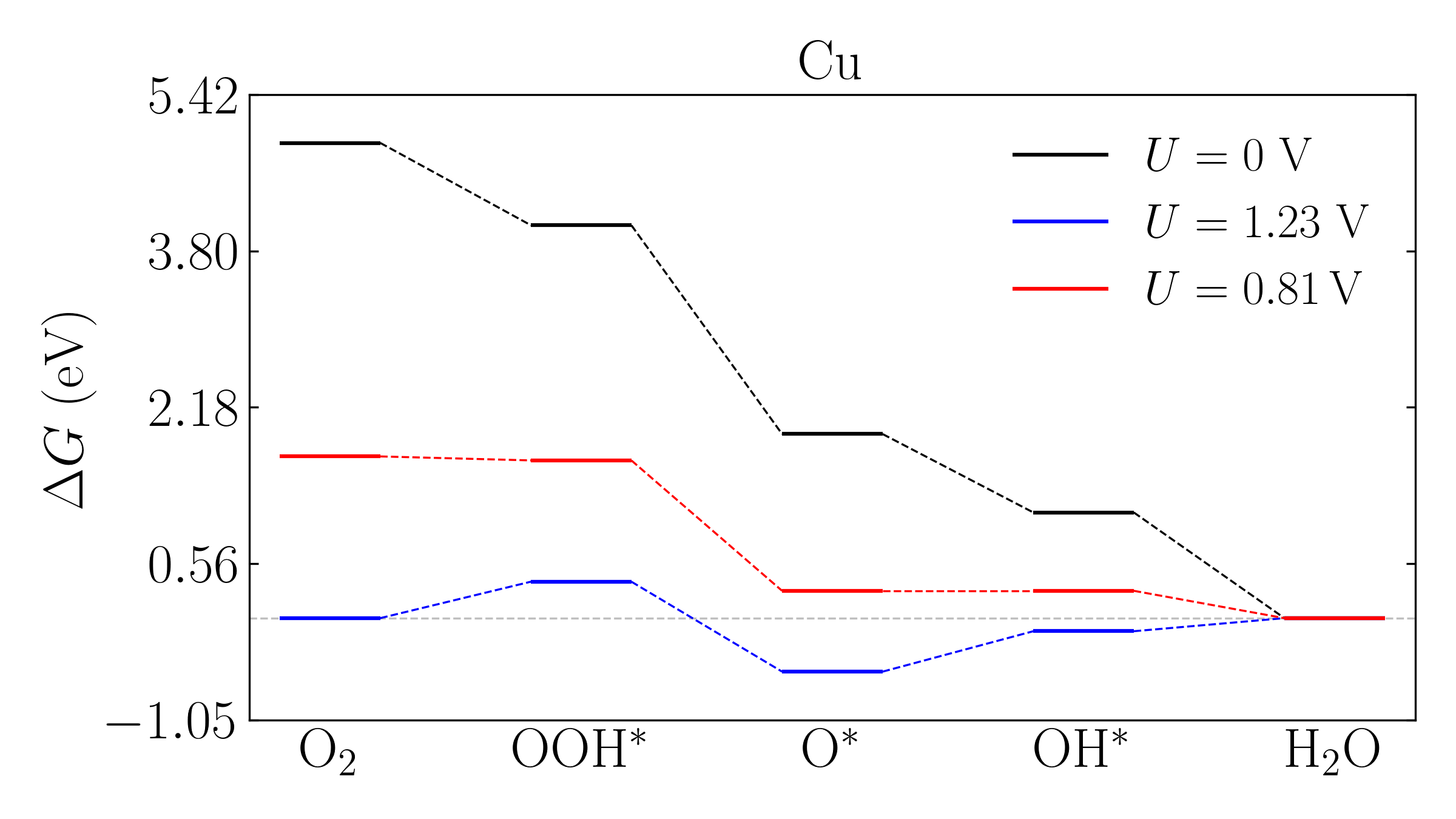}
  \end{subfigure}\hfill%
  \begin{subfigure}[b]{0.5\textwidth}
    \centering
    \includegraphics[width=\linewidth,height=0.4\textheight,keepaspectratio]{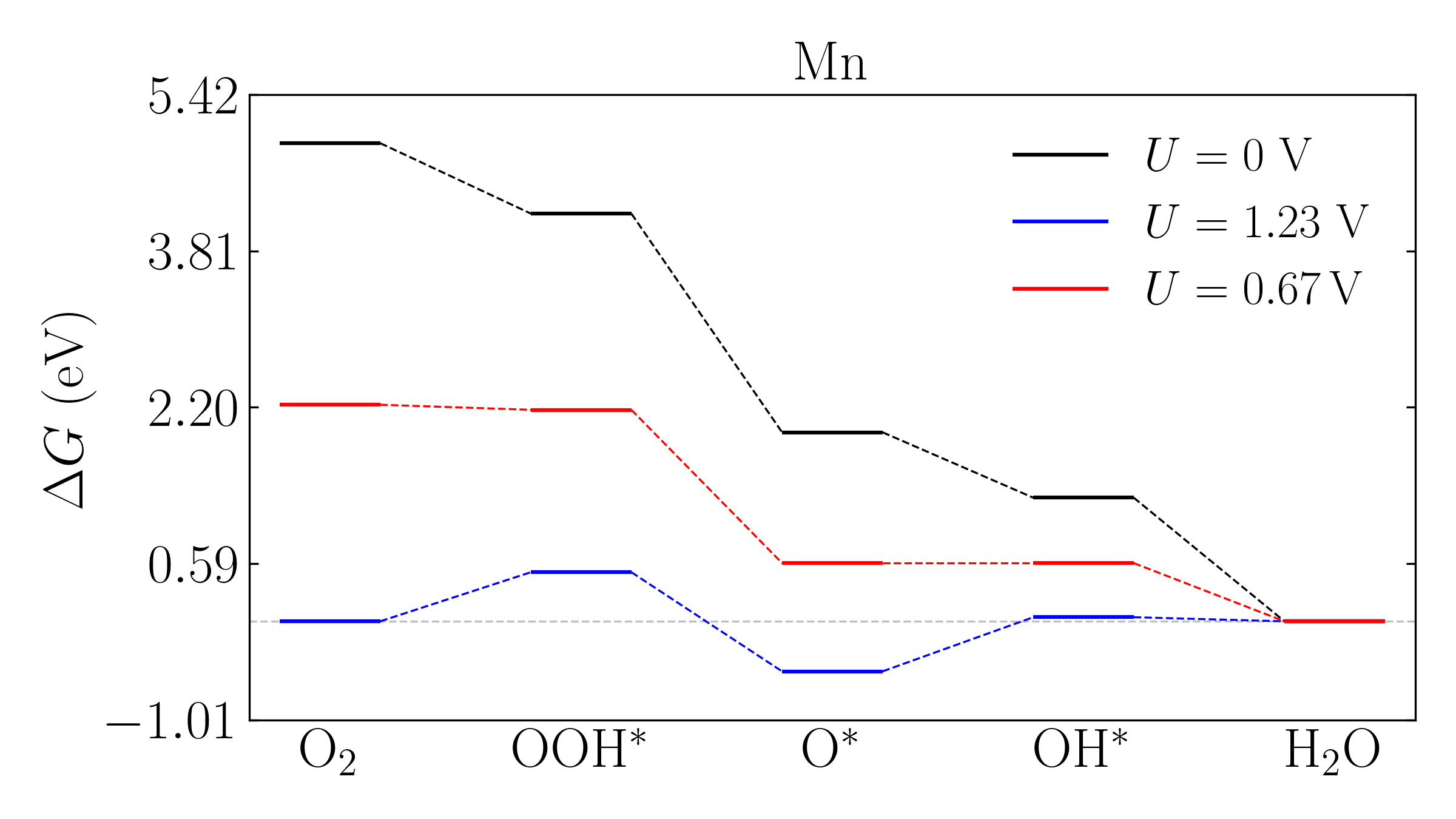}
  \end{subfigure}
  
  \caption{ORR Gibbs free energy for B/M/B (M $=$ Co, Cr, Cu and Mn) at different applied potentials: $U =0$ V (black), $U =1.23$ V (blue), and the onset potential $U =U_{onset}$ (red). }\label{fig:orr-all1}
\end{figure}

\begin{figure}[H]
  \centering
  \begin{subfigure}[b]{0.5\textwidth}
    \centering
    \includegraphics[width=\linewidth,height=0.4\textheight,keepaspectratio]{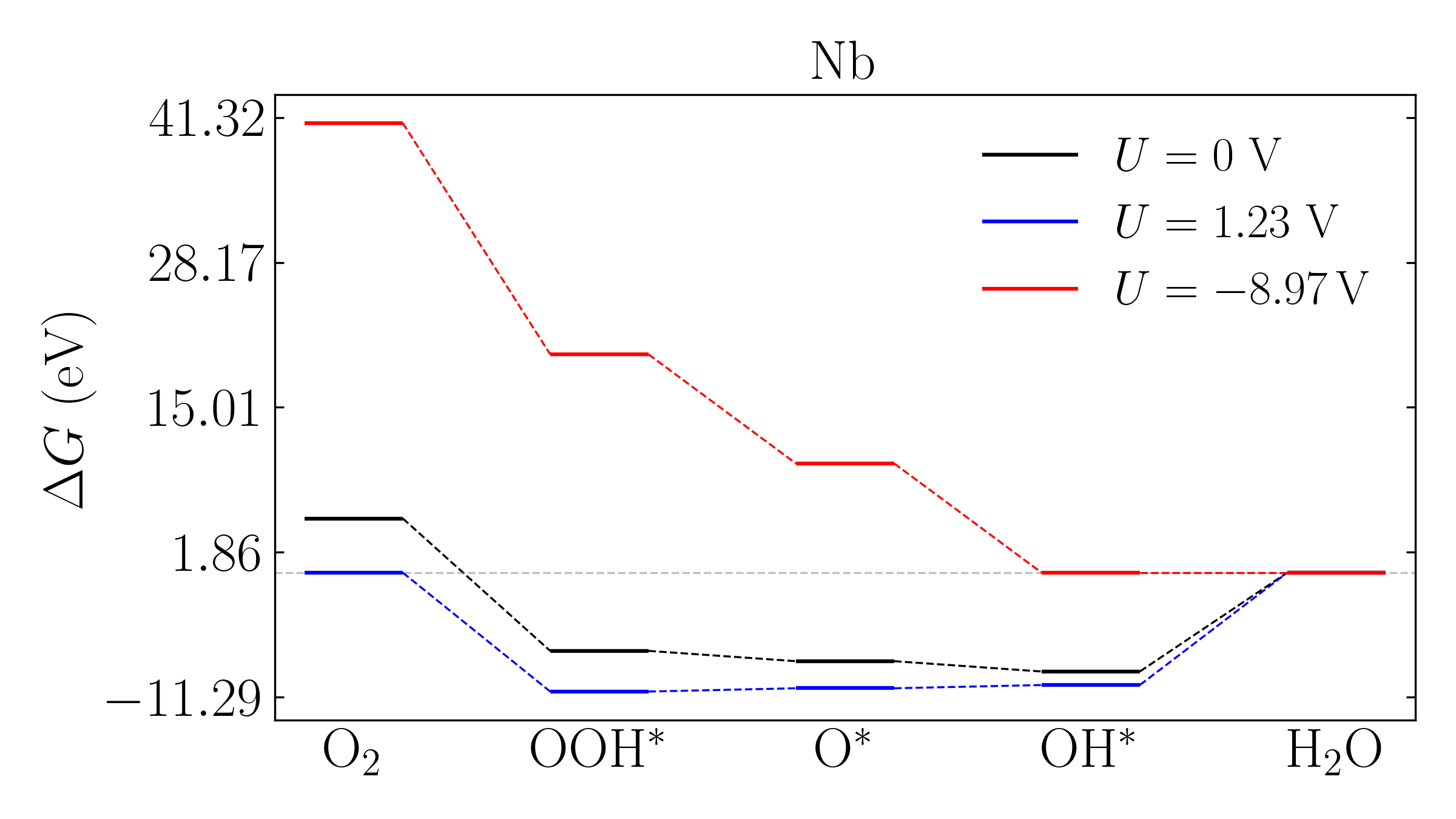}
  \end{subfigure}\hfill%
  \begin{subfigure}[b]{0.5\textwidth}
    \centering
    \includegraphics[width=\linewidth,height=0.4\textheight,keepaspectratio]{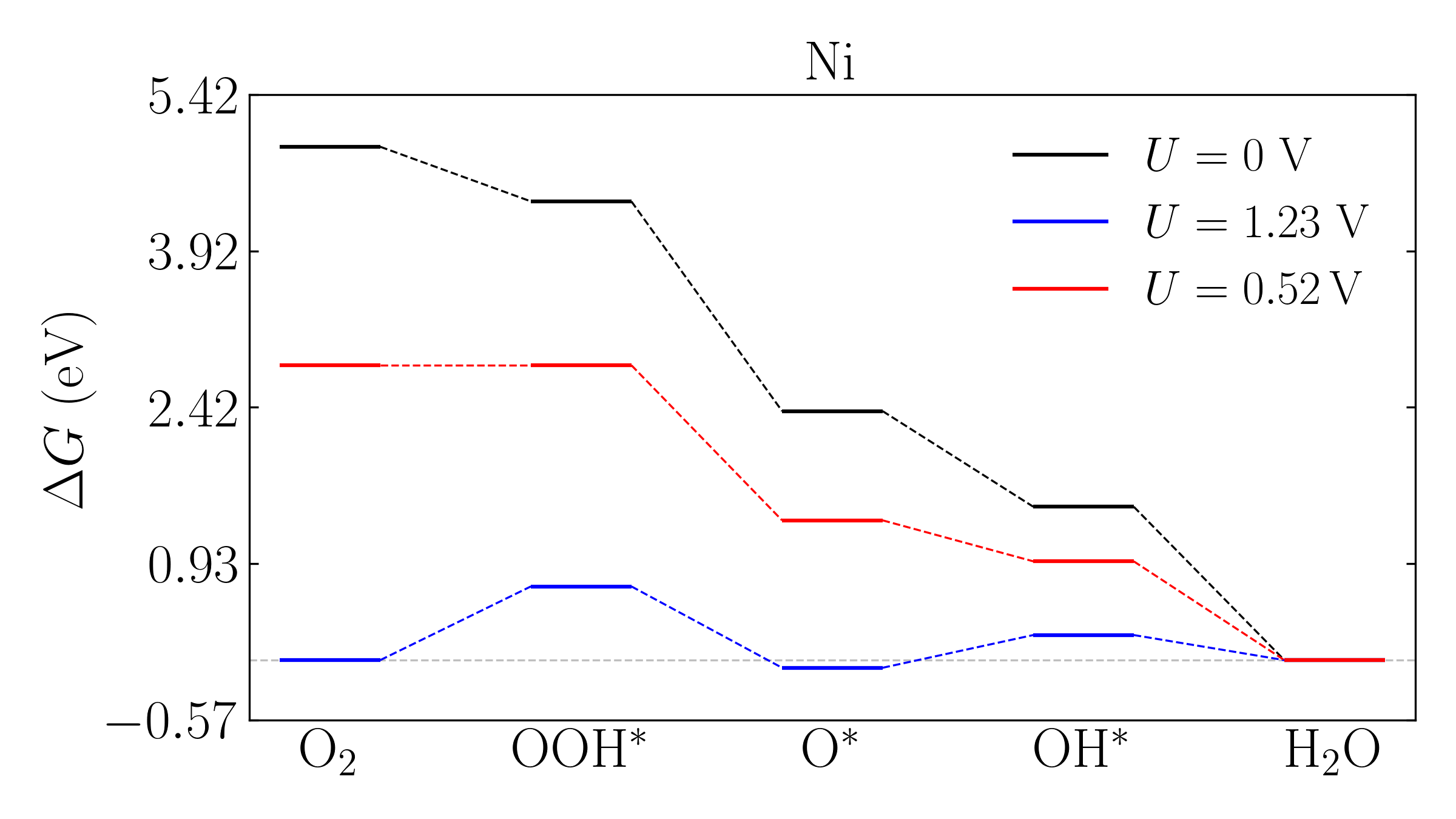}
  \end{subfigure}

  \par\vspace{-8pt}

  \begin{subfigure}[b]{0.5\textwidth}
    \centering
    \includegraphics[width=\linewidth,height=0.4\textheight,keepaspectratio]{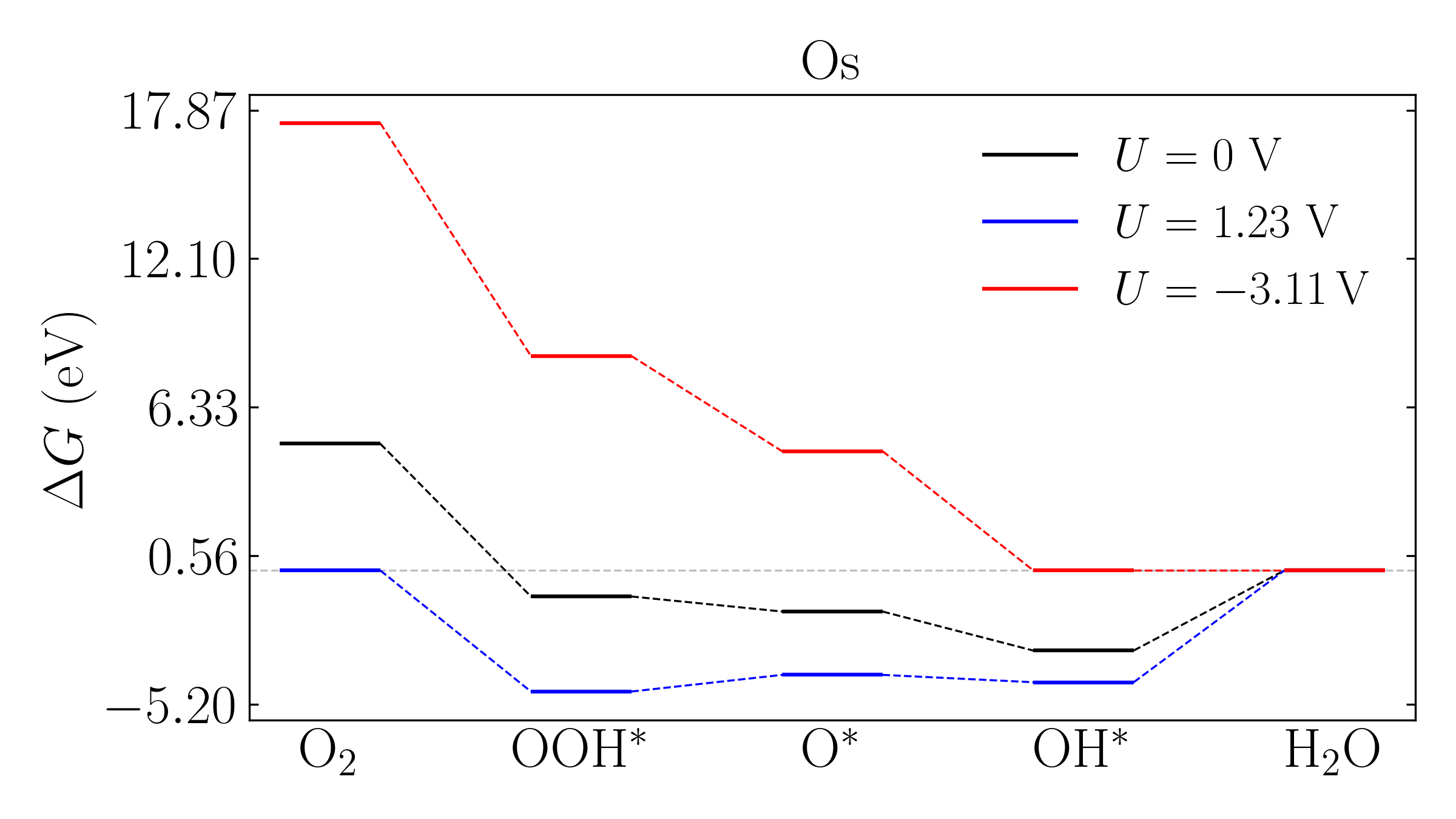}
  \end{subfigure}\hfill%
  \begin{subfigure}[b]{0.5\textwidth}
    \centering
    \includegraphics[width=\linewidth,height=0.4\textheight,keepaspectratio]{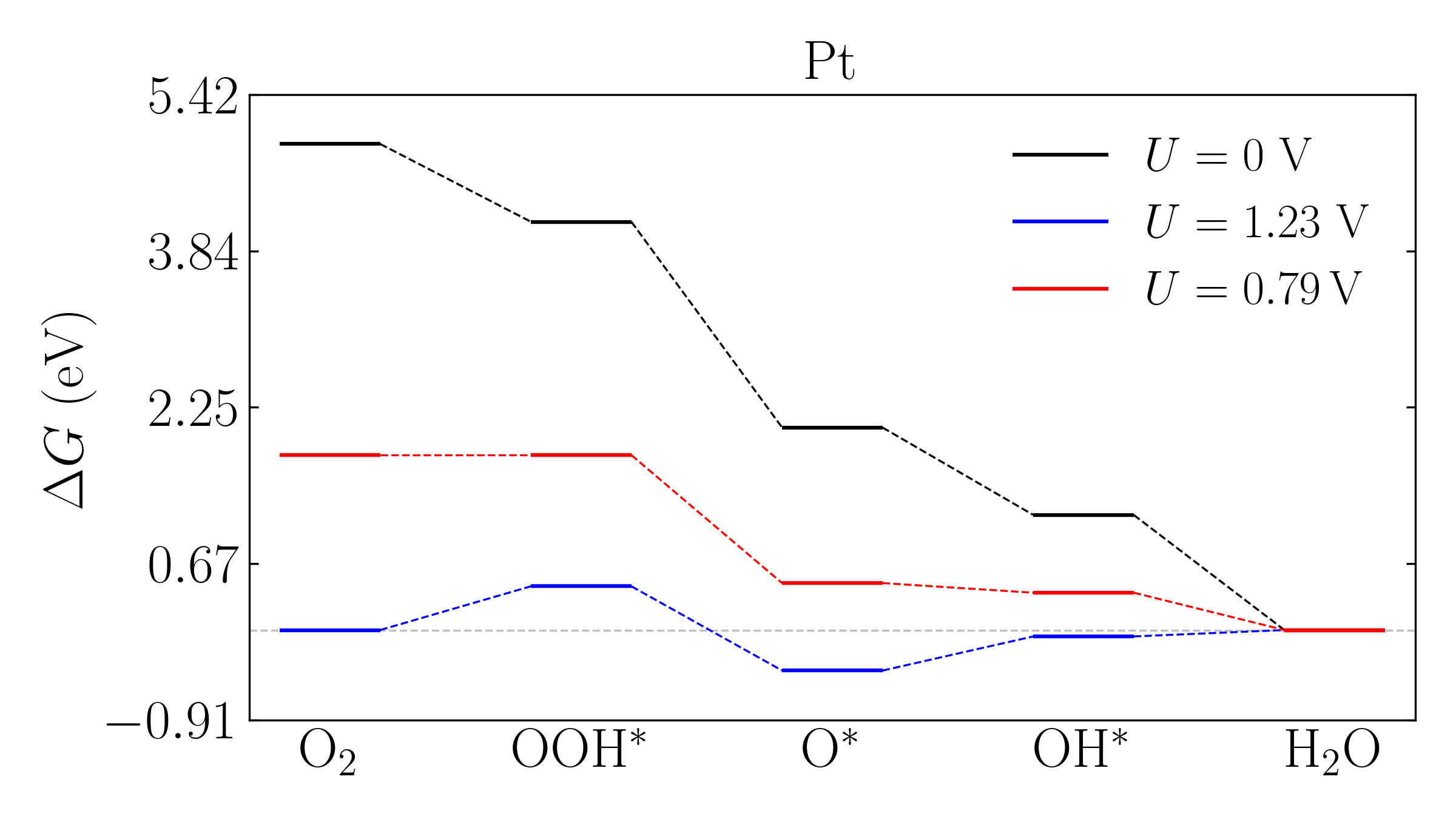}
  \end{subfigure}

\caption{ORR Gibbs free energy for B/M/B (M $=$ Nb, Ni, Os and Pt) at different applied potentials: $U =0$ V (black), $U =1.23$ V (blue), and the onset potential $U =U_{onset}$ (red). }\label{fig:orr-all2}
\end{figure}

\begin{figure}[H]
  \centering
  \begin{subfigure}[b]{0.5\textwidth}
    \centering
    \includegraphics[width=\linewidth,height=0.4\textheight,keepaspectratio]{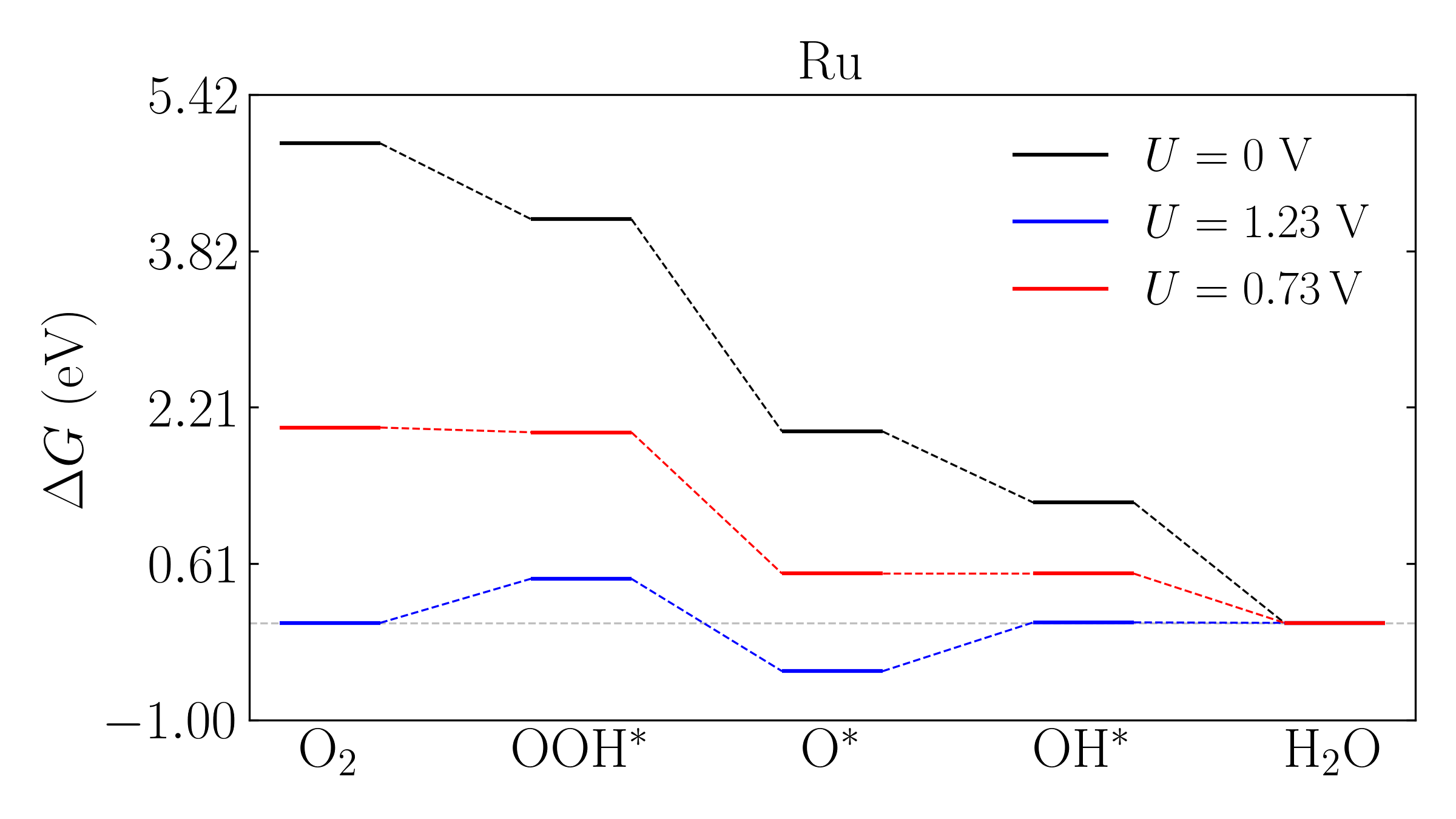}
  \end{subfigure}\hfill%
  \begin{subfigure}[b]{0.5\textwidth}
    \centering
    \includegraphics[width=\linewidth,height=0.4\textheight,keepaspectratio]{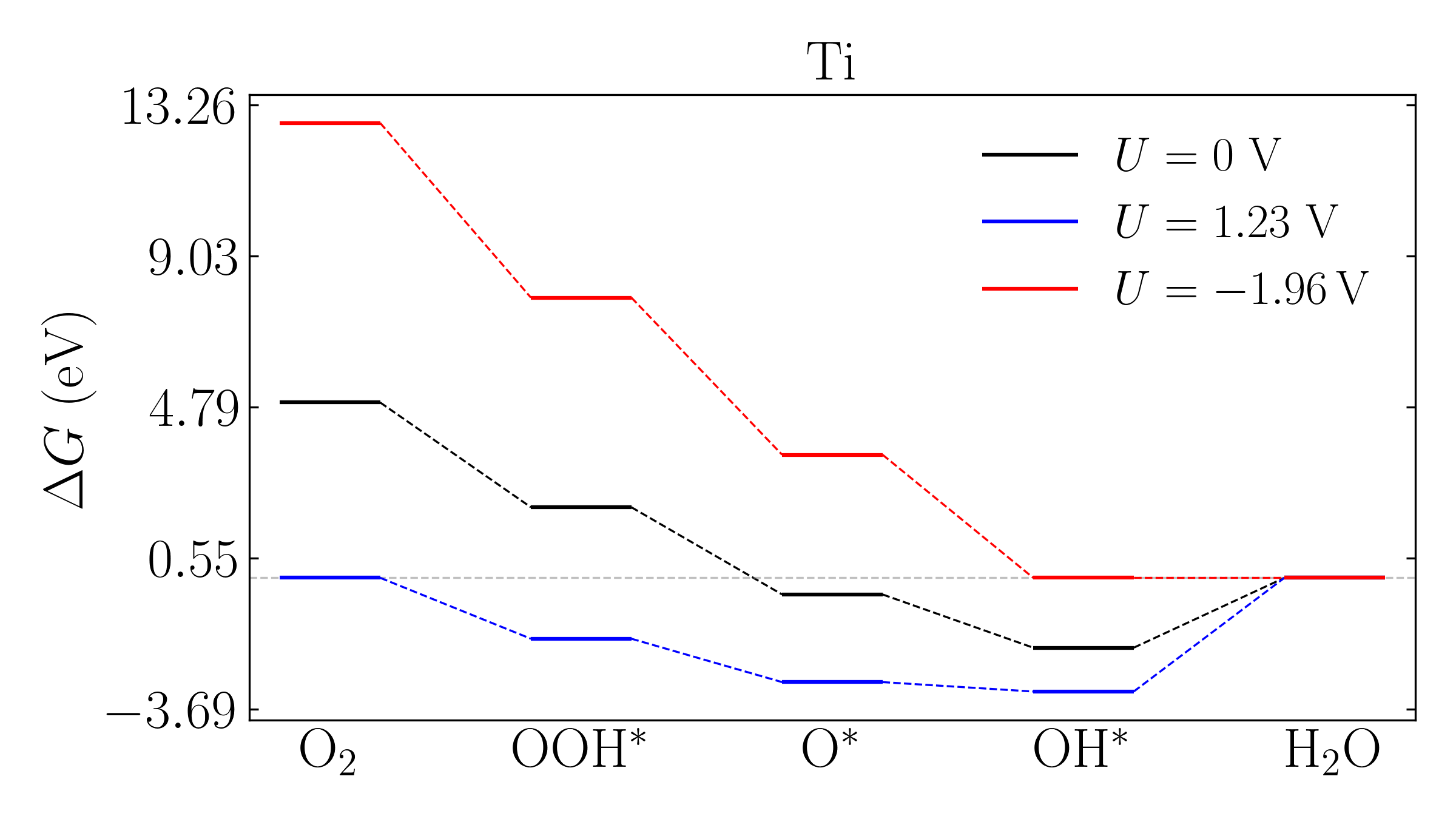}
  \end{subfigure}
  
  \par\vspace{-8pt}

  \begin{subfigure}[b]{0.5\textwidth}
    \centering
    \includegraphics[width=\linewidth,height=0.4\textheight,keepaspectratio]{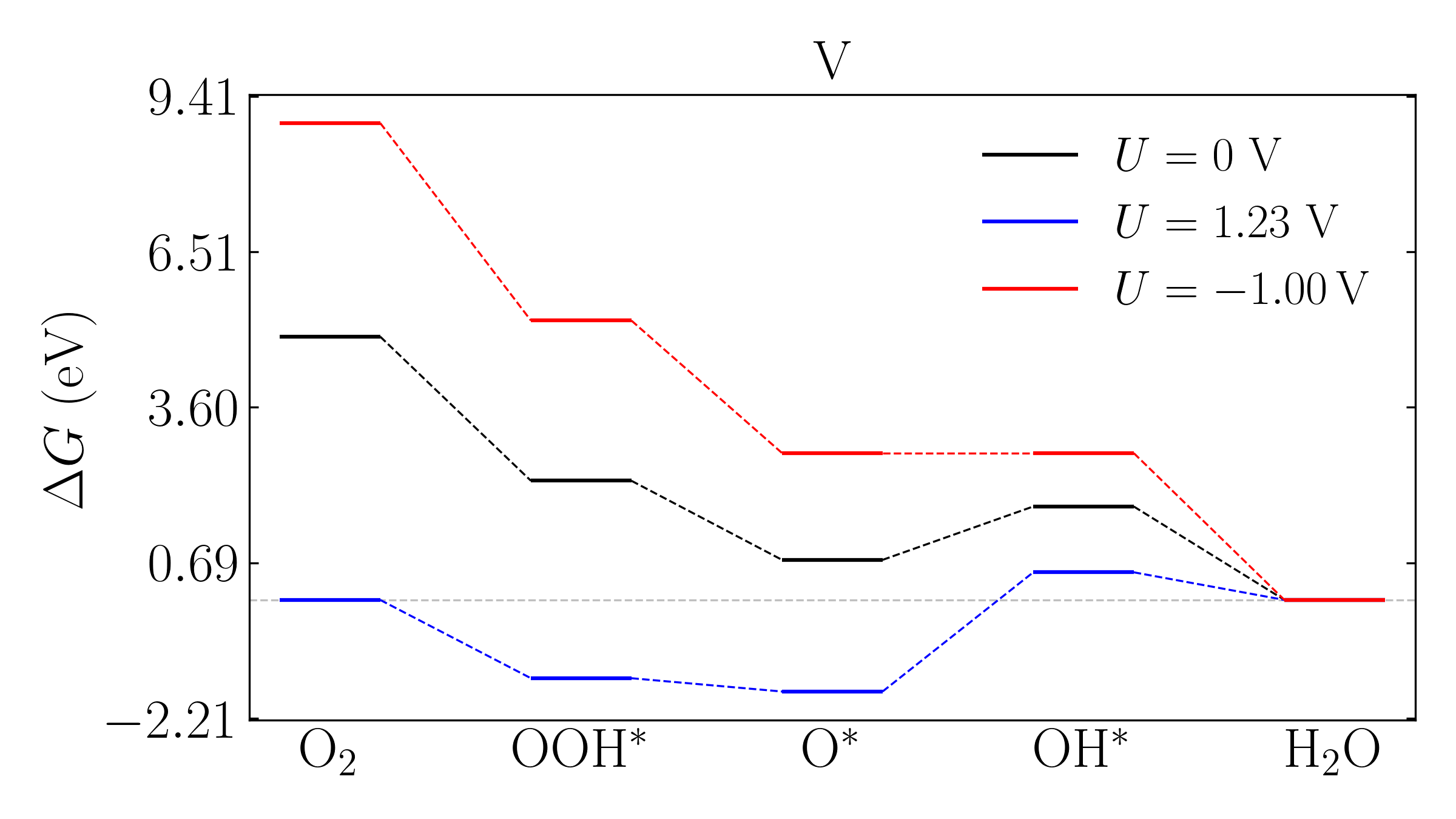}
  \end{subfigure}\hfill%
  \begin{subfigure}[b]{0.5\textwidth}
    \centering
    \includegraphics[width=\linewidth,height=0.4\textheight,keepaspectratio]{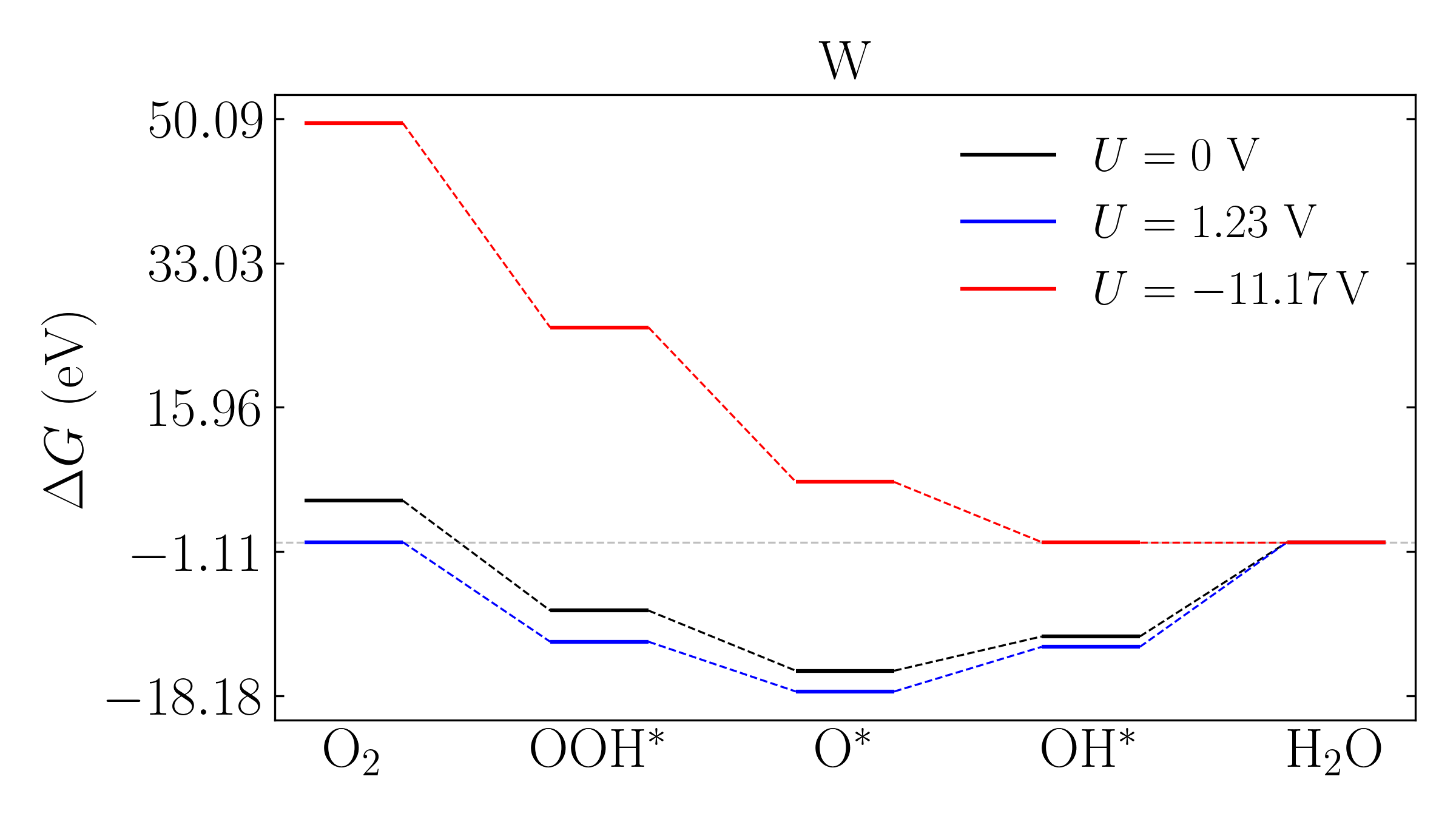}
  \end{subfigure}

  \caption{ORR Gibbs free energy for B/M/B (M $=$ Ru, Ti, V and W) at different applied potentials: $U =0$ V (black), $U =1.23$ V (blue), and the onset potential $U =U_{onset}$ (red). }\label{fig:orr-all3}
\end{figure}

\begin{figure}[H]
  \centering
  \begin{subfigure}[b]{0.5\textwidth}
    \centering
    \includegraphics[width=\linewidth,height=0.4\textheight,keepaspectratio]{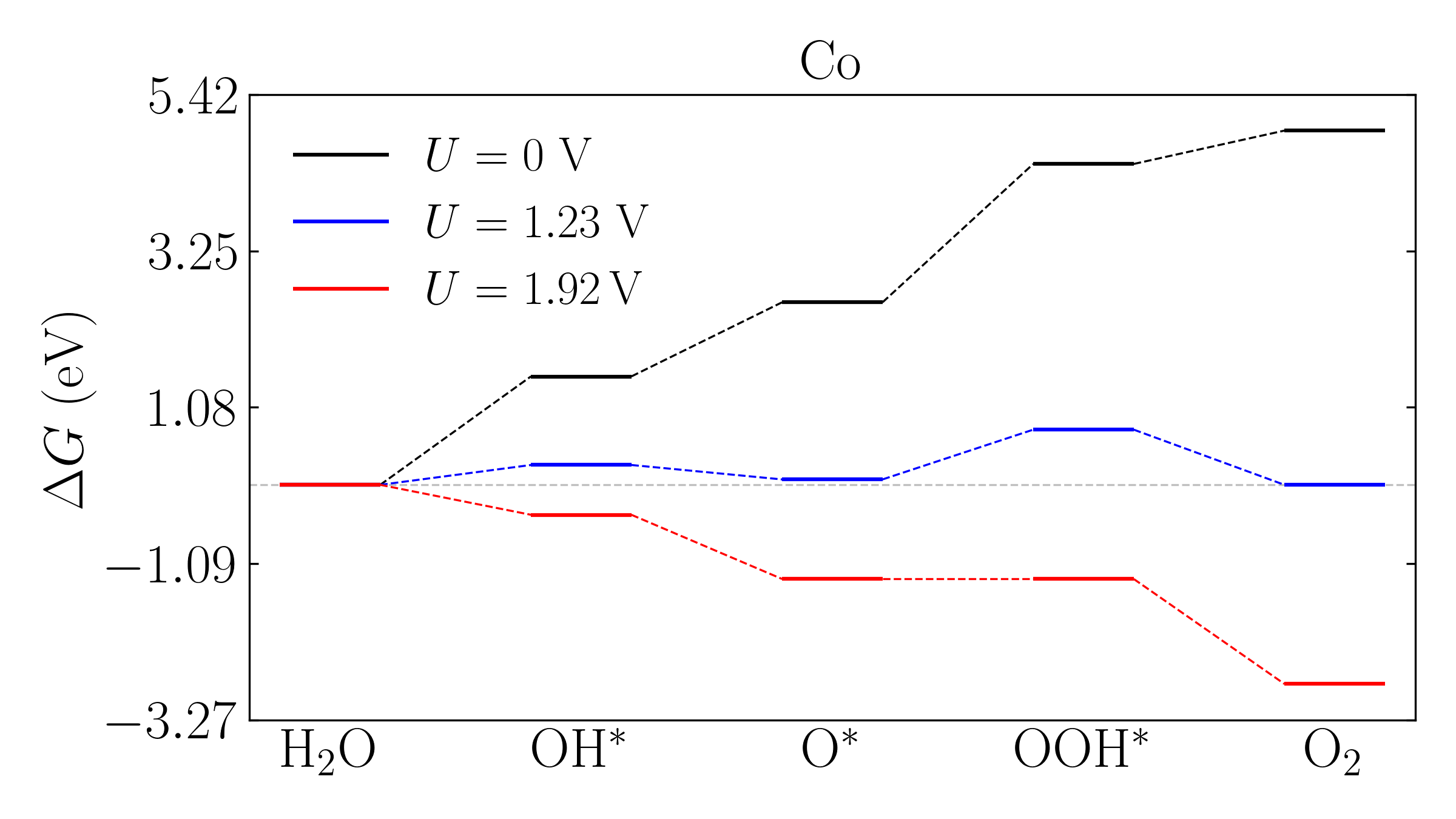}
  \end{subfigure}\hfill%
  \begin{subfigure}[b]{0.5\textwidth}
    \centering
    \includegraphics[width=\linewidth,height=0.4\textheight,keepaspectratio]{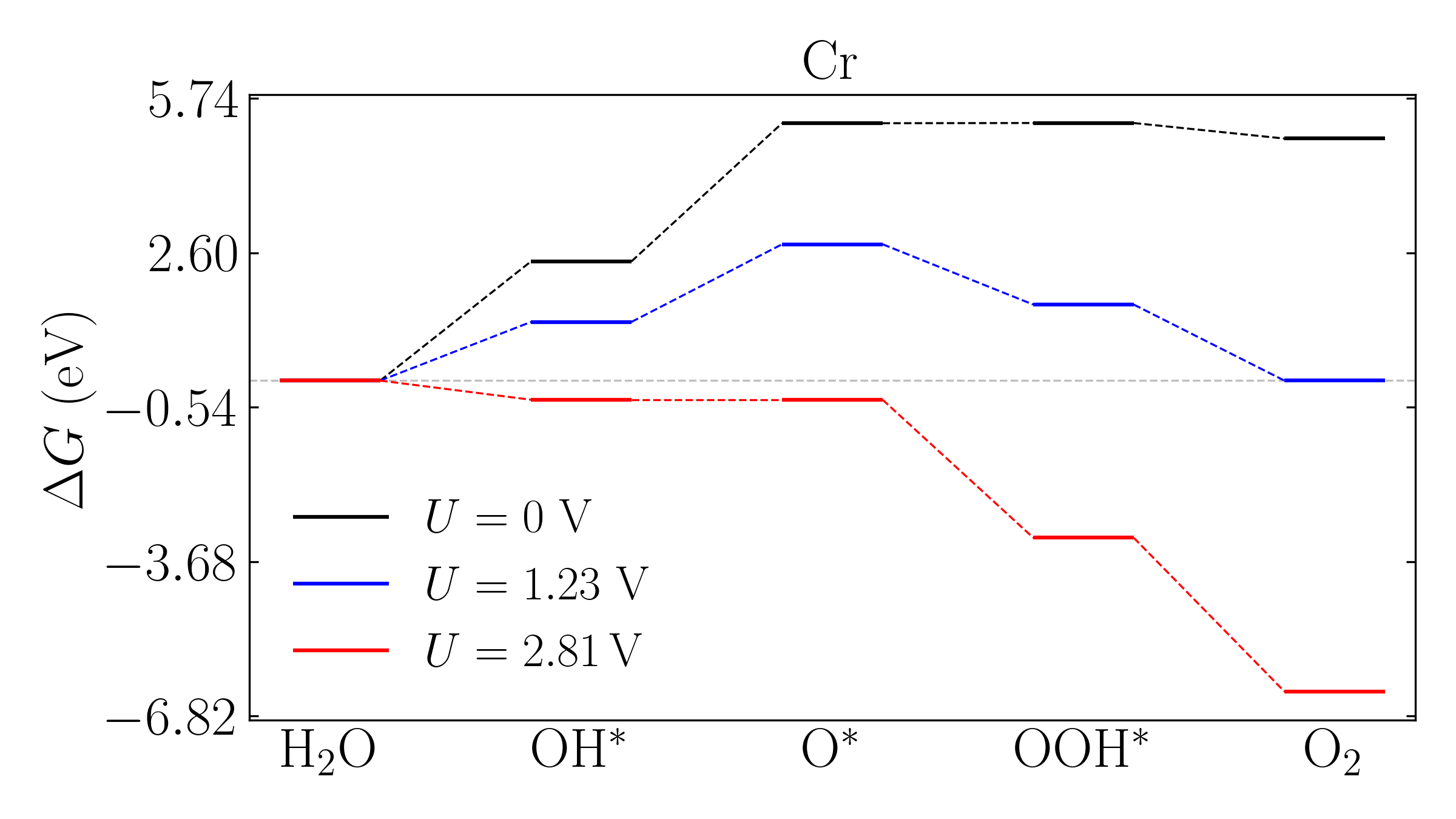}
  \end{subfigure}
  
  \par\vspace{-8pt}
  
  \begin{subfigure}[b]{0.5\textwidth}
    \centering
    \includegraphics[width=\linewidth,height=0.4\textheight,keepaspectratio]{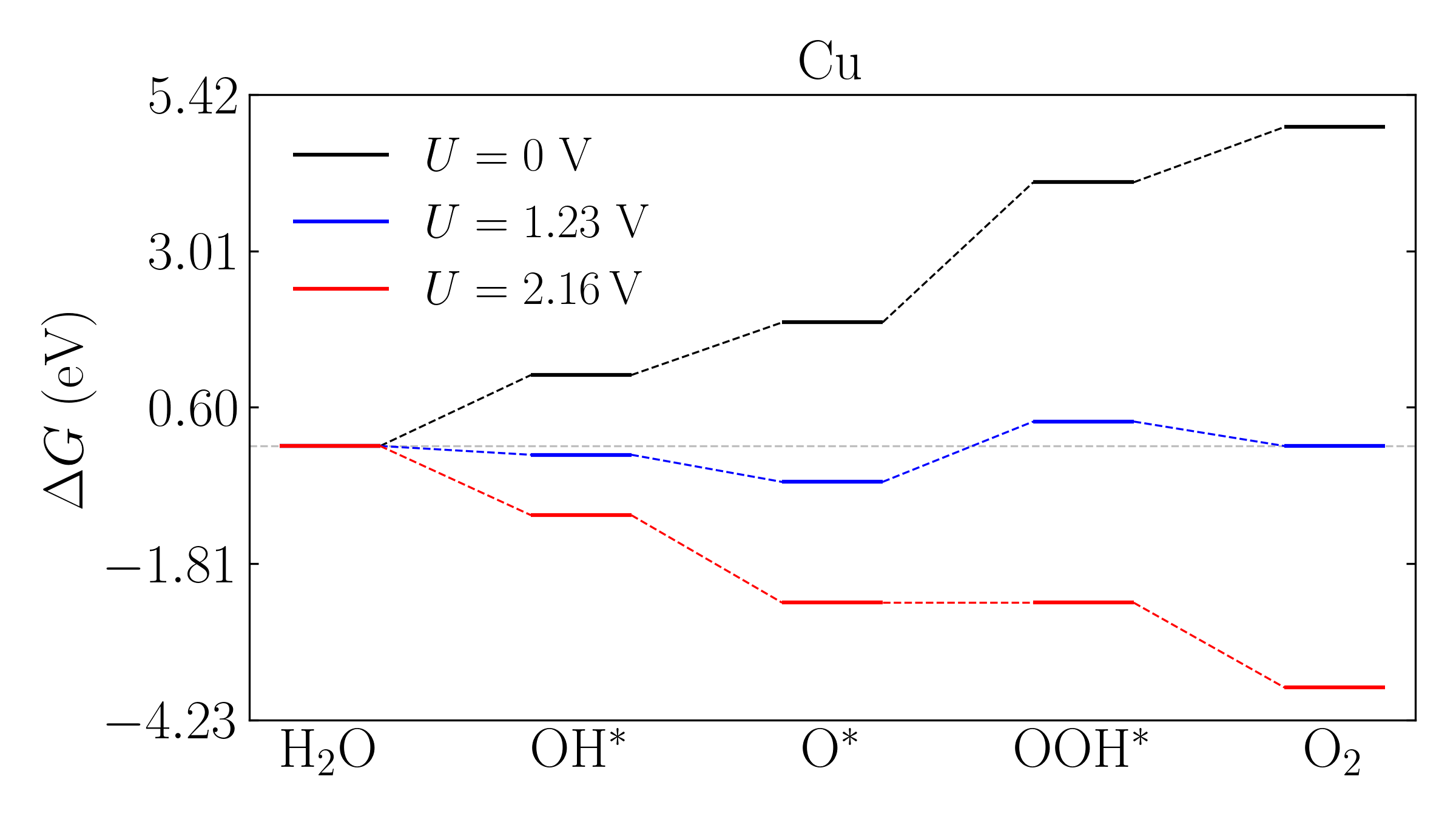}
  \end{subfigure}\hfill%
  \begin{subfigure}[b]{0.5\textwidth}
    \centering
    \includegraphics[width=\linewidth,height=0.4\textheight,keepaspectratio]{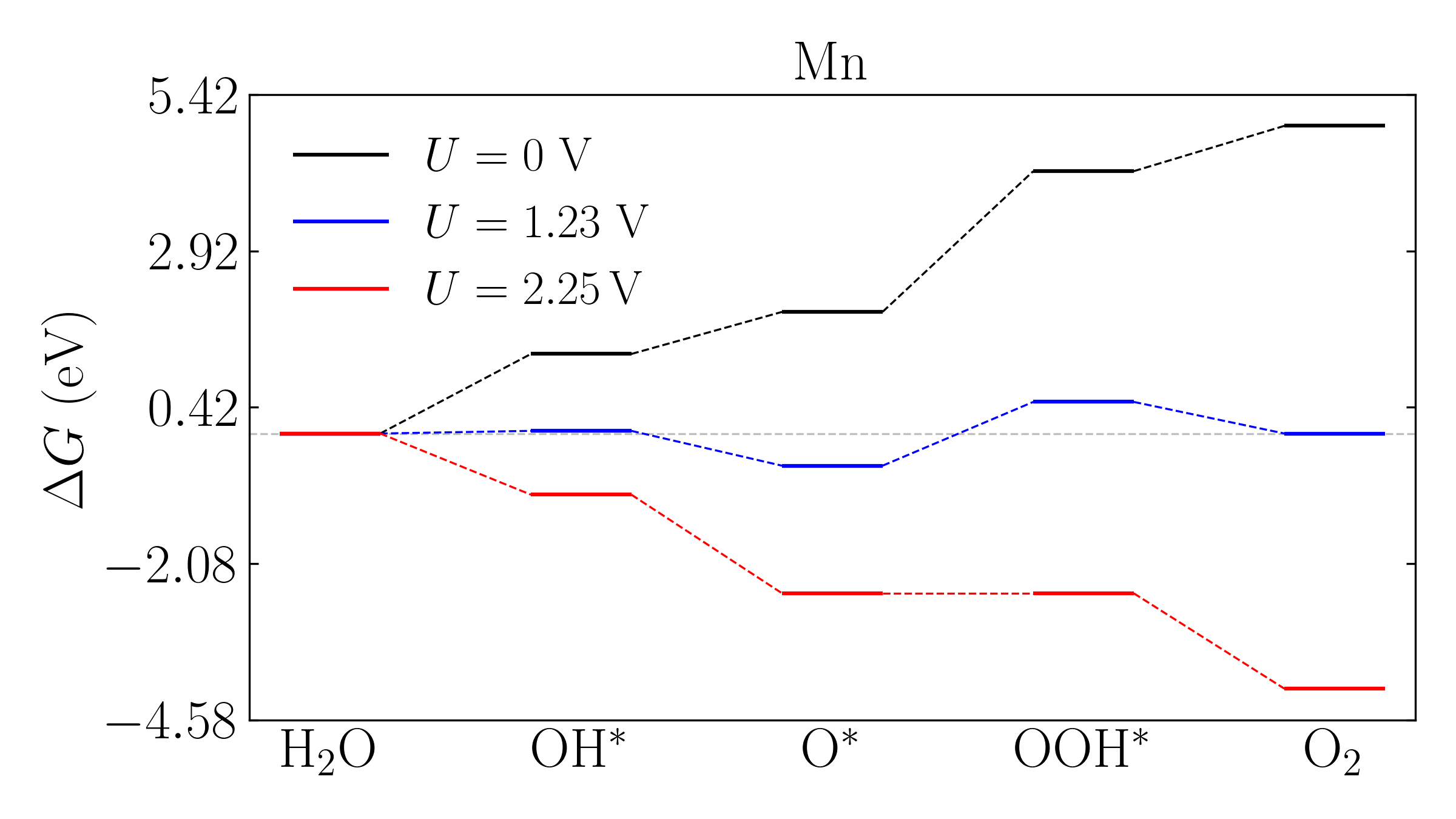}
  \end{subfigure}

  \caption{OER Gibbs free energy for B/M/B (M $=$ Co, Cr, Cu and Mn) at different applied potentials: $U =0$ V (black), $U =1.23$ V (blue), and the onset potential $U =U_{onset}$ (red). }\label{fig:oer-all1}
\end{figure}

\begin{figure}[H]
  \centering
  \begin{subfigure}[b]{0.5\textwidth}
    \centering
    \includegraphics[width=\linewidth,height=0.4\textheight,keepaspectratio]{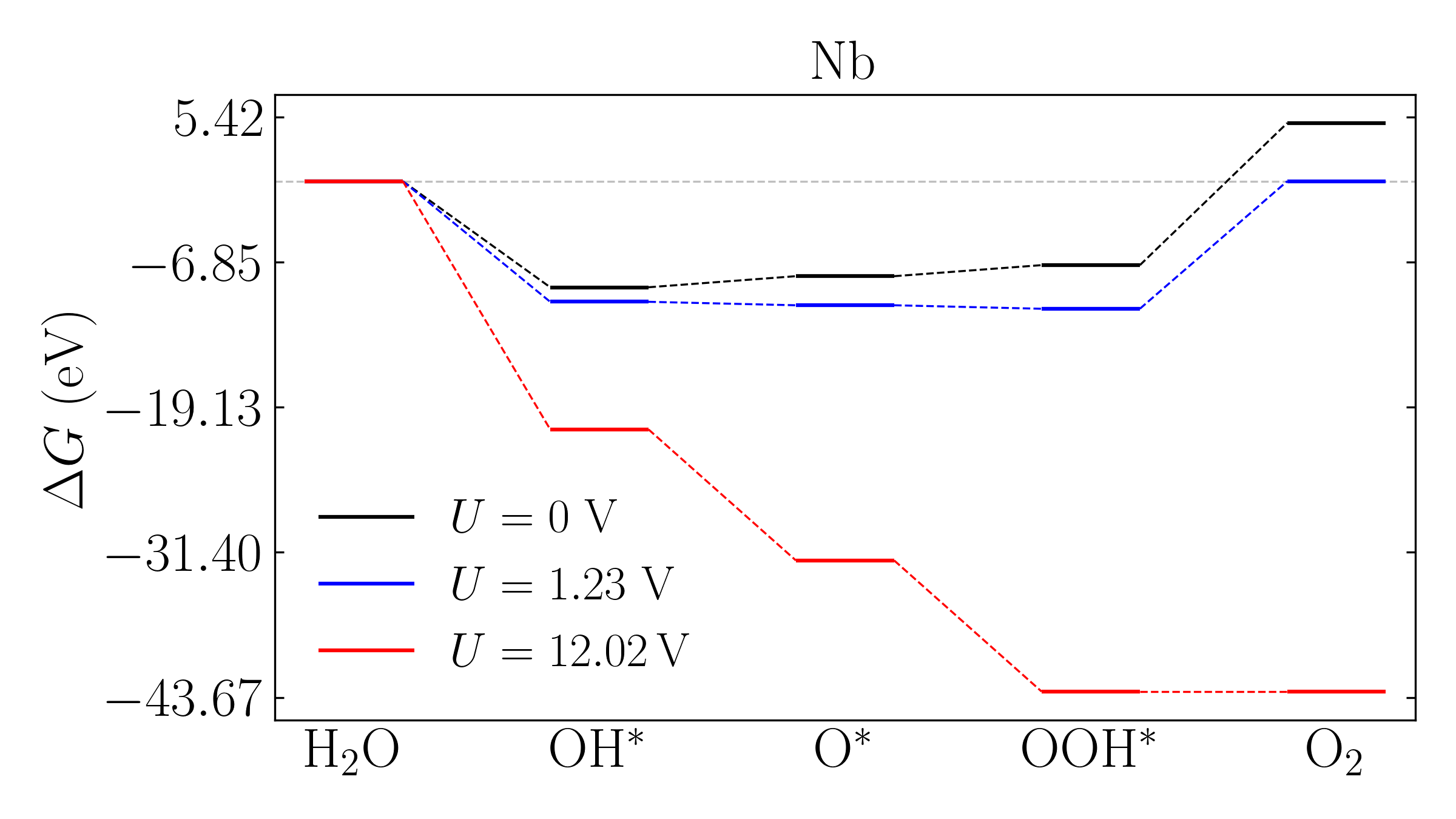}
  \end{subfigure}\hfill%
  \begin{subfigure}[b]{0.5\textwidth}
    \centering
    \includegraphics[width=\linewidth,height=0.4\textheight,keepaspectratio]{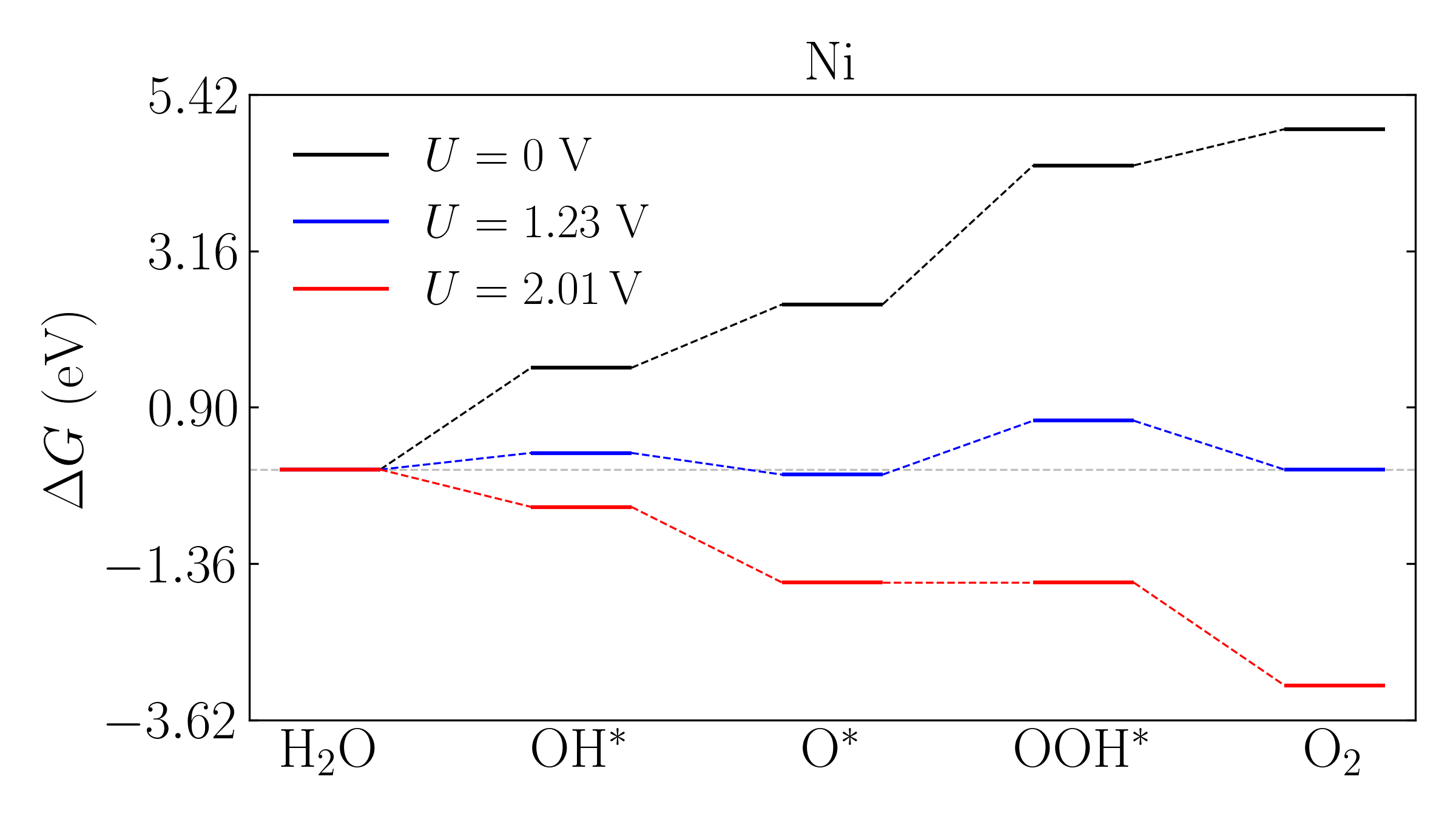}
  \end{subfigure}

  \par\vspace{-8pt}

  \begin{subfigure}[b]{0.5\textwidth}
    \centering
    \includegraphics[width=\linewidth,height=0.4\textheight,keepaspectratio]{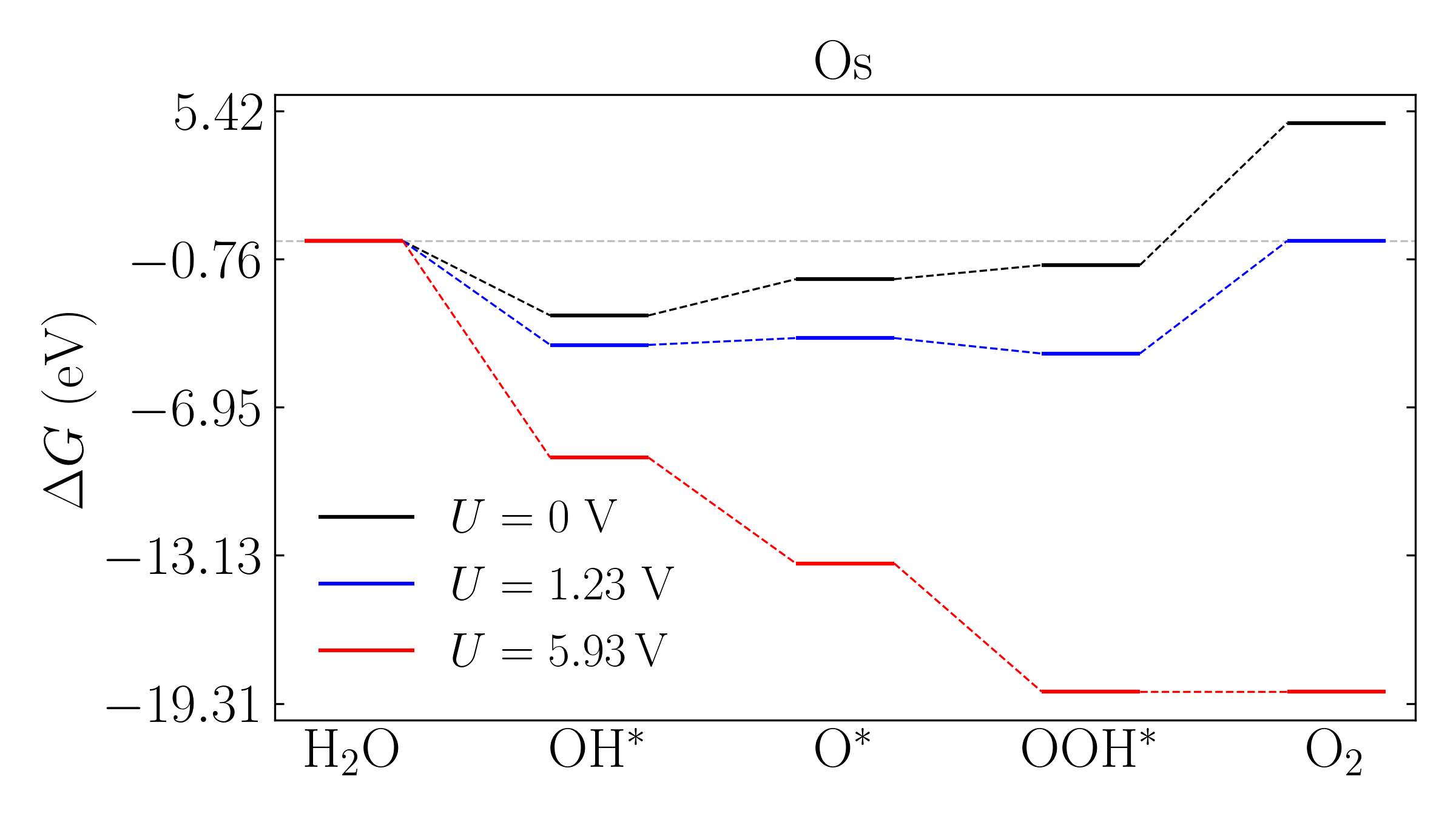}
  \end{subfigure}\hfill%
  \begin{subfigure}[b]{0.5\textwidth}
    \centering
    \includegraphics[width=\linewidth,height=0.4\textheight,keepaspectratio]{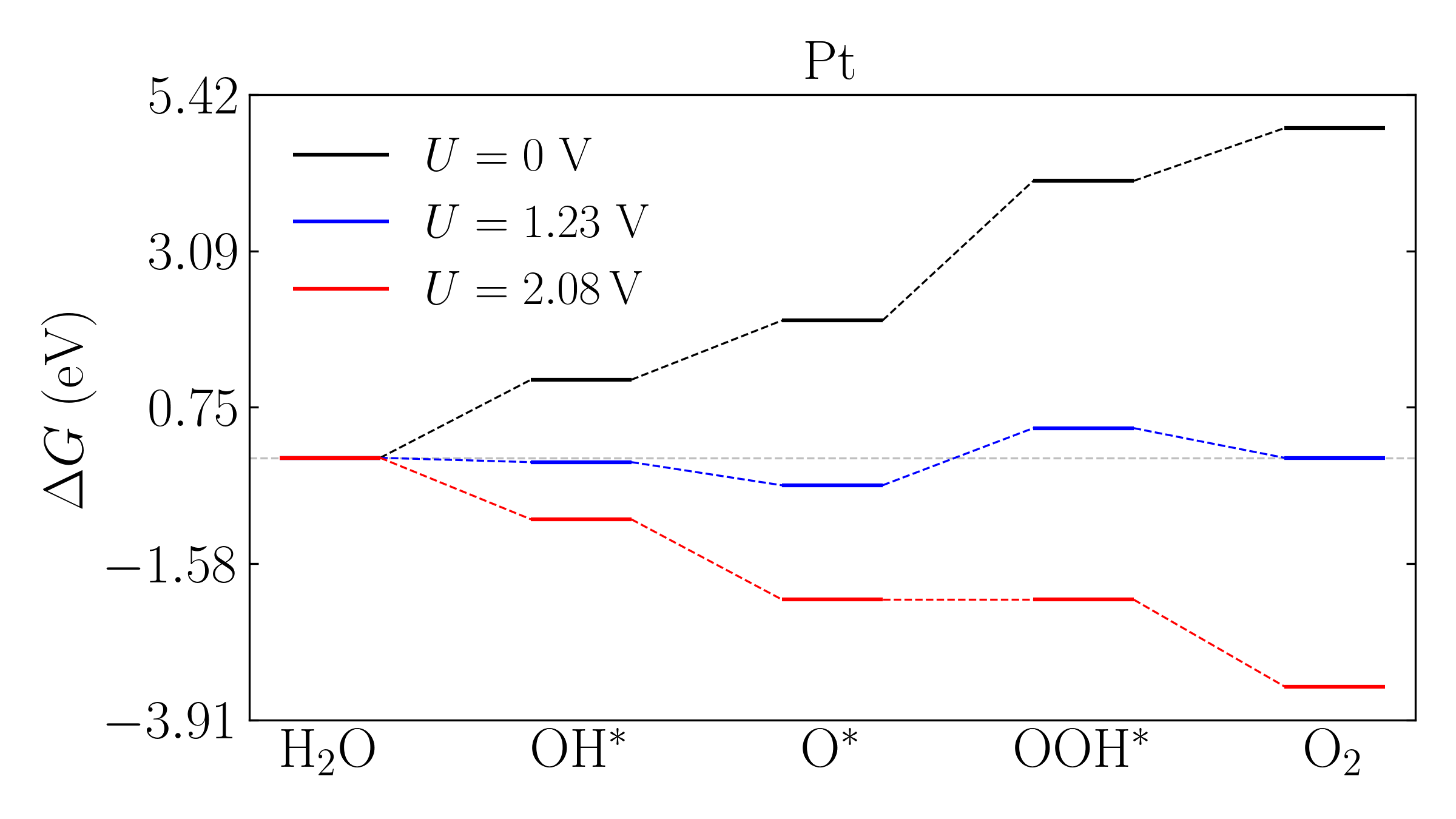}
  \end{subfigure}

  \caption{OER Gibbs free energy for B/M/B (M $=$ Nb, Ni, Os and Pt) at different applied potentials: $U =0$ V (black), $U =1.23$ V (blue), and the onset potential $U =U_{onset}$ (red). }\label{fig:oer-all2}
\end{figure}

\begin{figure}[H]
  \centering
  \begin{subfigure}[b]{0.5\textwidth}
    \centering
    \includegraphics[width=\linewidth,height=0.4\textheight,keepaspectratio]{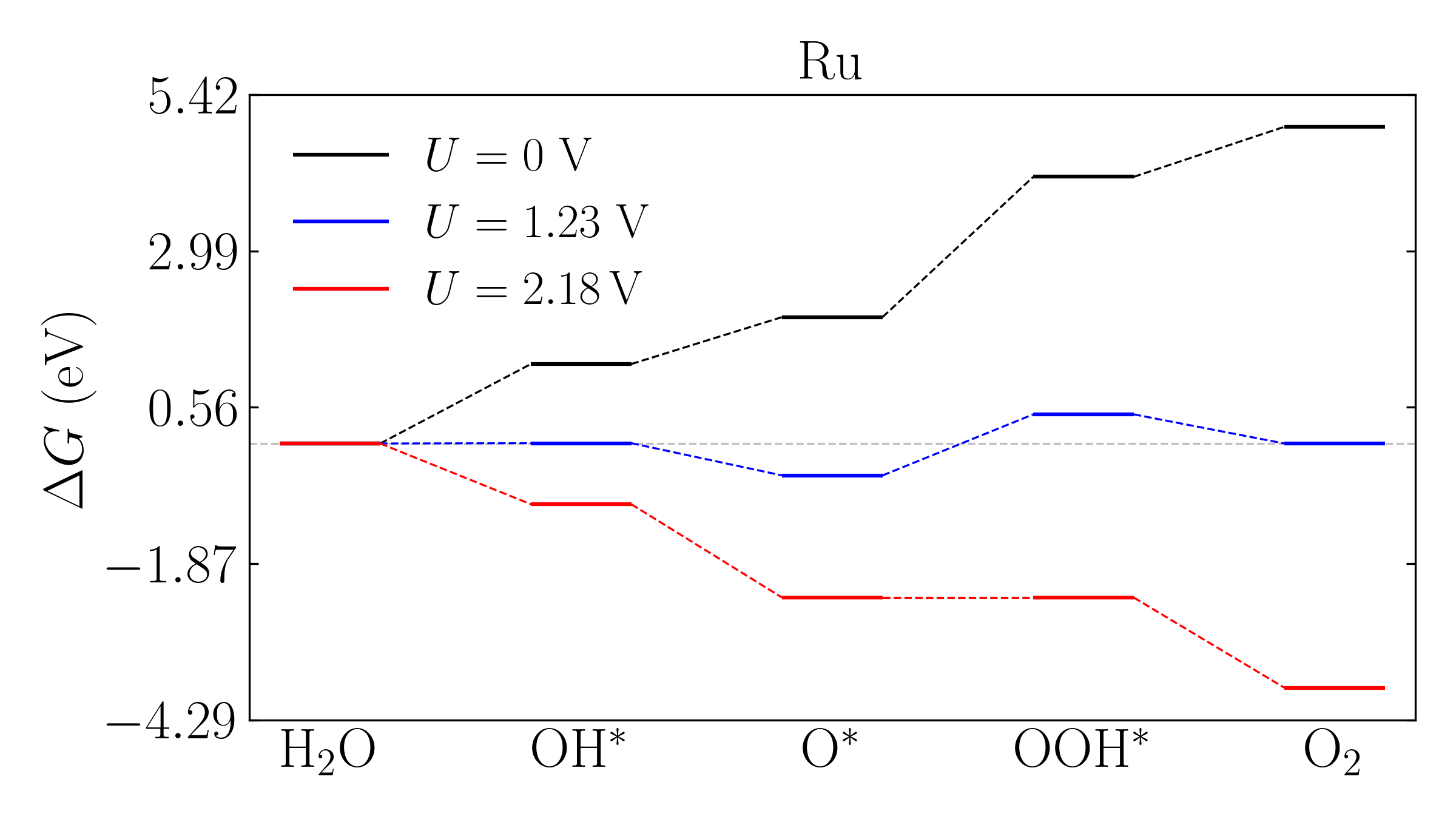}
  \end{subfigure}\hfill%
  \begin{subfigure}[b]{0.5\textwidth}
    \centering
    \includegraphics[width=\linewidth,height=0.4\textheight,keepaspectratio]{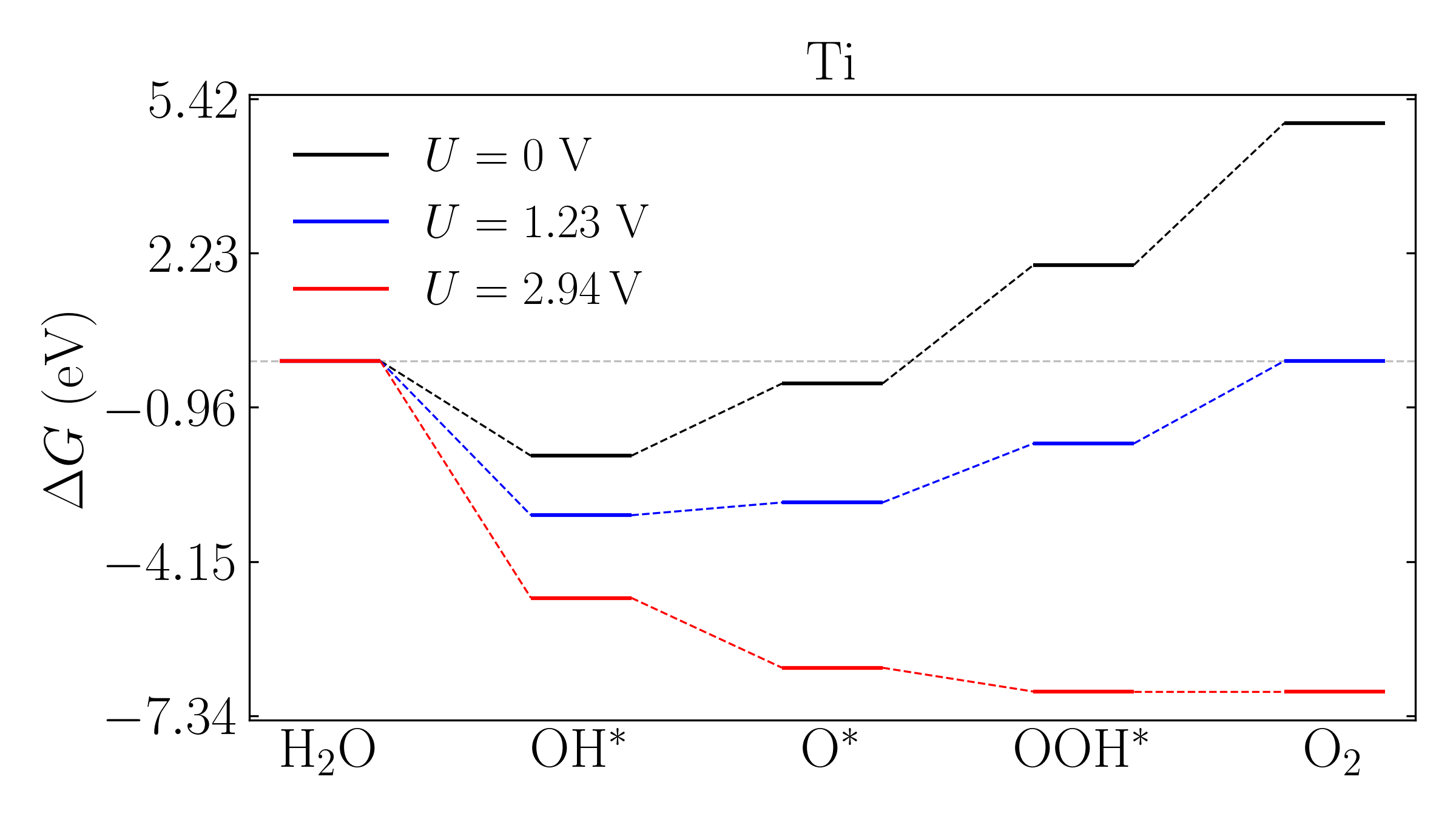}
  \end{subfigure}
  \par\vspace{-8pt}

  \begin{subfigure}[b]{0.5\textwidth}
    \centering
    \includegraphics[width=\linewidth,height=0.4\textheight,keepaspectratio]{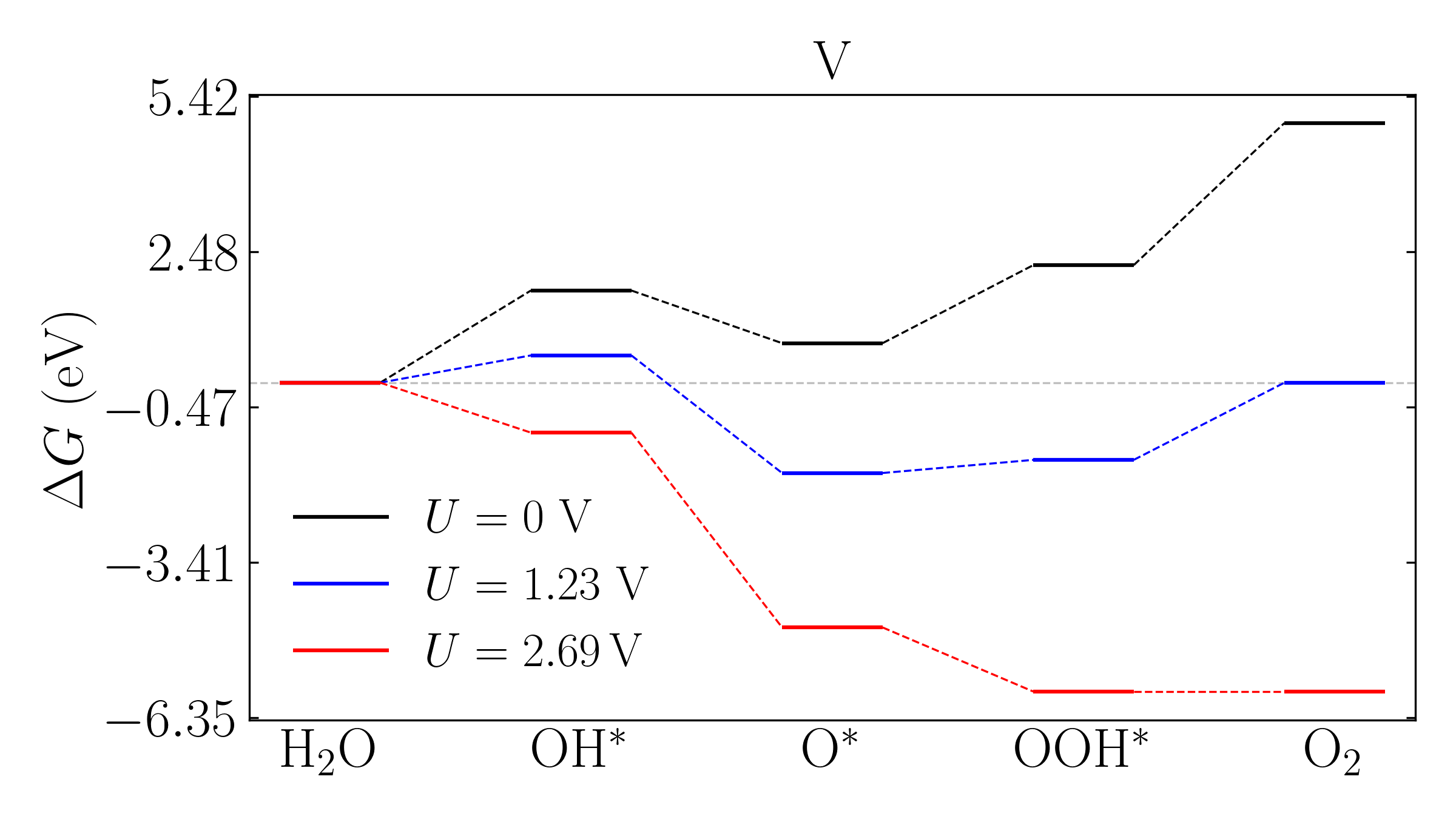}
  \end{subfigure}\hfill%
  \begin{subfigure}[b]{0.5\textwidth}
    \centering
    \includegraphics[width=\linewidth,height=0.4\textheight,keepaspectratio]{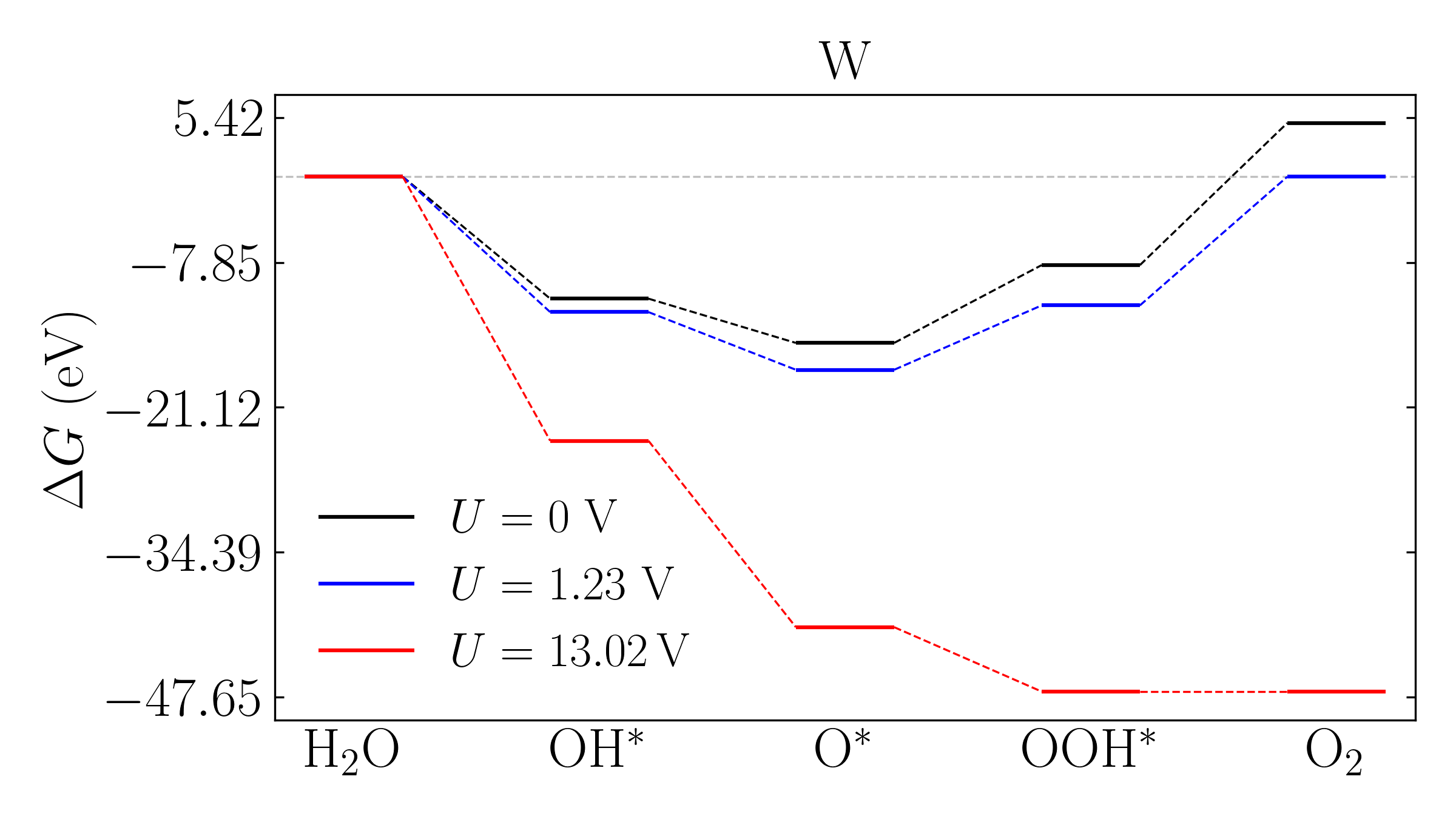}
  \end{subfigure}

  \caption{OER Gibbs free energy for B/M/B (M $=$ Ru, Ti, V and W) at different applied potentials: $U =0$ V (black), $U =1.23$ V (blue), and the onset potential $U =U_{onset}$ (red). }\label{fig:oer-all3}
\end{figure}

Here, we present the Gibbs free energies diagrams for each B/M/B, for the ORR, Figs.\,\ref{fig:orr-all1},\,\ref{fig:orr-all2} and \,\ref{fig:orr-all3} and OER, Fig.\,\ref{fig:oer-all1},\,\ref{fig:oer-all2} and \,\ref{fig:oer-all3}, reactions pathways. All energies are referenced to standard conditions.

\begin{figure}[H]
    \centering
    \includegraphics[width=1\linewidth]{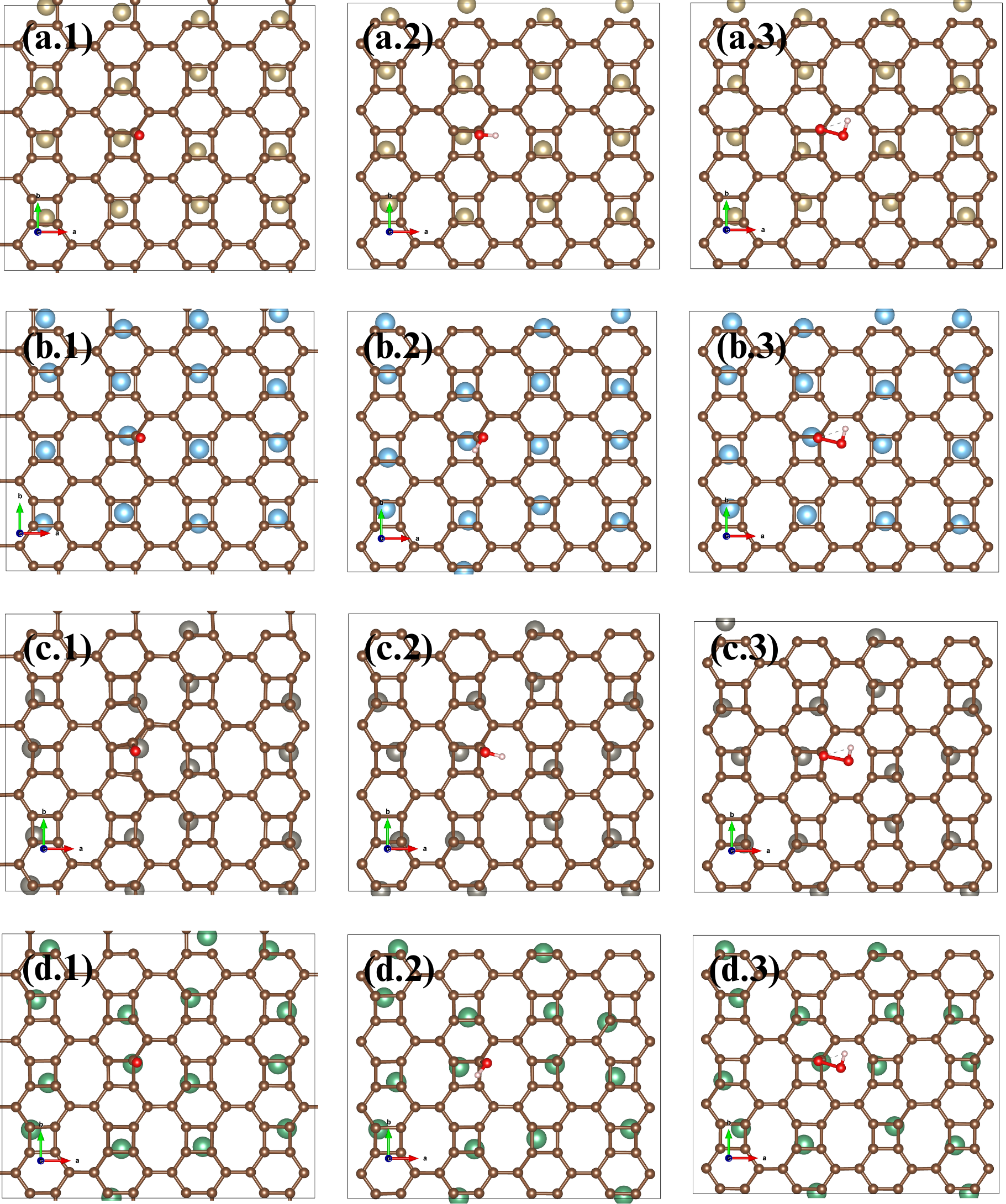}
    \caption{Top views of the optimized geometries for the B/Os/B ($a.x$), B/Ti/B ($b.x$), B/W/B ($c.x$) and B/Nb/B ($d.x$) systems following the adsorption of the O$^\ast$ ($z.1$), OH$^\ast$ ($z.2$), and OOH$^\ast$ ($z.3$) intermediates at the square-planar sites.}
    \label{fig:ti-os-w}
\end{figure}

Notably, the B/Os/B, B/Ti/B, B/W/B, and B/Nb/B systems exhibit the poorest catalytic performance among the B/M/B systems. InFig.\,\ref{fig:ti-os-w}, we illustrate a contributing factor to this low catalytic activity. Adsorption of any species-O$^\ast$, OH$^\ast$, or OOH$^\ast$-onto these specific systems induces a displacement of the M$=$Os,Ti,W and Nb atoms from the C$^{468}$ sites. This general shift of the metal atoms leads to a considerable increase in the energy of the $\ast +$O$^\ast$ (or $\ast +$OH$^\ast$ or $\ast +$OOH$^\ast$) system, thereby raising the energy barriers and consequently increasing the overpotential for both the ORR and OER.

\begin{figure}[H]
    \centering
    \includegraphics[width=1\linewidth]{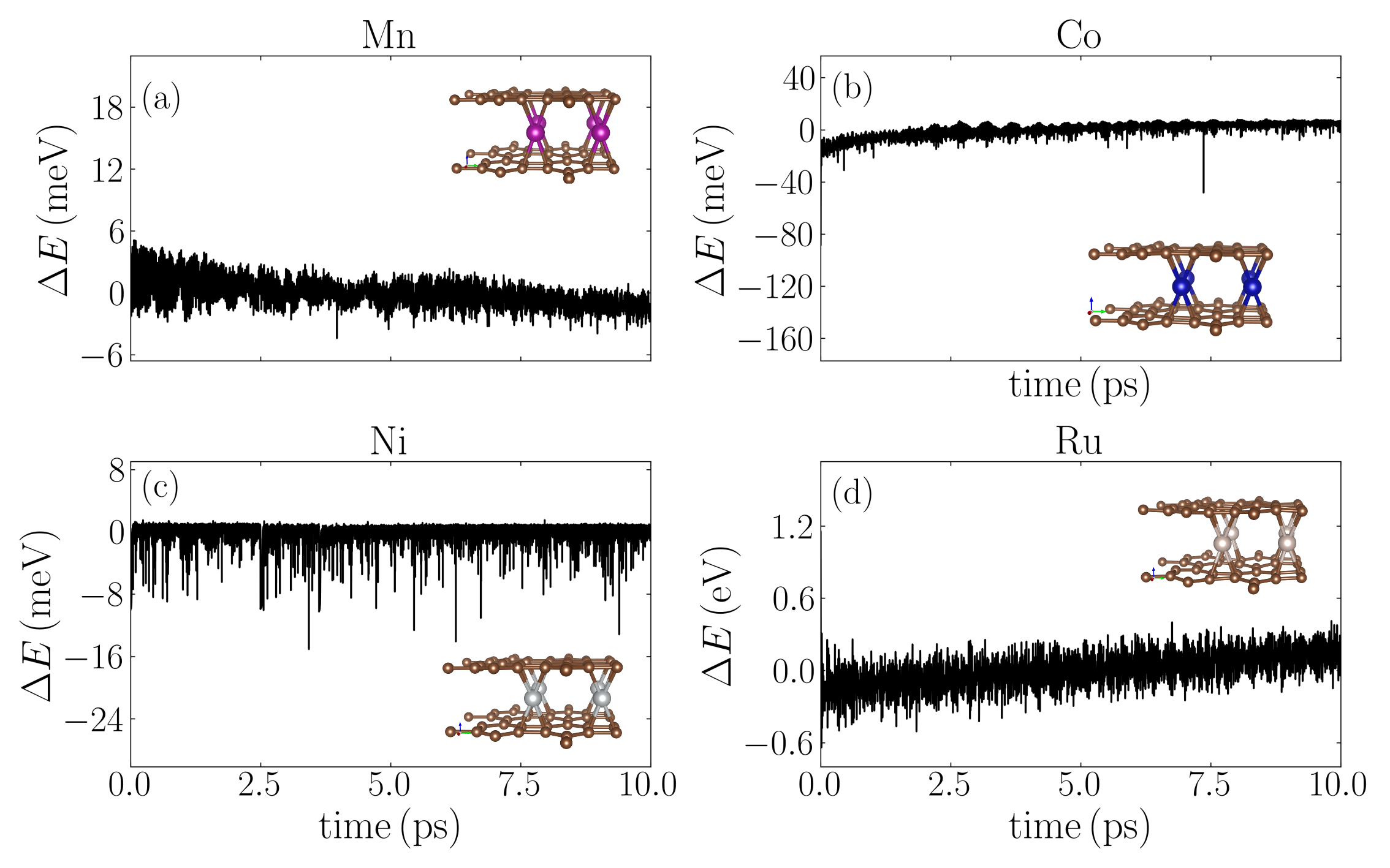}
    \caption{AIMD simulation of B/M/B with M $=$ Mn, Co, Ni and Ru for $10$\,ps.}
    \label{fig:fig_others_aimd}
\end{figure}

We also show the AIMD for the B/M/B systems with M = Mn, Co, Ni, and Ru in ~\ref{fig:fig_others_aimd}, i.e., other B/M/B high-catalytic-performance systems for ORR and OER. With these results, it is possible to consider that the B/M/B systems with relatively high catalytic performance are stable at $T=400$\,K. Additionally, we can infer that the intercalated metals do not cluster and remain in the square-planar site of the B/B framework.

Additionally, we performed the AIMD of B/Fe/B in the presence of water, at the experimental concentration (1\,g\,cm$^{-3}$). These results indicate that the biphenylene bilayer effectively acts as a protective barrier, shielding the intercalated metal centers from direct interaction with the aqueous environment where ORR/OER occur.

\begin{figure}[H]
    \centering
    \includegraphics[width=1\linewidth]{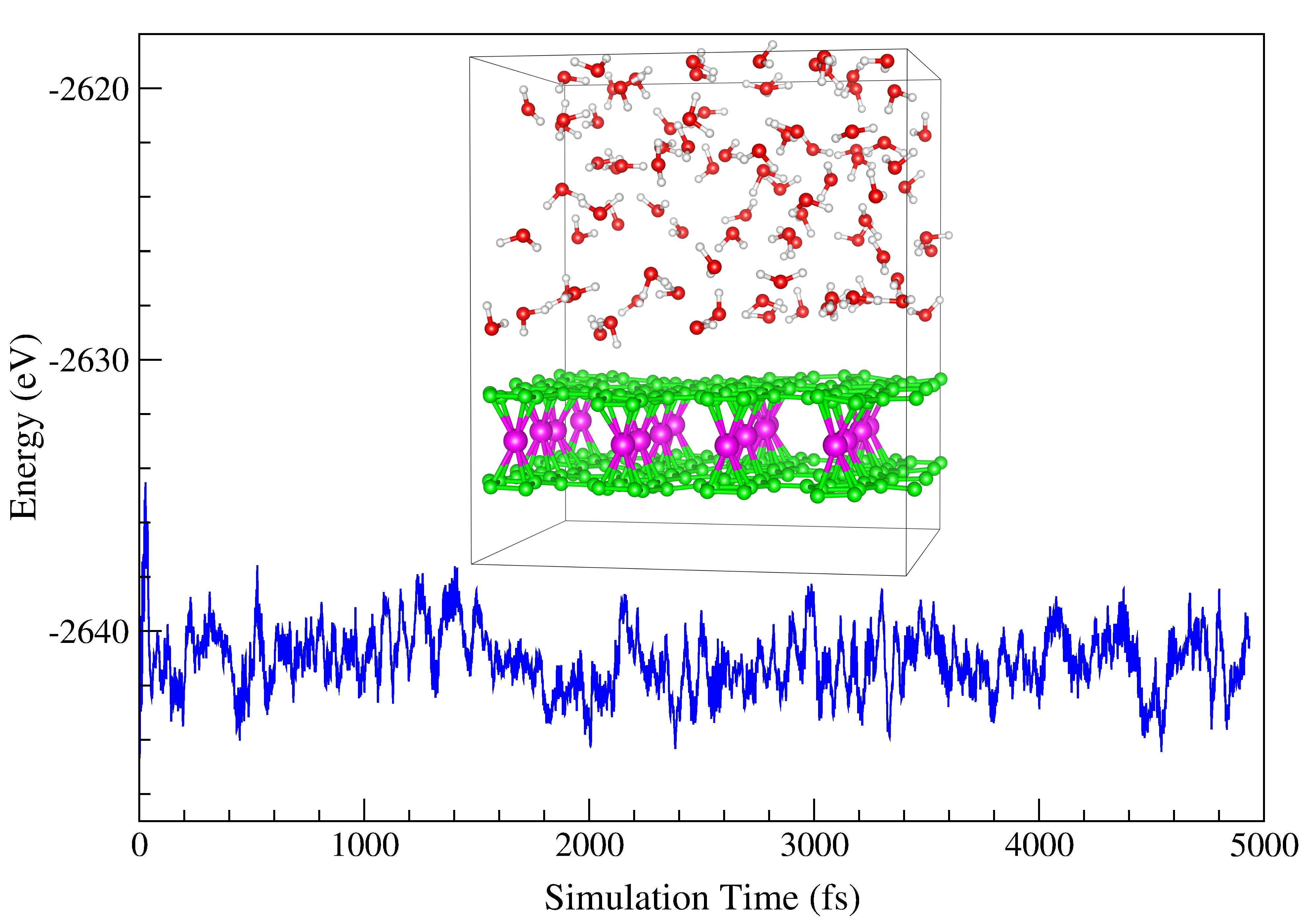}
    \caption{AIMD simulation of B/Fe/B in the presence of H$_2$O for 5\,ps with experimental density of water 1\,g\,cm$^{-3}$.}
    \label{fig:fig_h2o_aimd}
\end{figure}

\newpage
\section{Electronic Structure}
\begin{figure}[H]
  \centering

\begin{subfigure}{\textwidth}
\centering
\includegraphics[width=0.7\linewidth]{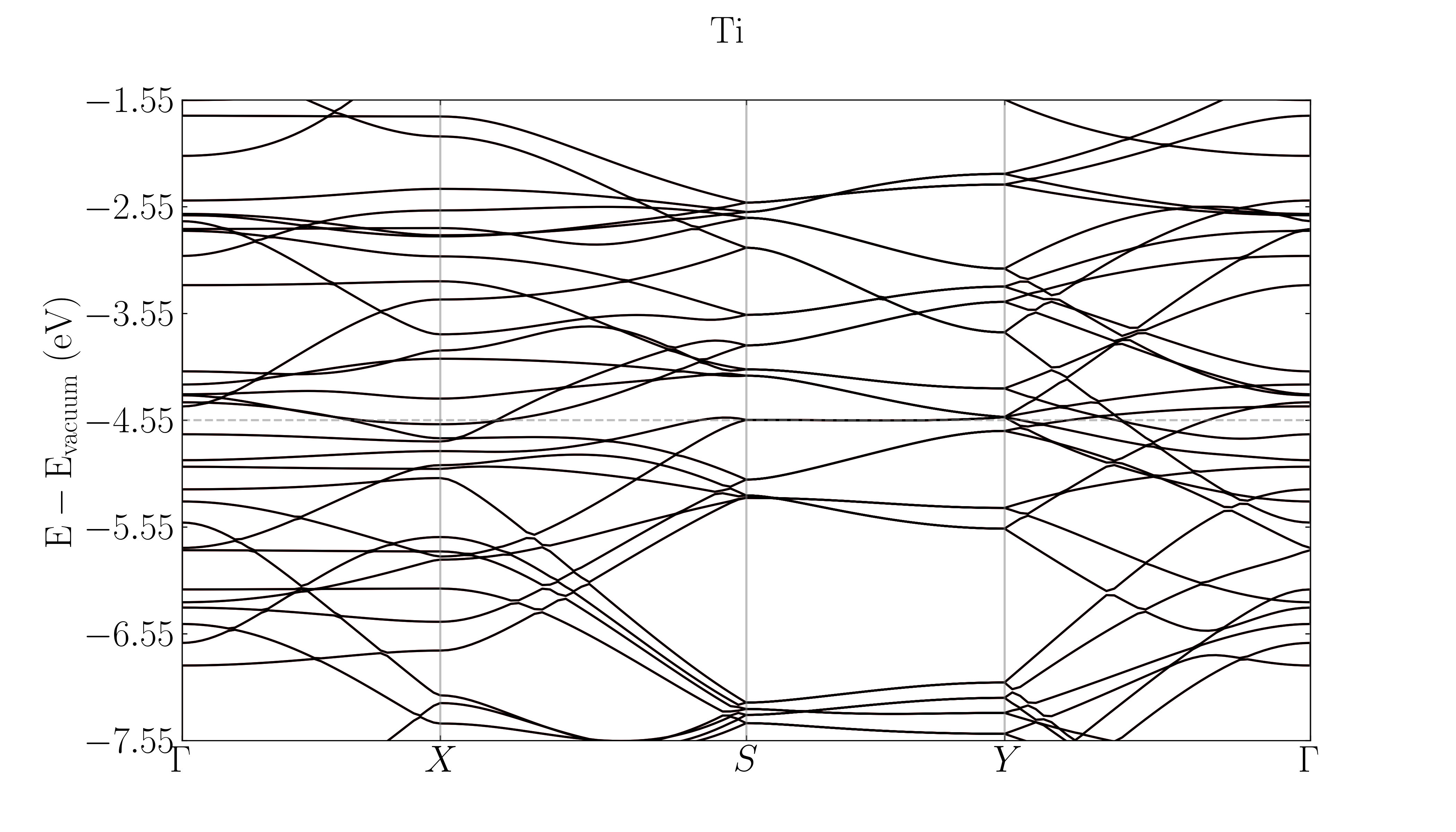}
\end{subfigure}

\vspace{4pt}

\begin{subfigure}{\textwidth}
\centering
\includegraphics[width=0.7\linewidth]{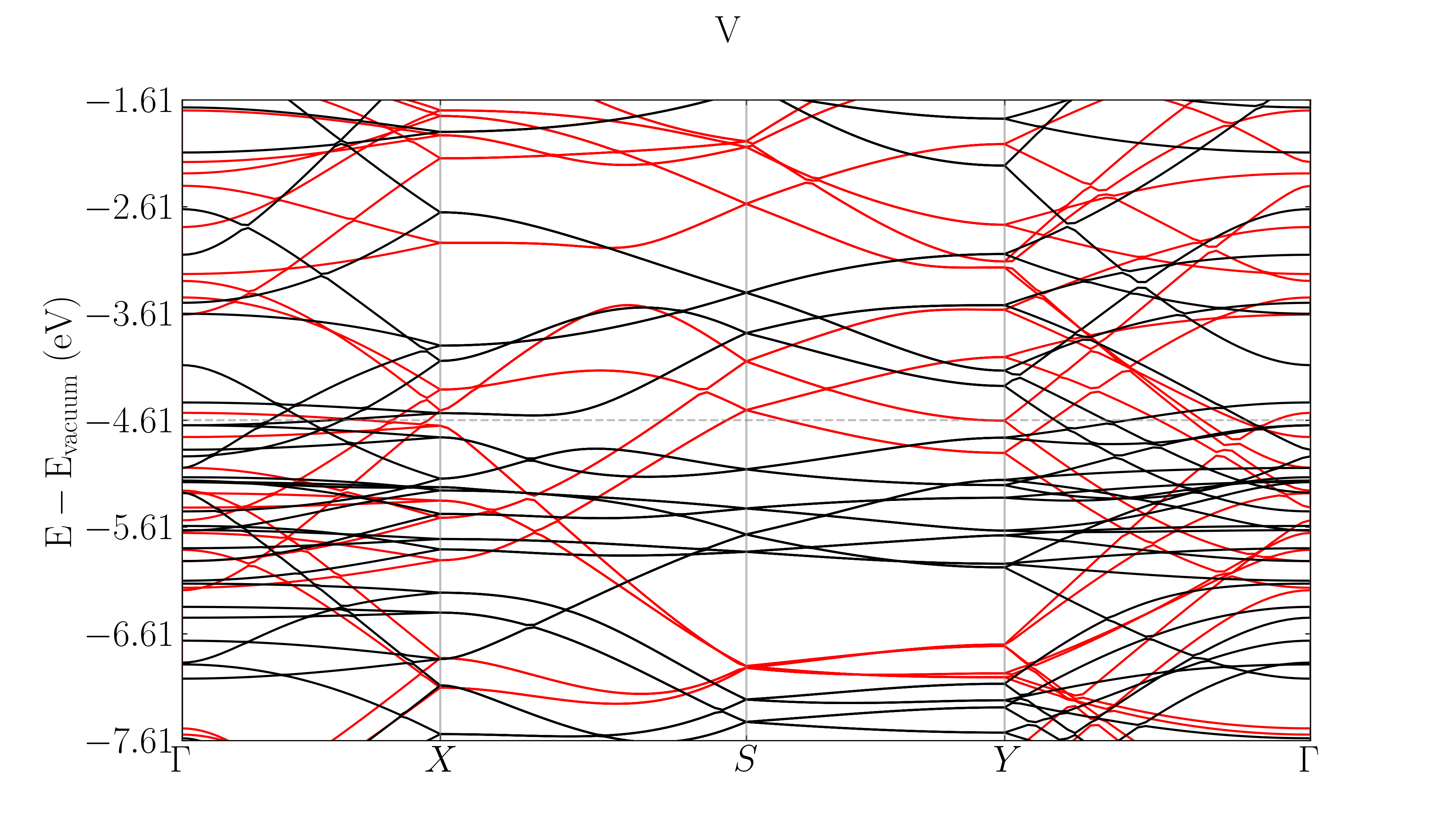}
\end{subfigure}

\vspace{4pt}

\begin{subfigure}{\textwidth}
\centering
\includegraphics[width=0.7\linewidth]{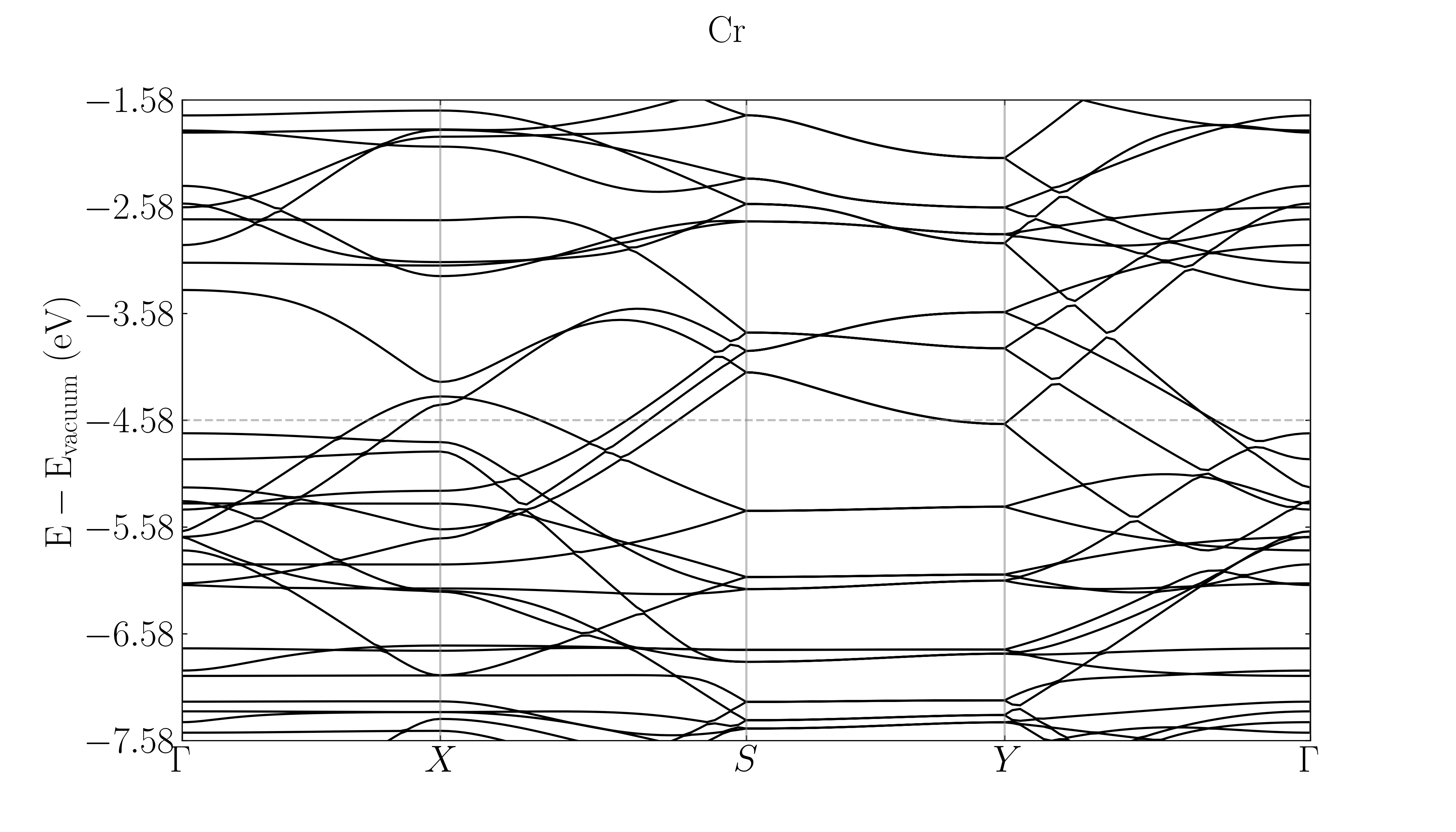}
\end{subfigure}

  \caption{The spin-polarized band structures of B/M/B with M = Ti, V and Cr, relative to the vacuum level. The dashed line indicates the Fermi energy level.}\label{fig:band-all1}
\end{figure}

\begin{figure}[H]
  \centering

\begin{subfigure}{\textwidth}
\centering
\includegraphics[width=0.7\linewidth]{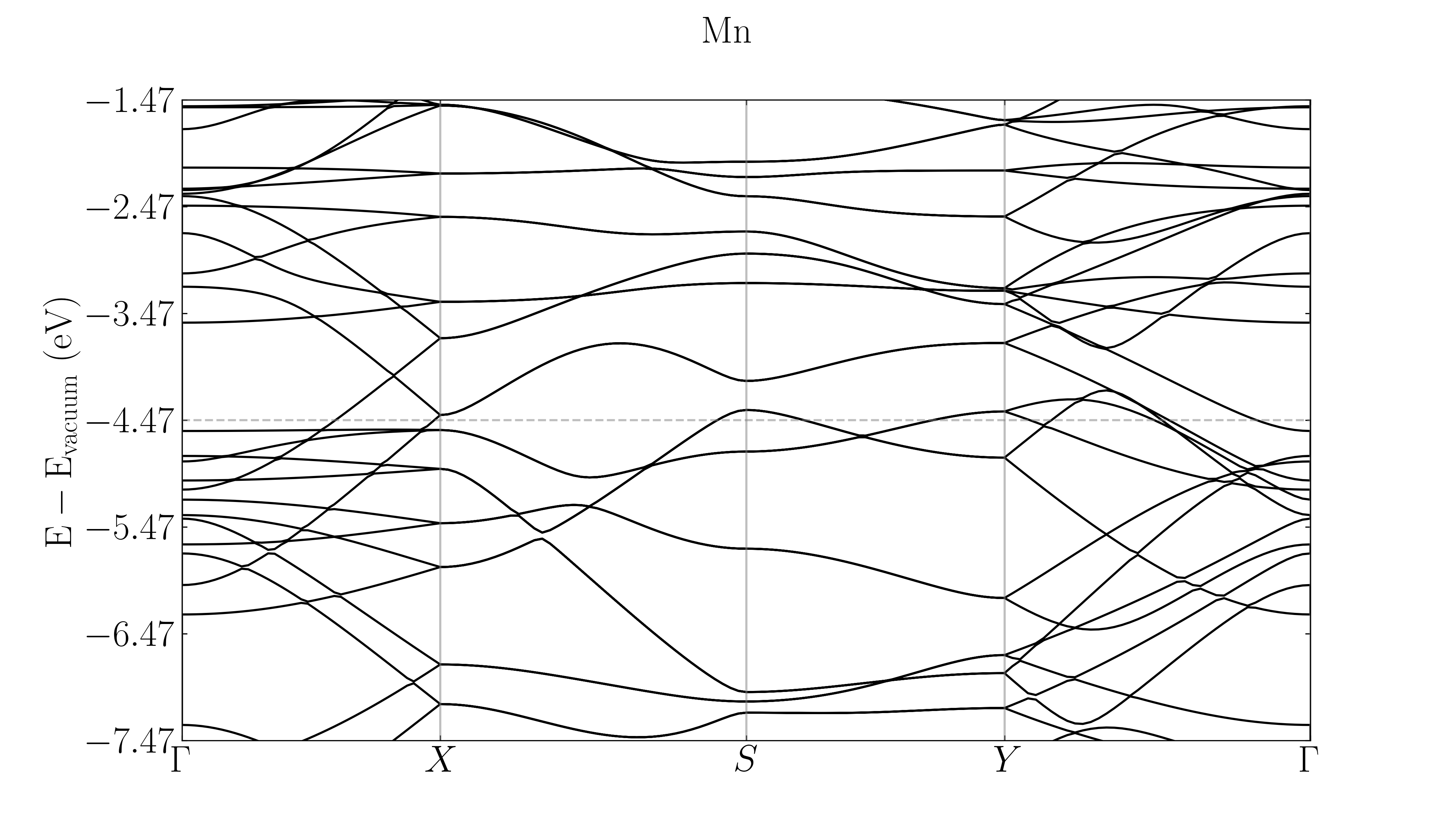}
\end{subfigure}

\vspace{4pt}

\begin{subfigure}{\textwidth}
\centering
\includegraphics[width=0.7\linewidth]{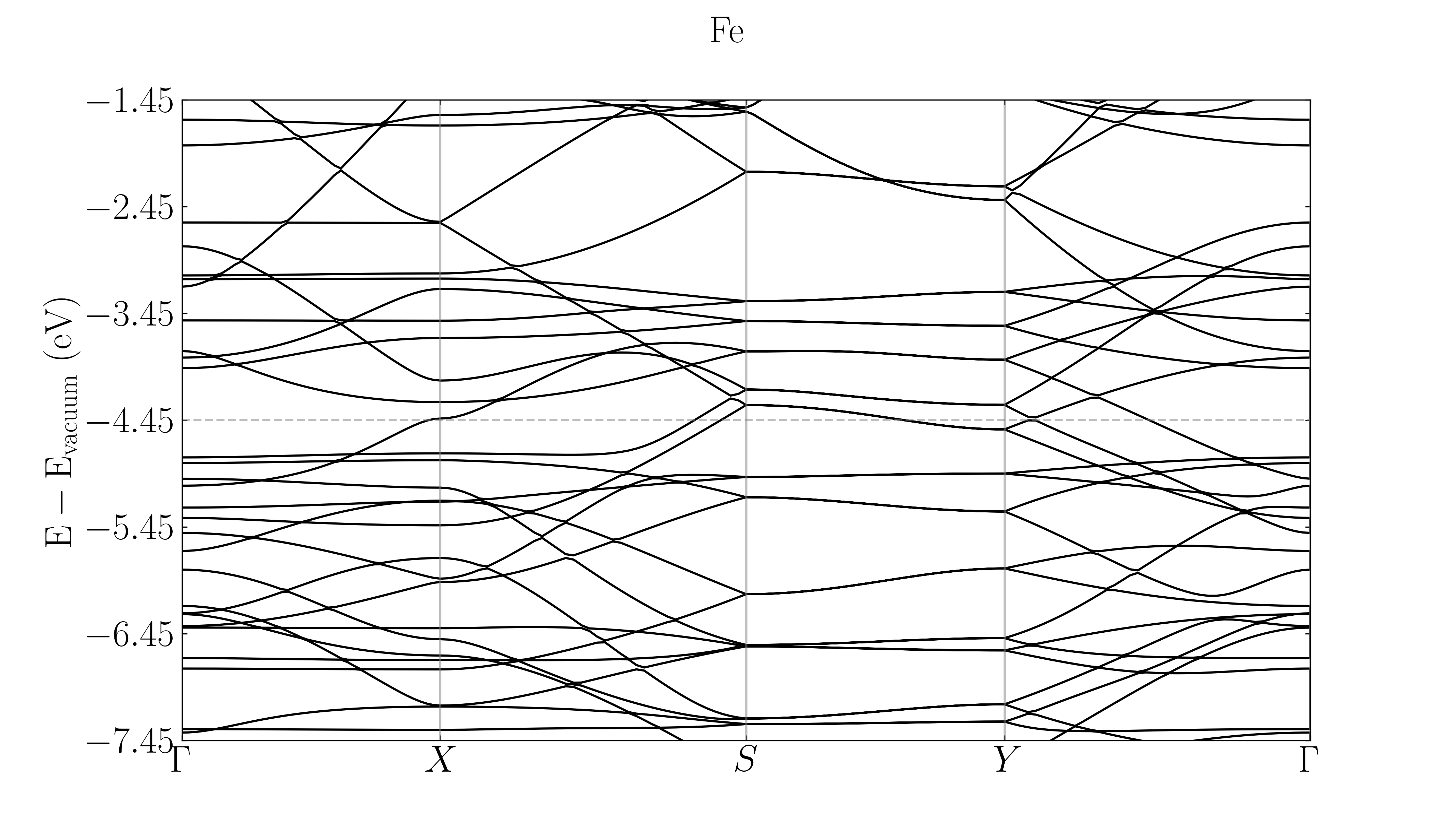}
\end{subfigure}

\vspace{4pt}

\begin{subfigure}{\textwidth}
\centering
\includegraphics[width=0.7\linewidth]{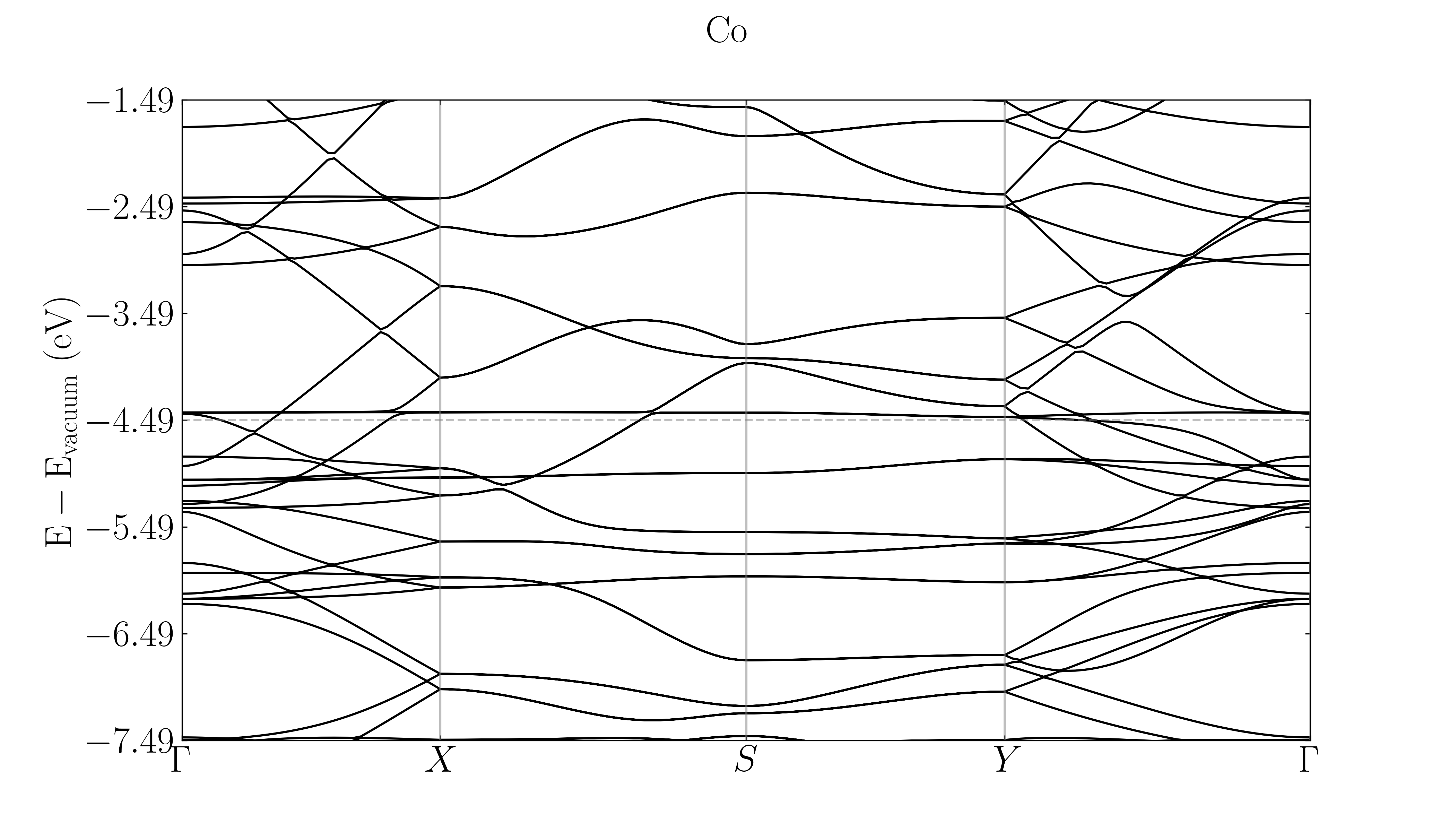}
\end{subfigure}

  \caption{The spin-polarized band structures B/M/B with M = Mn, Fe and Co, relative to the vacuum level. The dashed line indicates the Fermi energy level.}\label{fig:band-all2}
\end{figure}

\begin{figure}[H]
  \centering

\begin{subfigure}{\textwidth}
\centering
\includegraphics[width=0.7\linewidth]{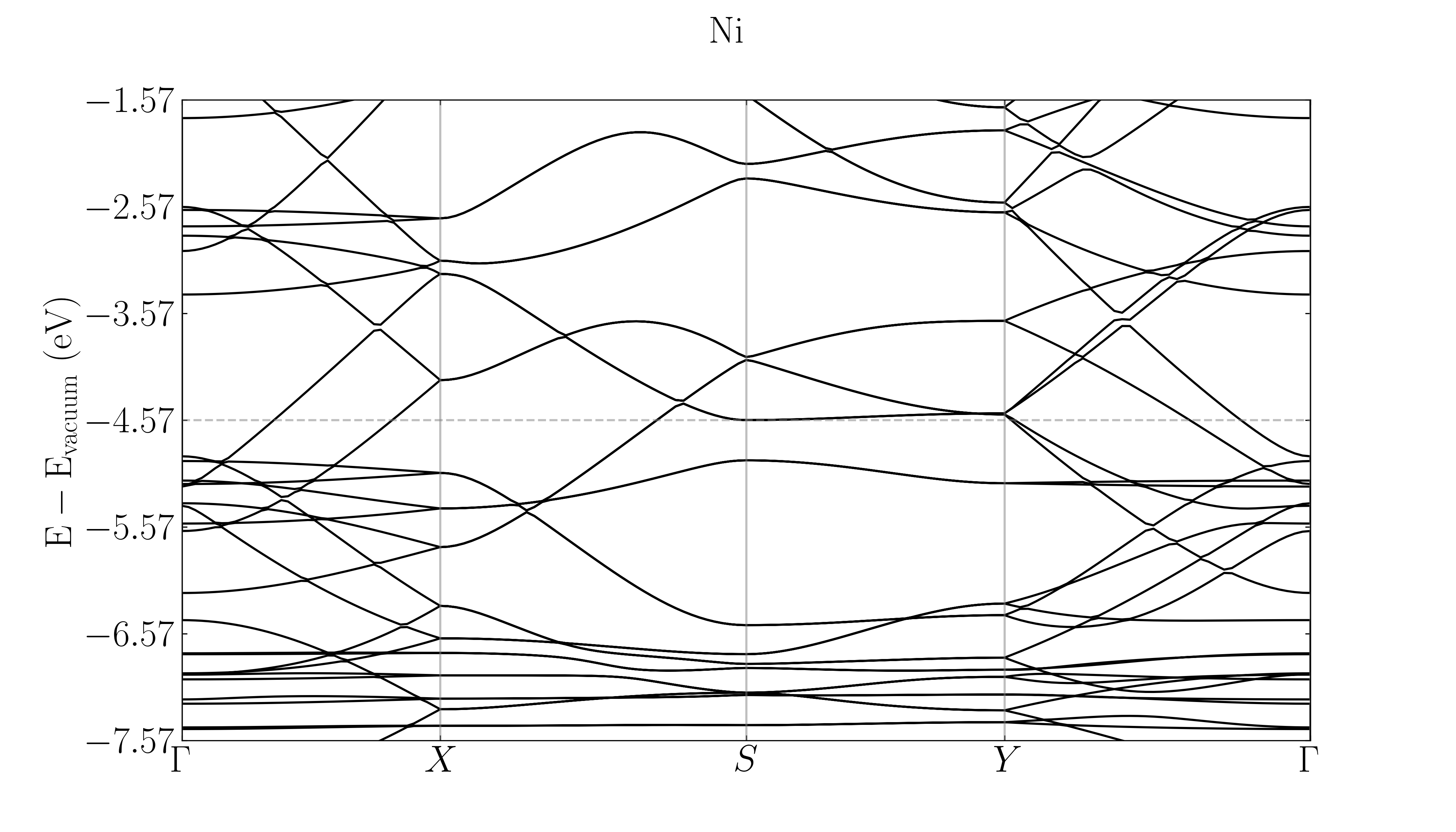}
\end{subfigure}

\vspace{4pt}

\begin{subfigure}{\textwidth}
\centering
\includegraphics[width=0.7\linewidth]{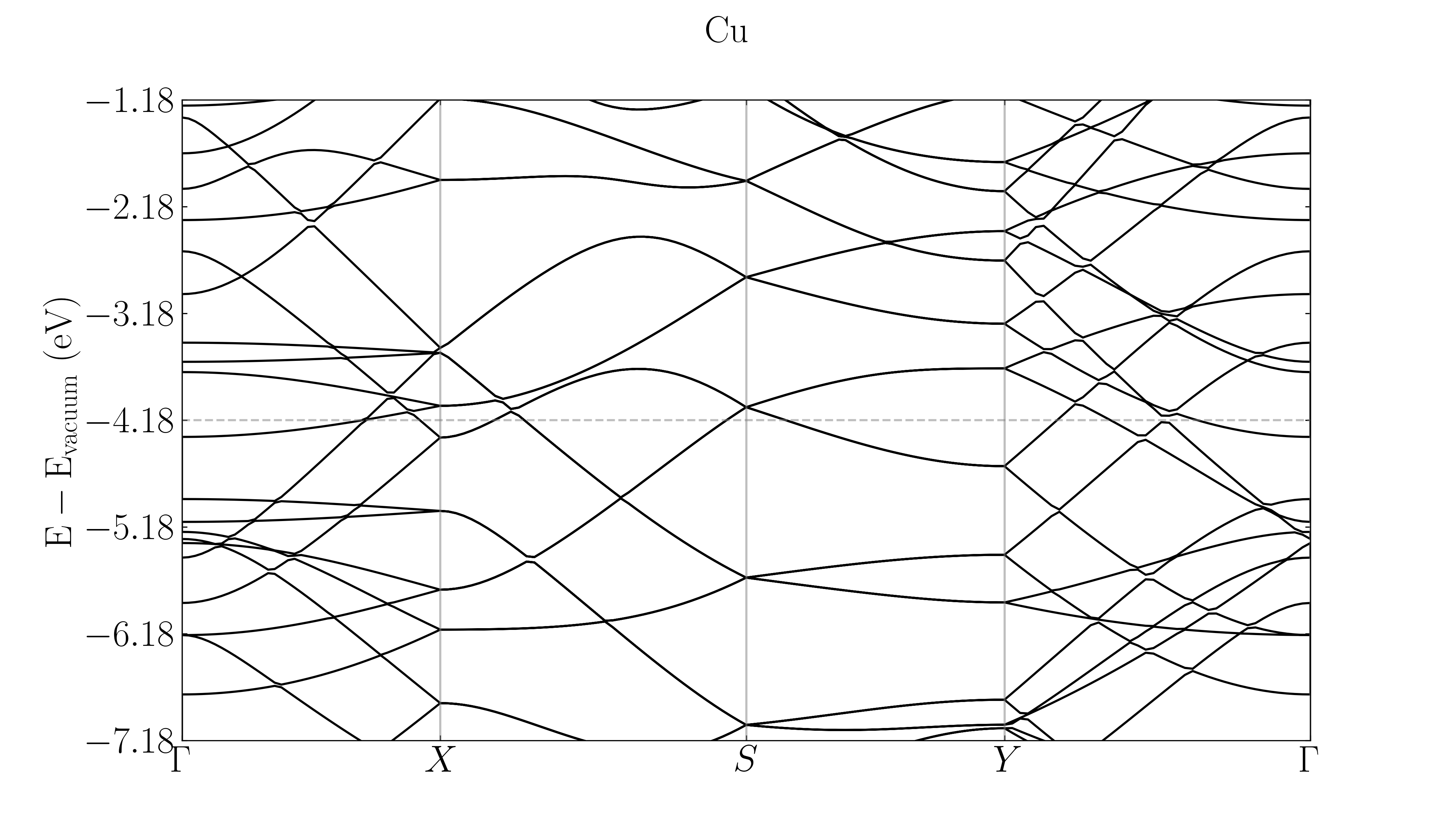}
\end{subfigure}

\vspace{4pt}

\begin{subfigure}{\textwidth}
\centering
\includegraphics[width=0.7\linewidth]{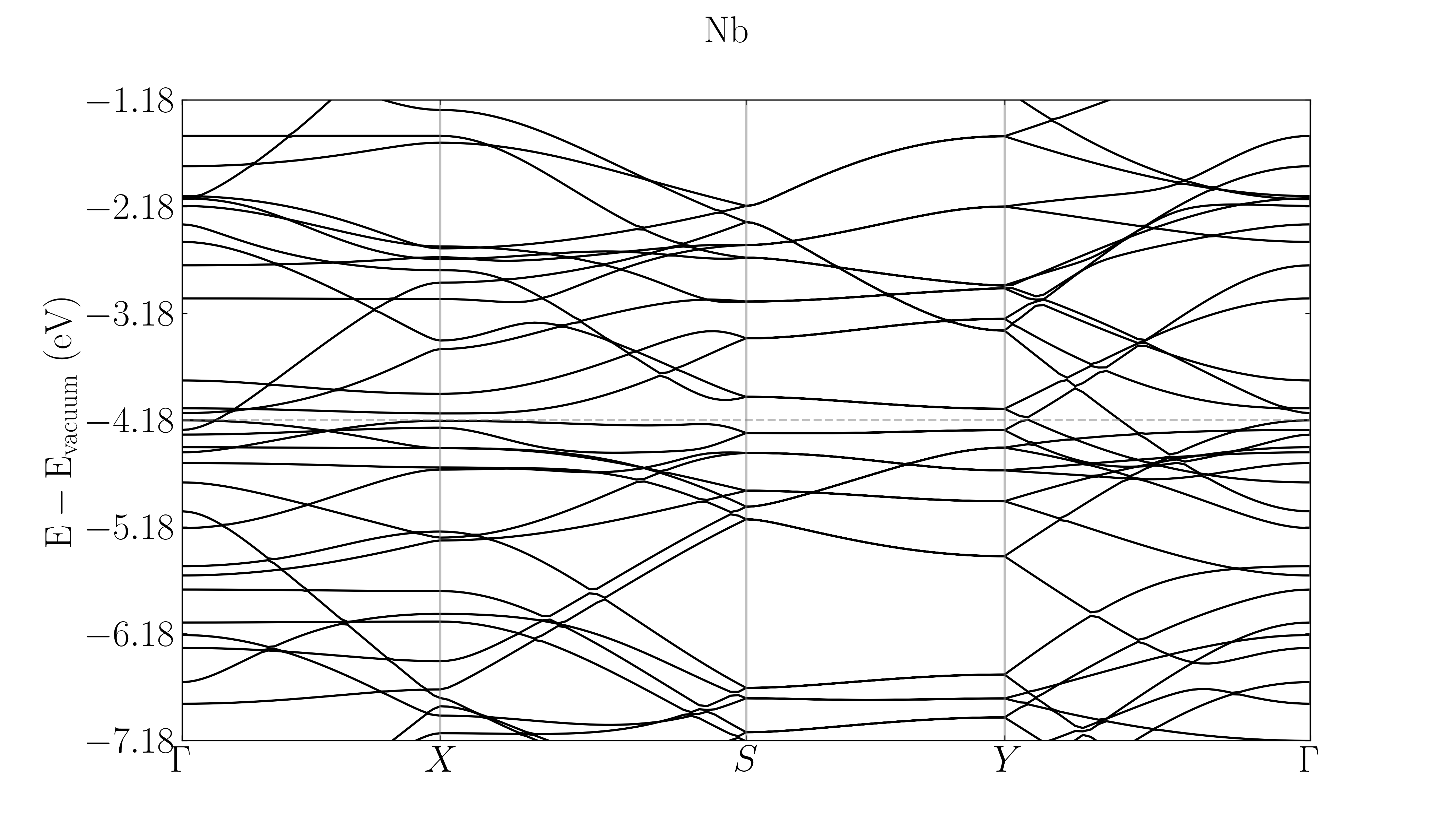}
\end{subfigure}

  \caption{The spin-polarized band structures B/M/B with M = Ni, Cu and Nb, relative to the vacuum level. The dashed line indicates the Fermi energy level.}\label{fig:band-all3}
\end{figure}

\begin{figure}[H]
  \centering

\begin{subfigure}{\textwidth}
\centering
\includegraphics[width=0.7\linewidth]{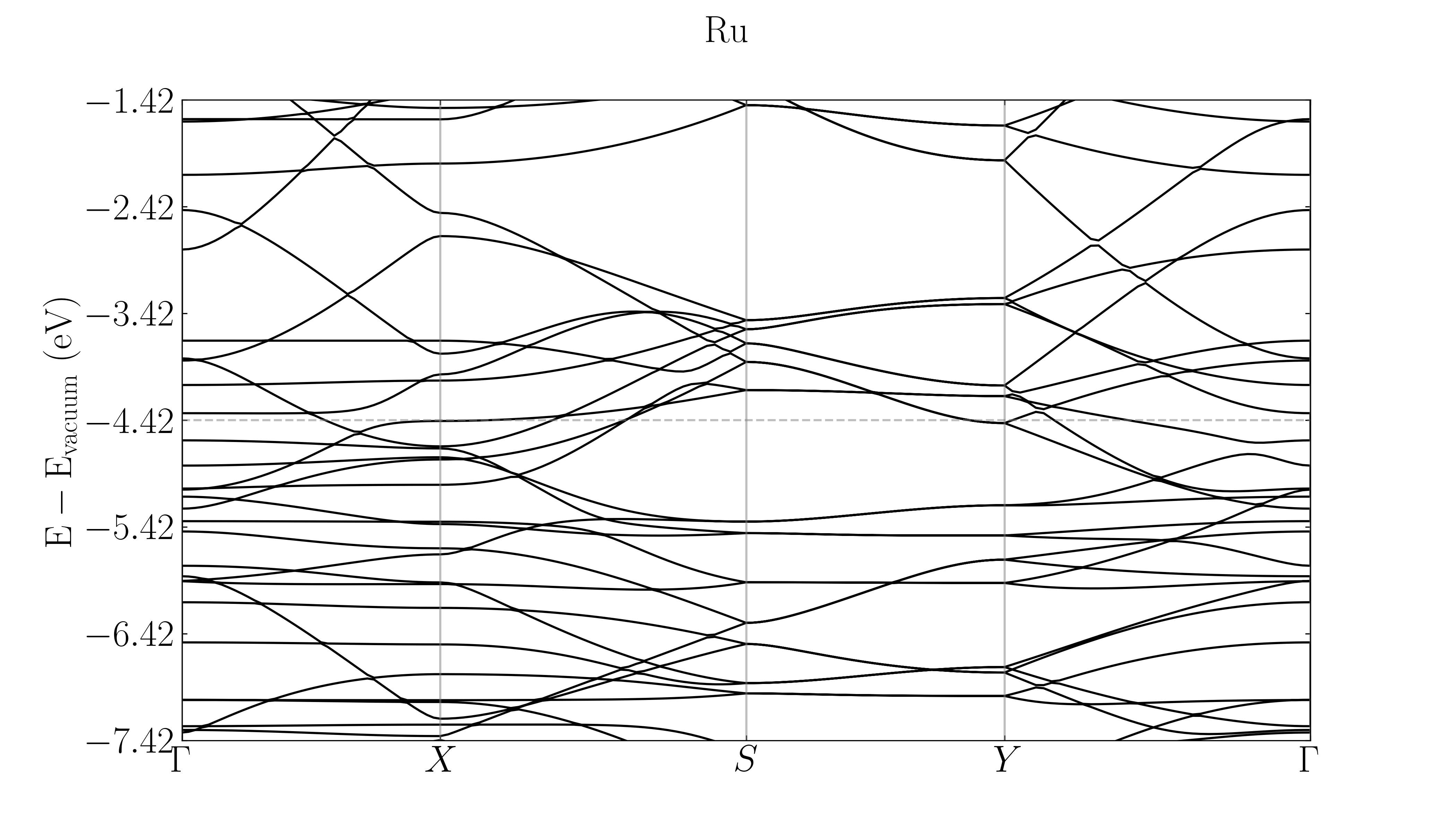}
\end{subfigure}

\vspace{4pt}

\begin{subfigure}{\textwidth}
\centering
\includegraphics[width=0.7\linewidth]{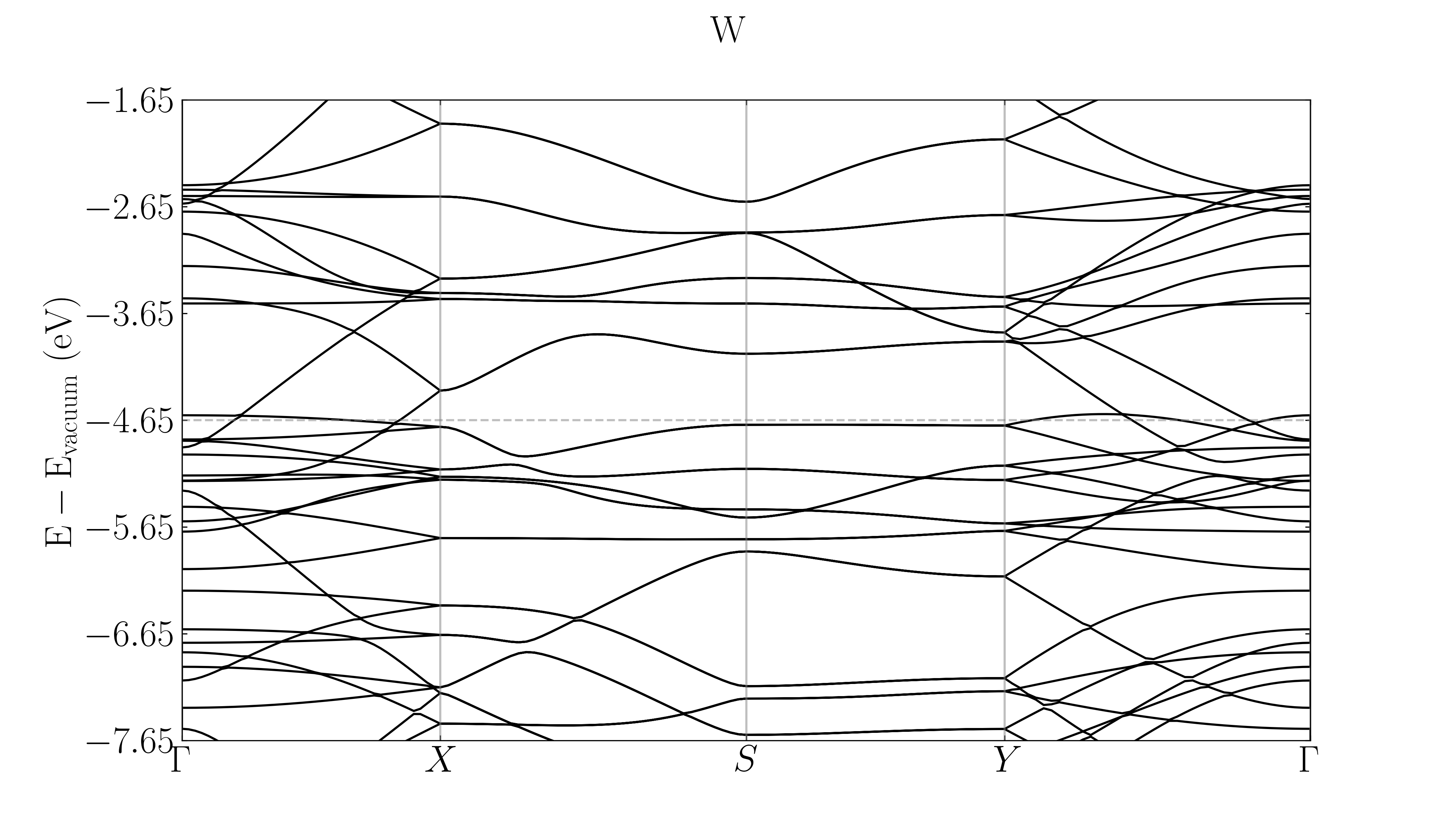}
\end{subfigure}

\vspace{4pt}

\begin{subfigure}{\textwidth}
\centering
\includegraphics[width=0.7\linewidth]{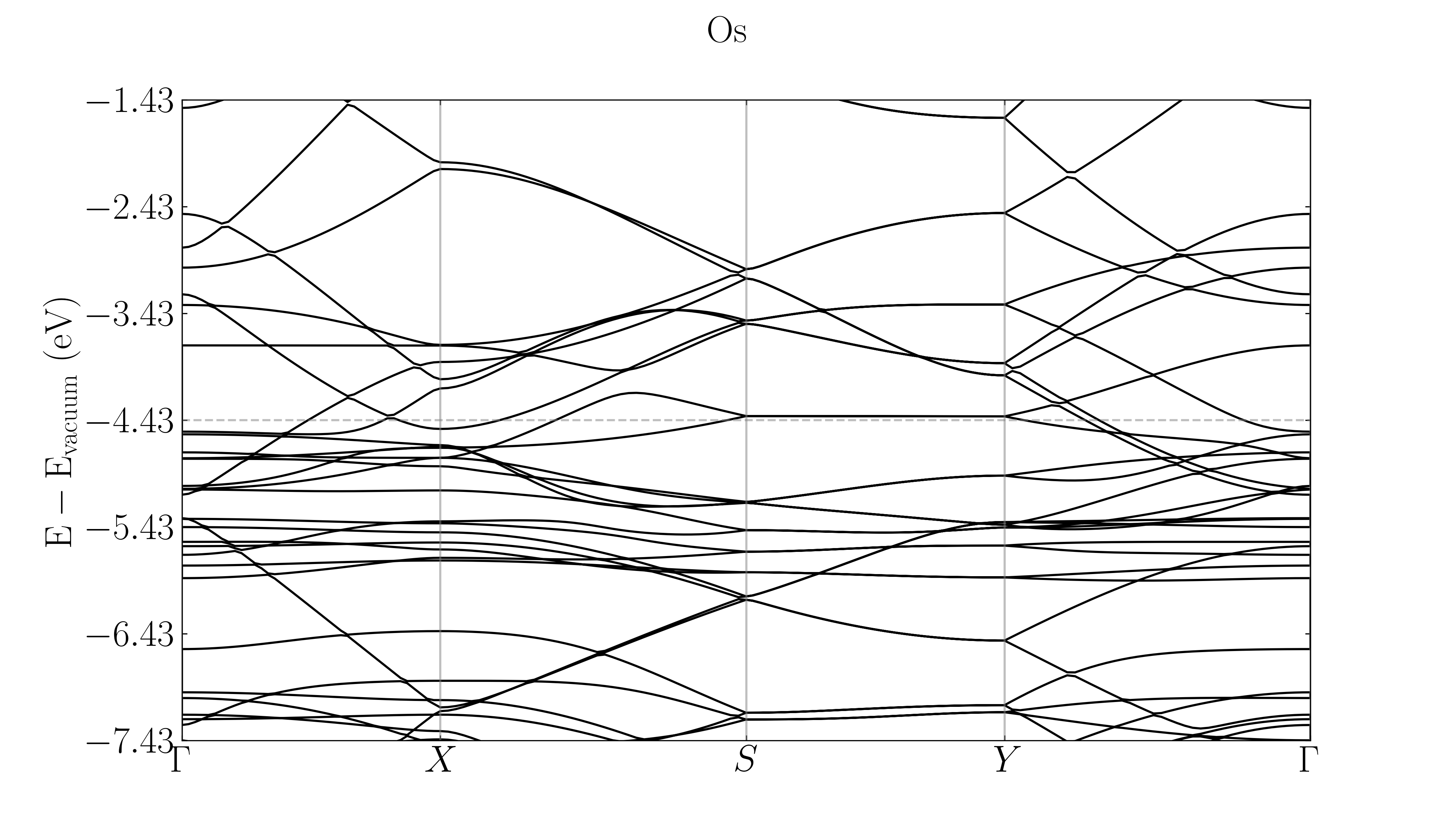}
\end{subfigure}

  \caption{The spin-polarized band structures B/M/B with M = Ru, W and Os, relative to the vacuum level. The dashed line indicates the Fermi energy level.}\label{fig:band-all4}
\end{figure}

\begin{figure}[H]
    \centering
    \includegraphics[width=0.7\linewidth]{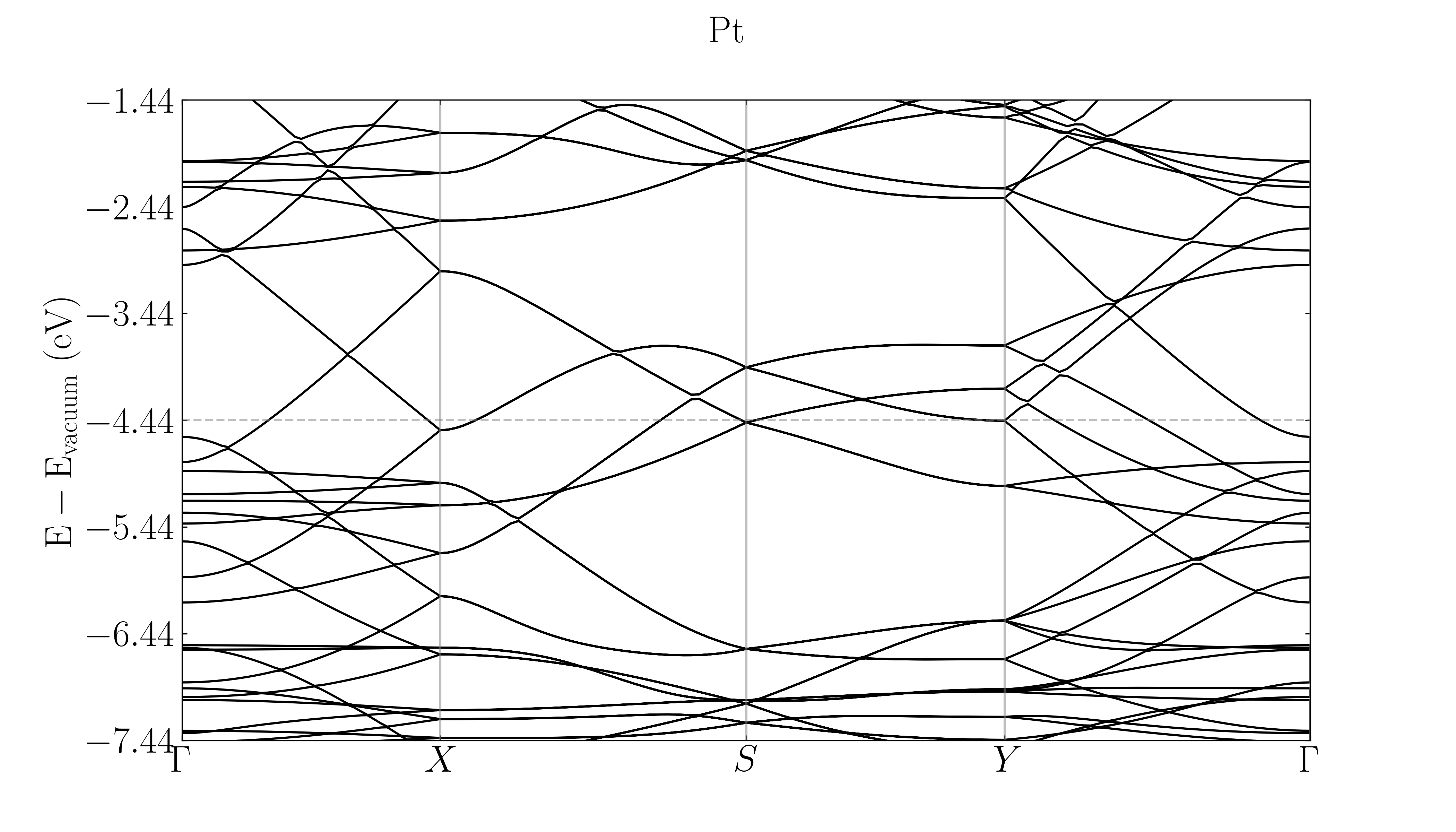}
    \caption{The spin-polarized band structures B/M/B with M = Pt, relative to the vacuum level. The dashed line indicates the Fermi energy level.}
    \label{fig:pt_band}
\end{figure}

In this section, we present the electronic properties of the B/M/B systems. Figs.\,~\ref{fig:band-all1}~\ref{fig:band-all2}~\ref{fig:band-all3}~\ref{fig:band-all4} and ~\ref{fig:pt_band} shows the spin-polarized band structures, black lines are the spin-up component, while red lines are the spin-down components, for all B/M/B systems in their most stable magnetic phase with M positioned at the C$^{468}$ site. The magnetic phase is essentially localized on the M atoms, whereas for the carbon atoms, the magnetization is essentially zero, with no spin component distinction for the carbon atoms. The choice of M at the C$^{468}$ site is justified because if M were positioned at the hexagonal or octagonal site, the electronic density loses localization at the square sites, hindering surface adsorption\,\cite{Chen2023}.

\newpage


\begin{figure}[H]
  \centering
  \begin{subfigure}[b]{0.5\textwidth}
    \centering
    \includegraphics[width=\linewidth,height=0.5\textheight,keepaspectratio]{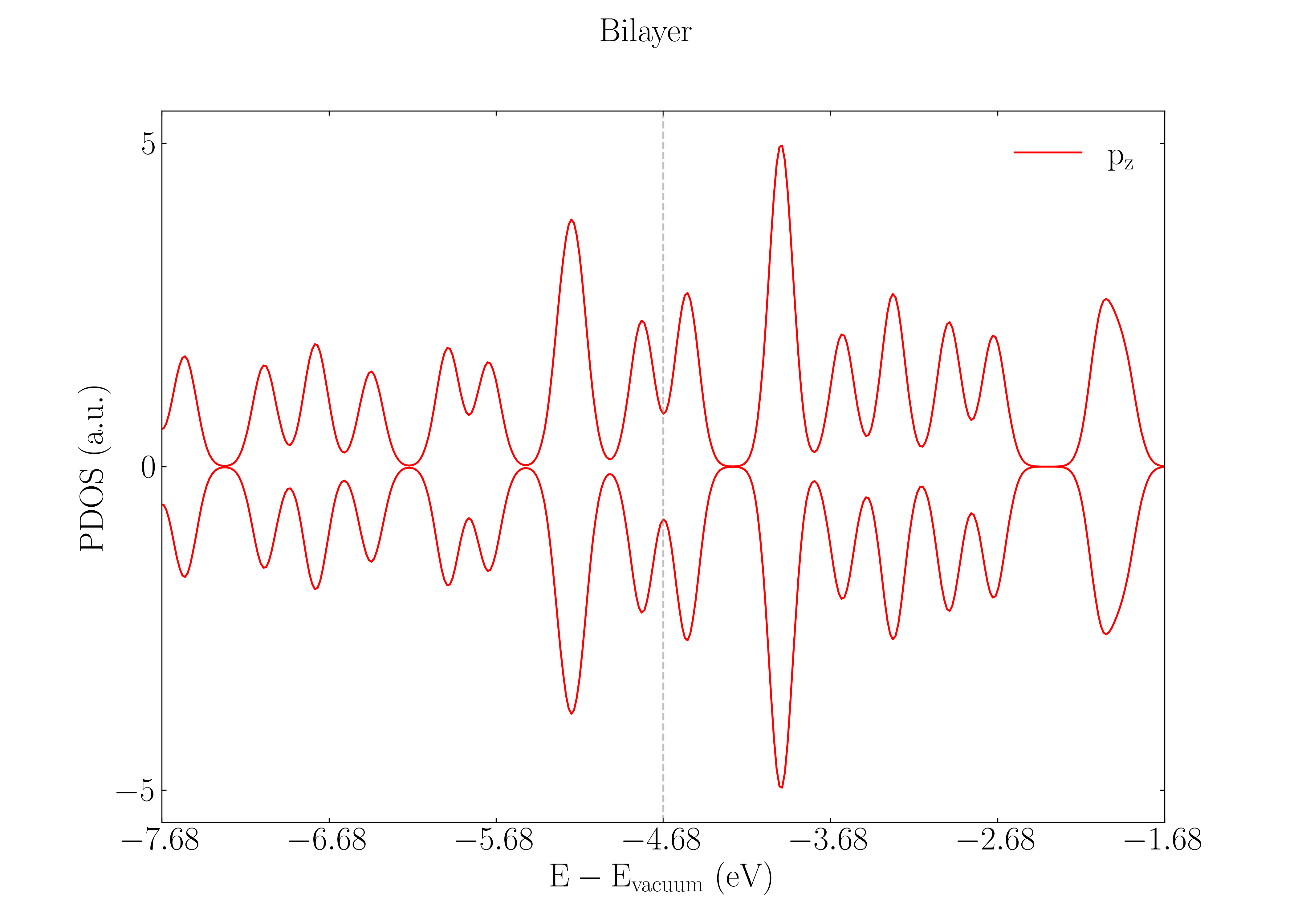}
  \end{subfigure}\hfill%
  \begin{subfigure}[b]{0.5\textwidth}
    \centering
    \includegraphics[width=\linewidth,height=0.5\textheight,keepaspectratio]{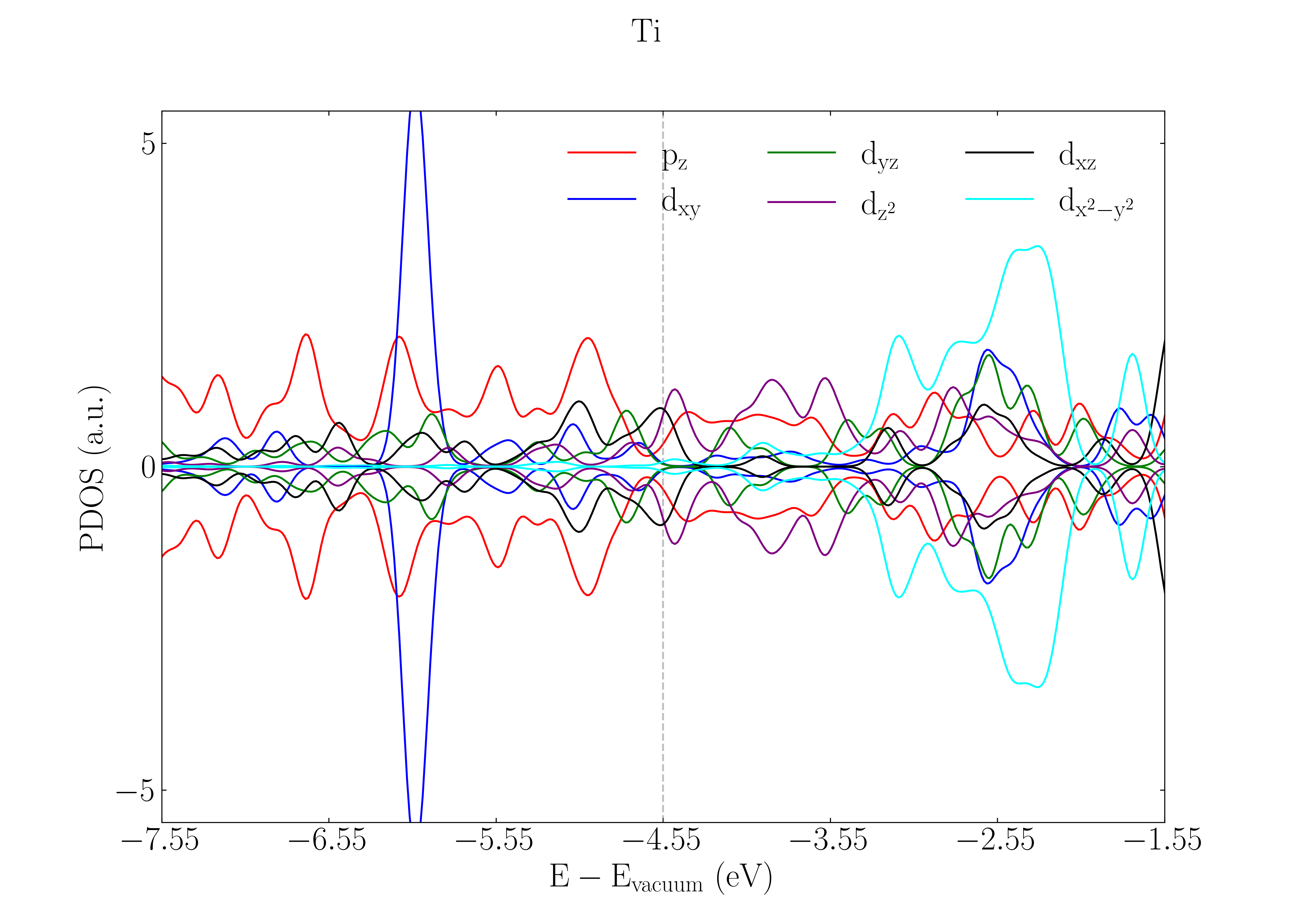}
  \end{subfigure}
  \par\vspace{-2pt}
  \begin{subfigure}[b]{0.5\textwidth}
    \centering
    \includegraphics[width=\linewidth,height=0.5\textheight,keepaspectratio]{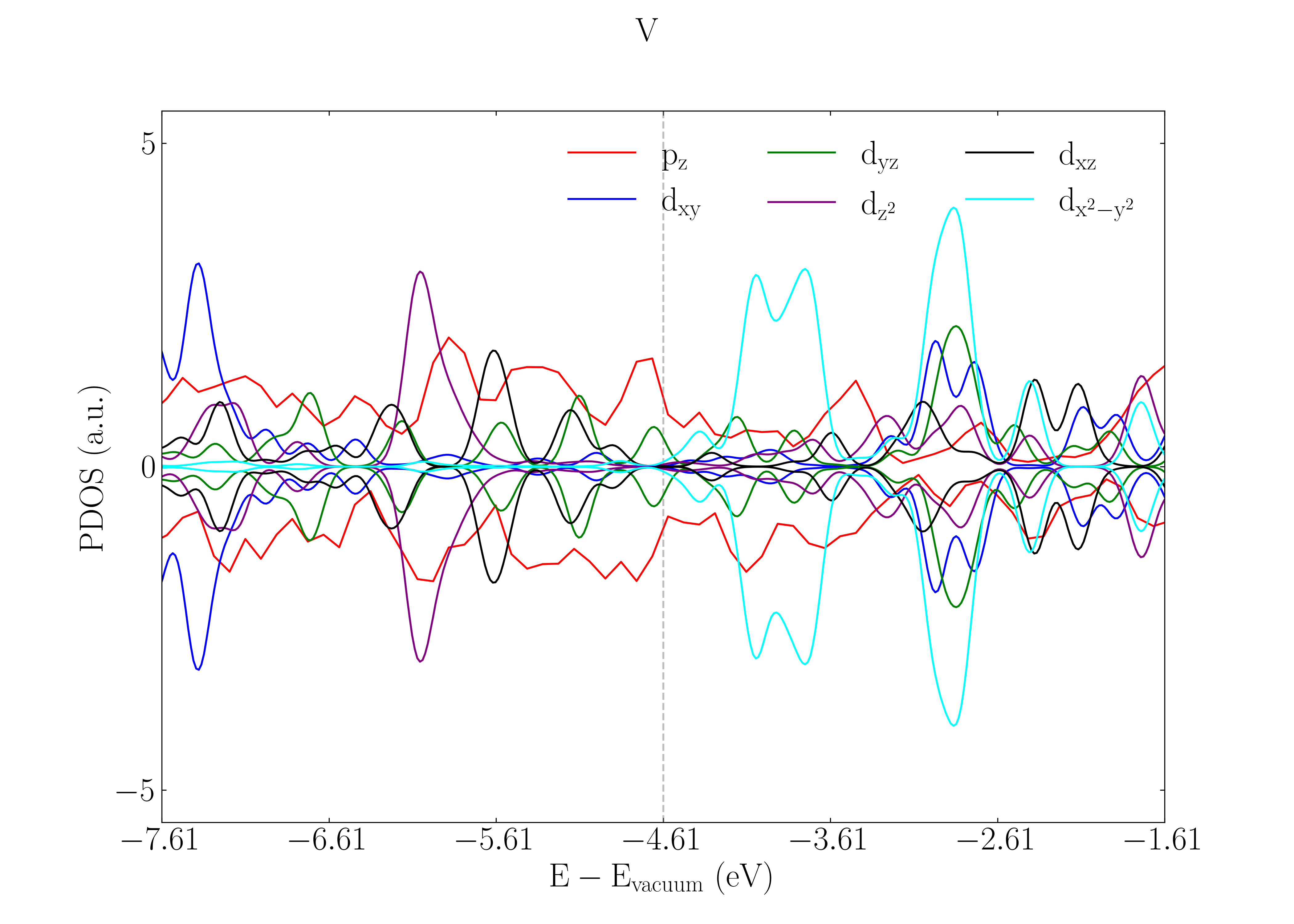}
  \end{subfigure}\hfill%
  \begin{subfigure}[b]{0.5\textwidth}
    \centering
    \includegraphics[width=\linewidth,height=0.5\textheight,keepaspectratio]{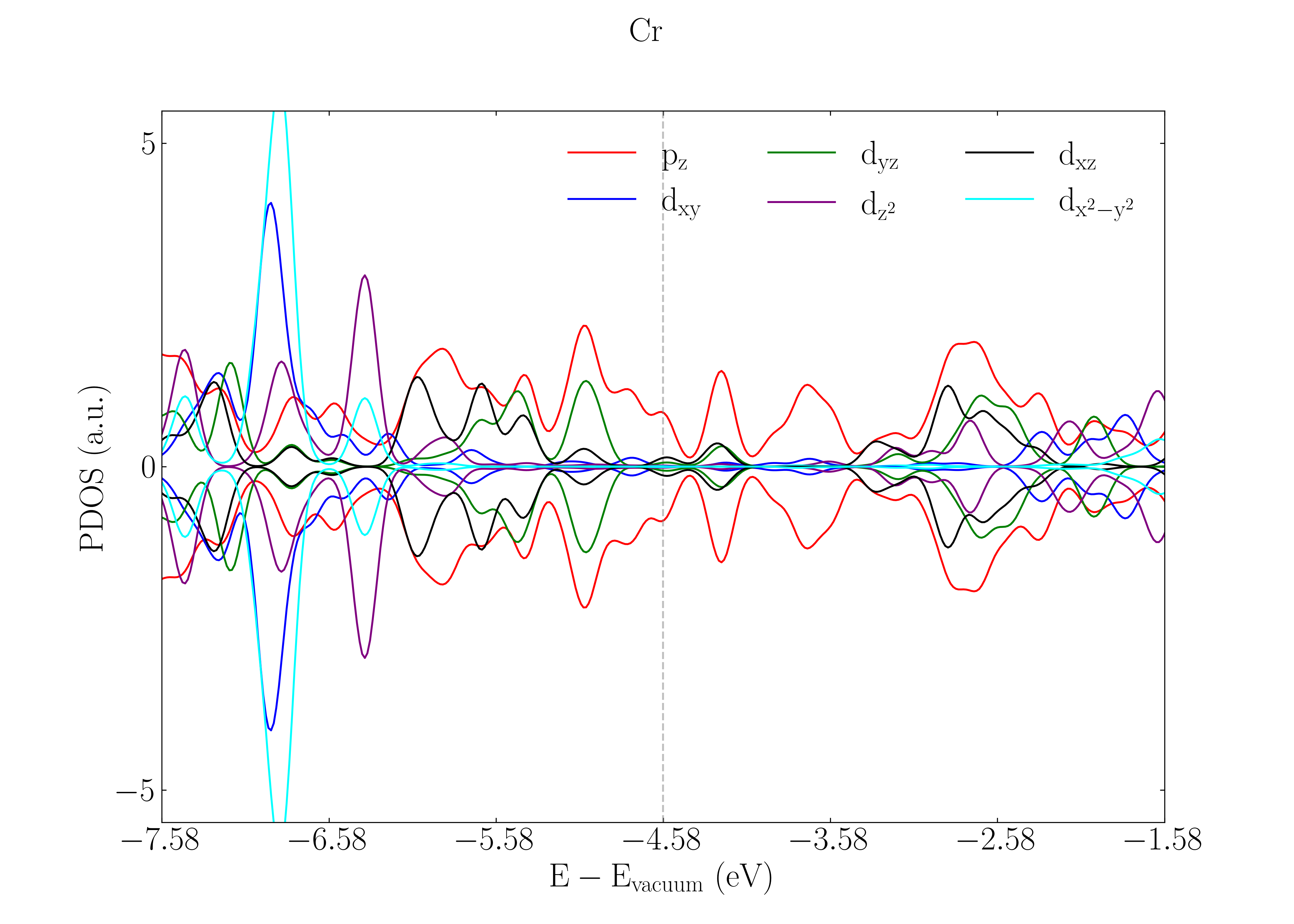}
  \end{subfigure}

  \caption{The PDOS of $p_z$ and $d$-orbitals for B/B and B/M/B for M = Ti, V and Cr, where the lines blue is d$_{xy}$-orbital, green is d$_{yz}$-orbital, purple is d$_{z^2}$-orbital, black is d$_{xy}$-orbital, cyan is d$_{x^2-y^2}$-orbital, and the red line is $p_z$-orbital from C$^{468}$ atoms. The dashed line indicates the Fermi energy level.}\label{fig:dos-all1}
\end{figure}

\begin{figure}[H]
  \centering
  \begin{subfigure}[b]{0.5\textwidth}
    \centering
    \par\vspace{-8pt}
    \includegraphics[width=\linewidth,height=0.5\textheight,keepaspectratio]{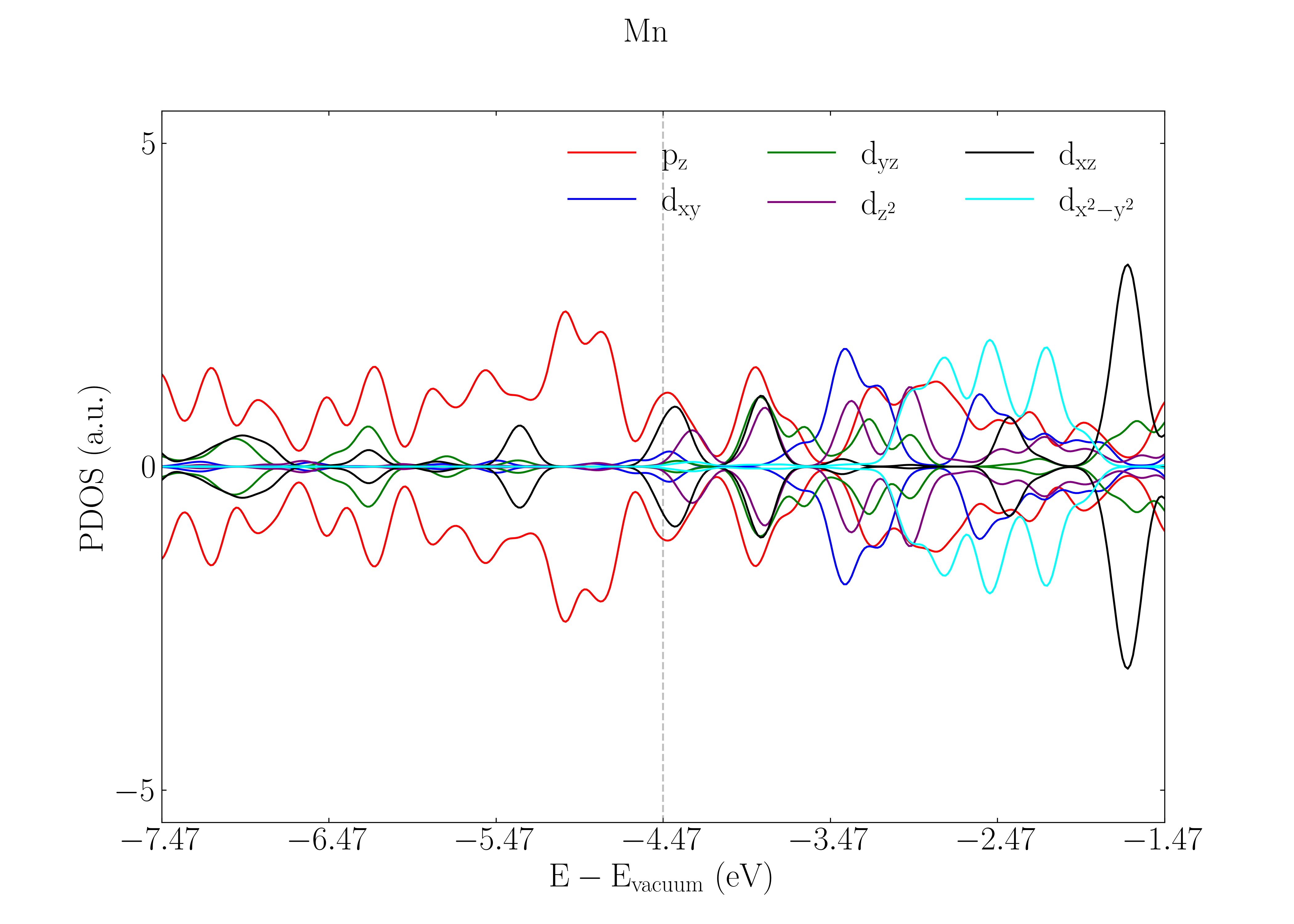}
  \end{subfigure}\hfill%
  \begin{subfigure}[b]{0.5\textwidth}
    \centering
    \includegraphics[width=\linewidth,height=0.5\textheight,keepaspectratio]{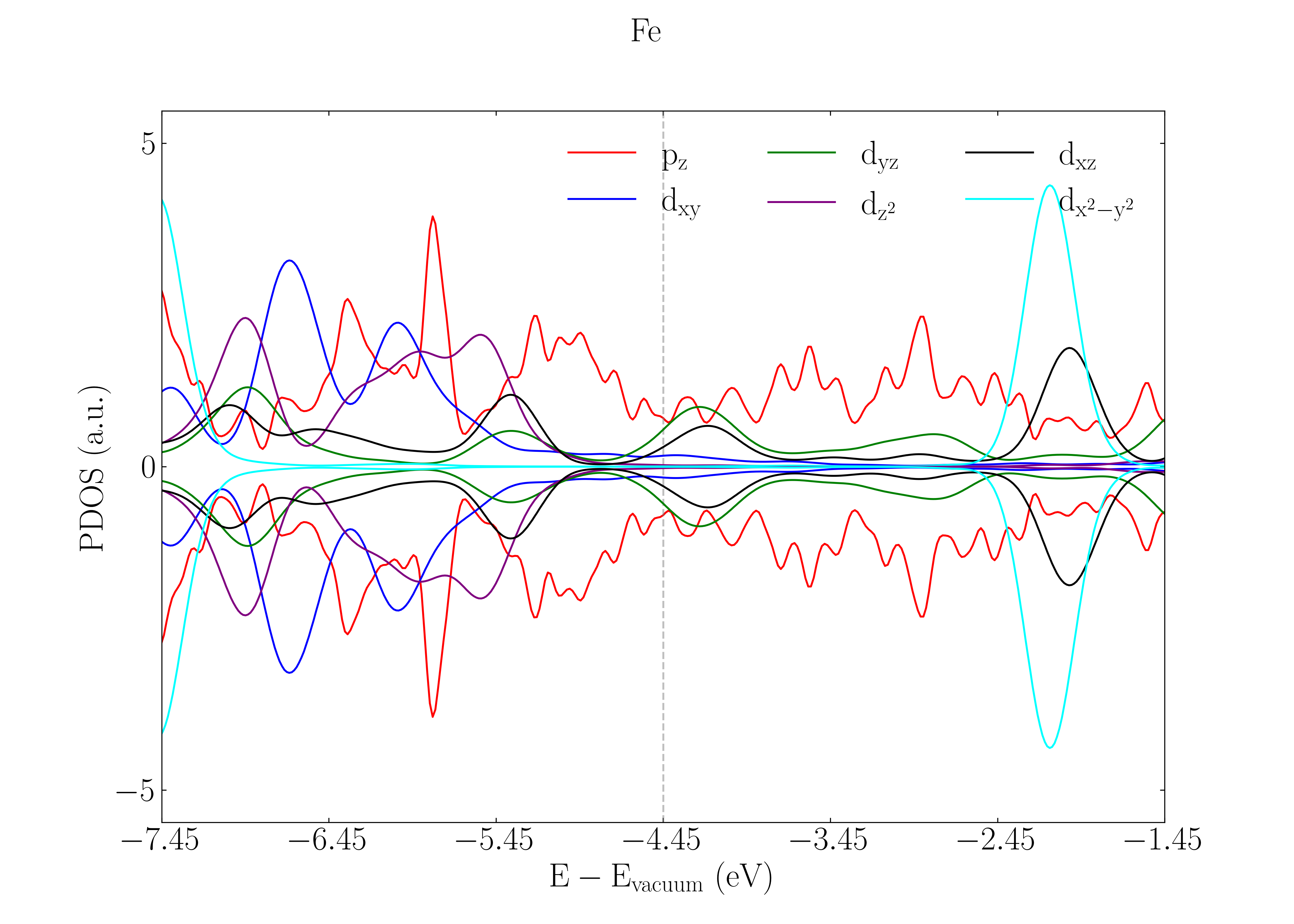}
  \end{subfigure}
  \par\vspace{-2pt}
  \begin{subfigure}[b]{0.5\textwidth}
    \centering
    \includegraphics[width=\linewidth,height=0.5\textheight,keepaspectratio]{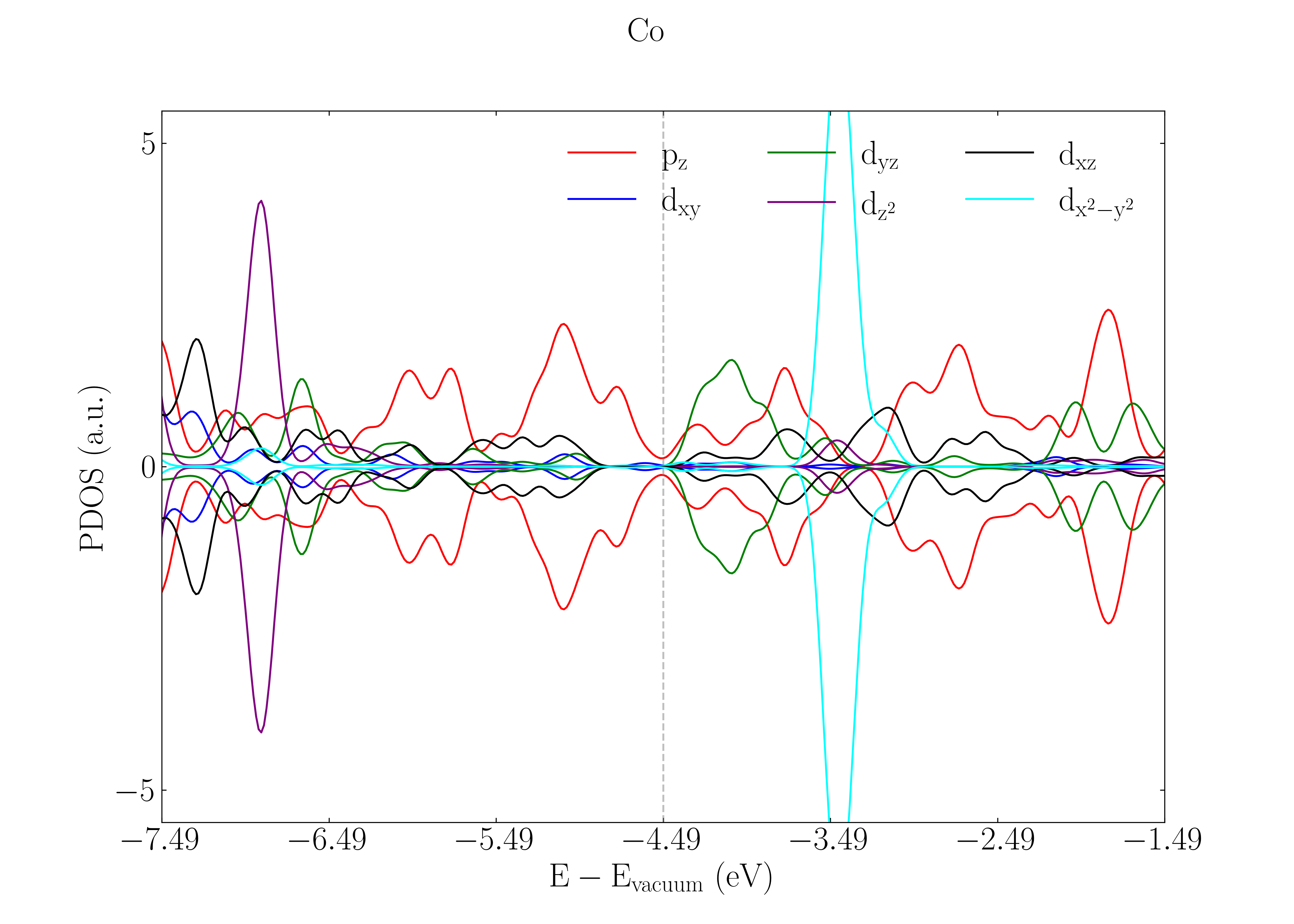}
  \end{subfigure}\hfill%
  \begin{subfigure}[b]{0.5\textwidth}
    \centering
    \includegraphics[width=\linewidth,height=0.5\textheight,keepaspectratio]{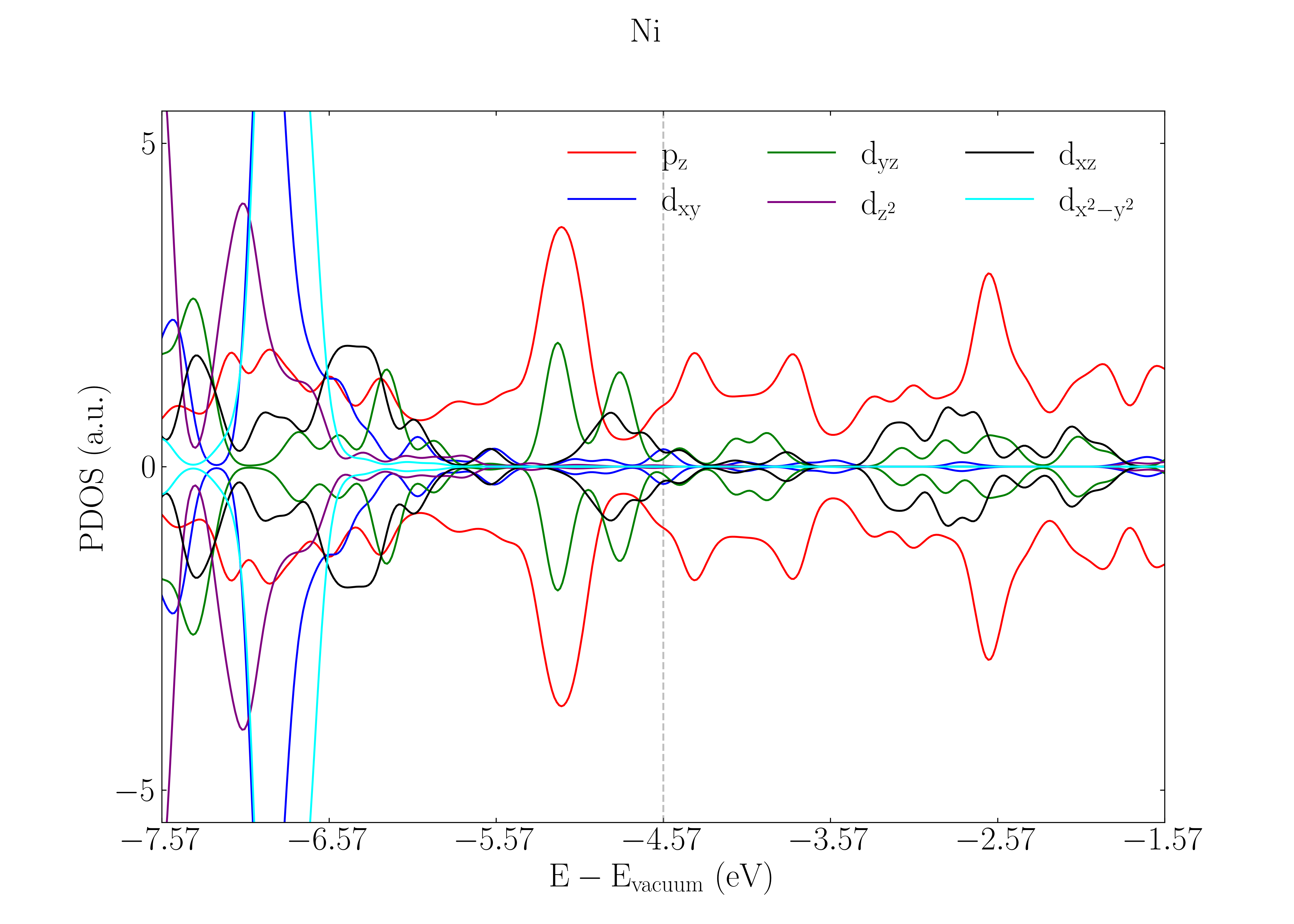}
  \end{subfigure}
\caption{The PDOS of $p_z$ and $d$-orbitals for B/M/B for M = Mn, Fe, Co and Ni, where the lines blue is d$_{xy}$-orbital, green is d$_{yz}$-orbital, purple is d$_{z^2}$-orbital, black is d$_{xy}$-orbital, cyan is d$_{x^2-y^2}$-orbital, and the red line is $p_z$-orbital from C$^{468}$ atoms. The dashed line indicates the Fermi energy level.}\label{fig:dos-all2}
\end{figure}

\begin{figure}[H]
  \centering
  \begin{subfigure}[b]{0.5\textwidth}
    \centering
    \includegraphics[width=\linewidth,height=0.5\textheight,keepaspectratio]{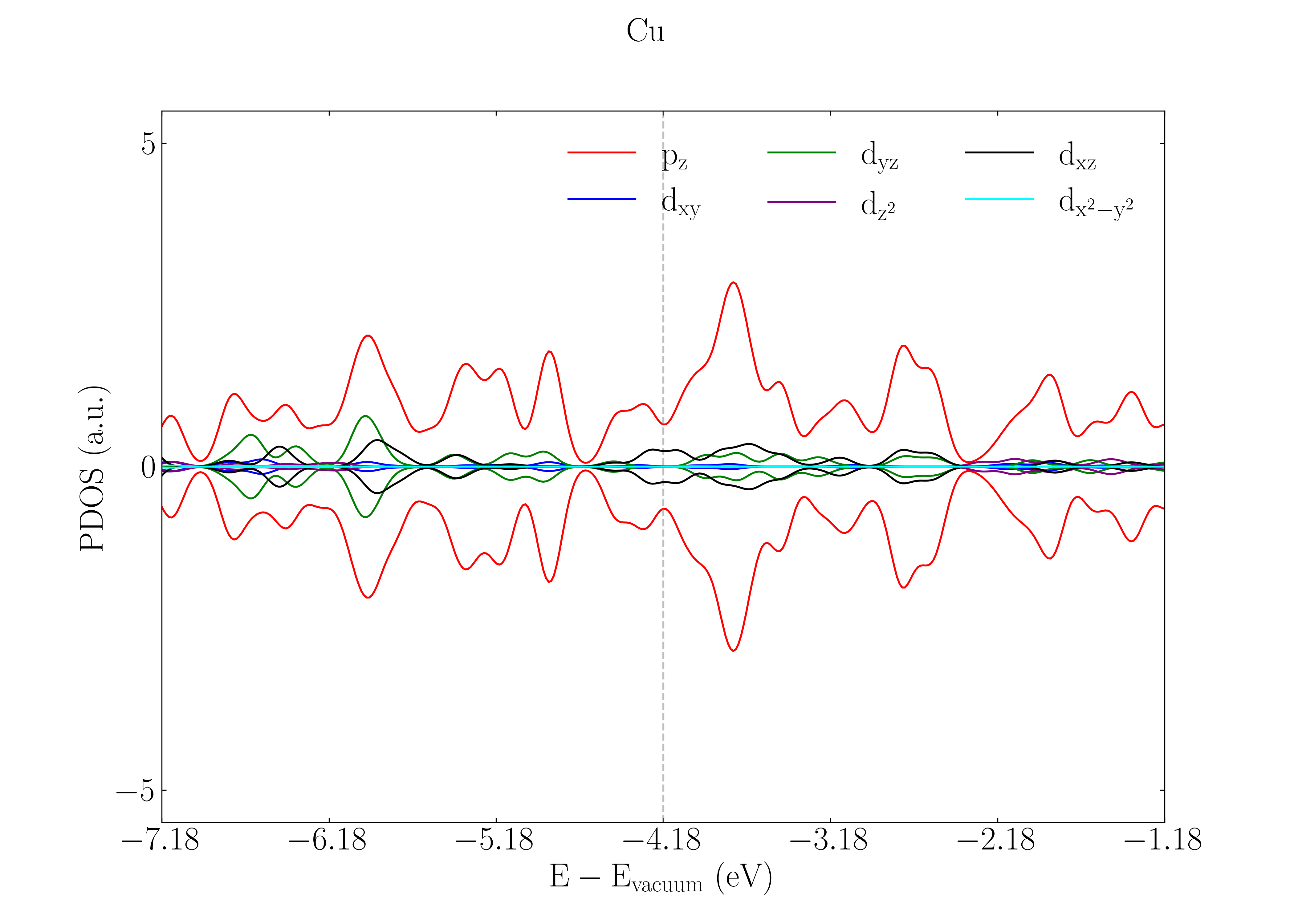}
  \end{subfigure}\hfill
  \begin{subfigure}[b]{0.5\textwidth}
    \centering
    \includegraphics[width=\linewidth,height=0.5\textheight,keepaspectratio]{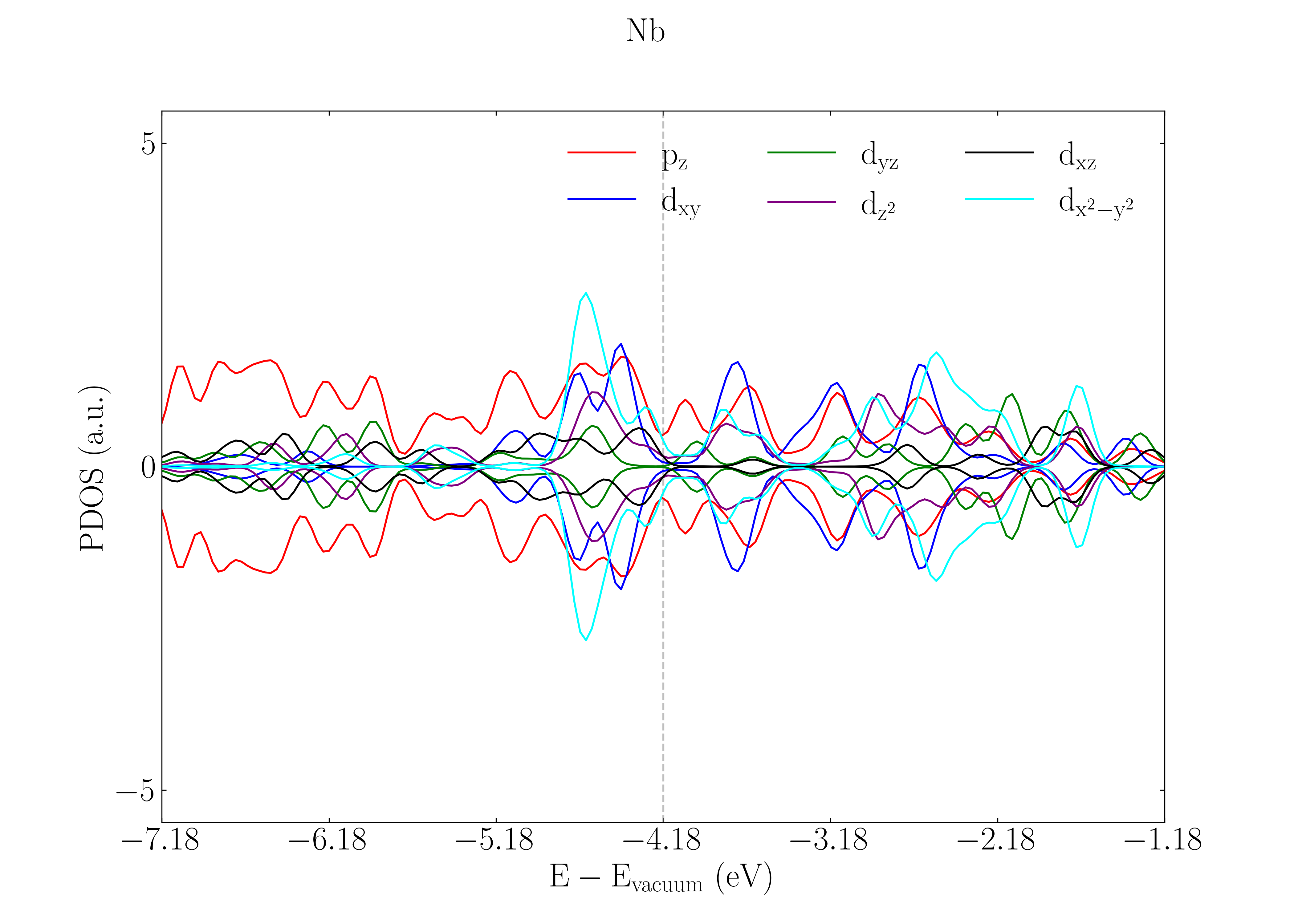}
  \end{subfigure}
  \par\vspace{-2pt}

  \begin{subfigure}[b]{0.5\textwidth}
    \centering
    \includegraphics[width=\linewidth,height=0.5\textheight,keepaspectratio]{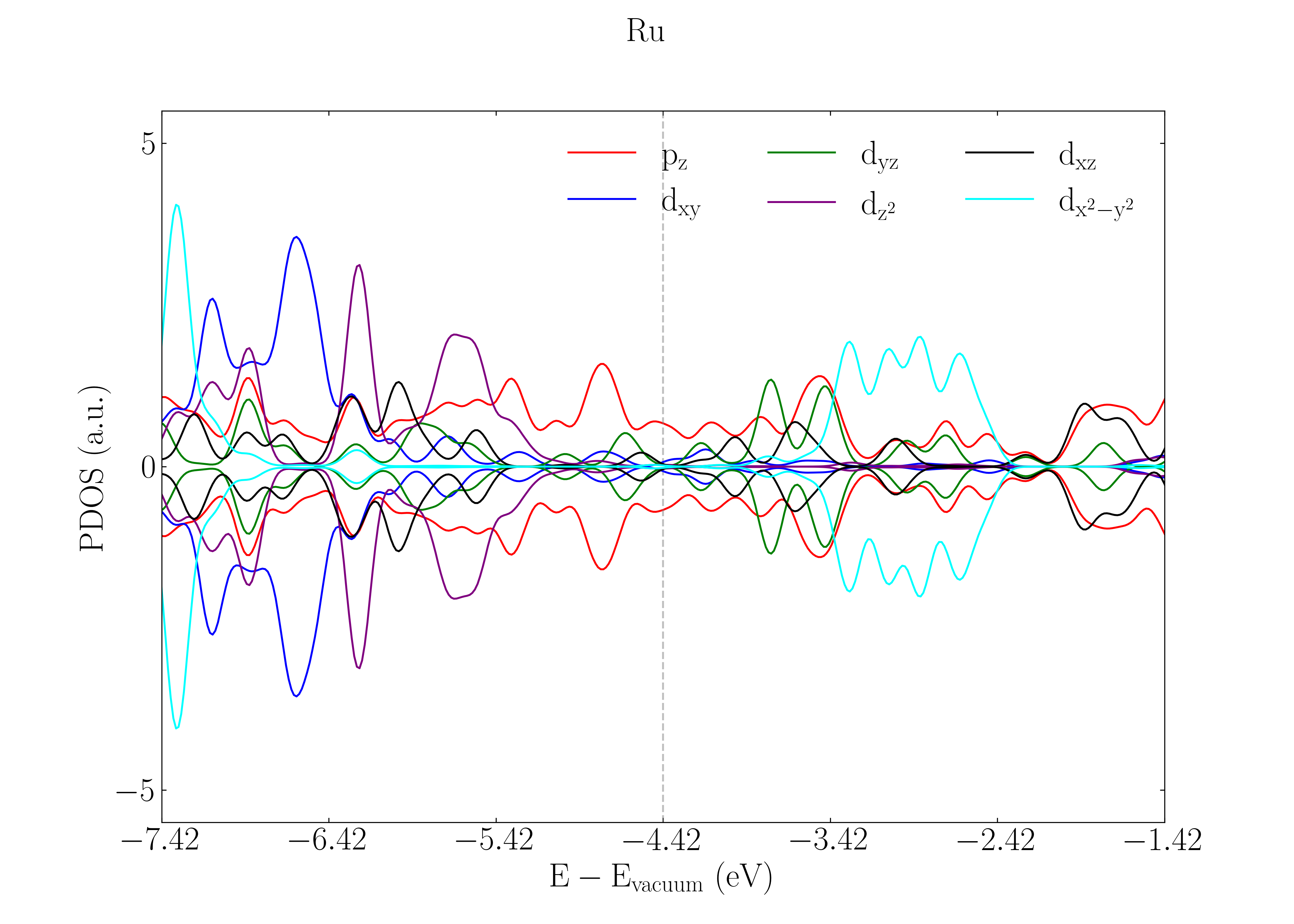}
  \end{subfigure}\hfill%
  \begin{subfigure}[b]{0.5\textwidth}
    \centering
    \includegraphics[width=\linewidth,height=0.5\textheight,keepaspectratio]{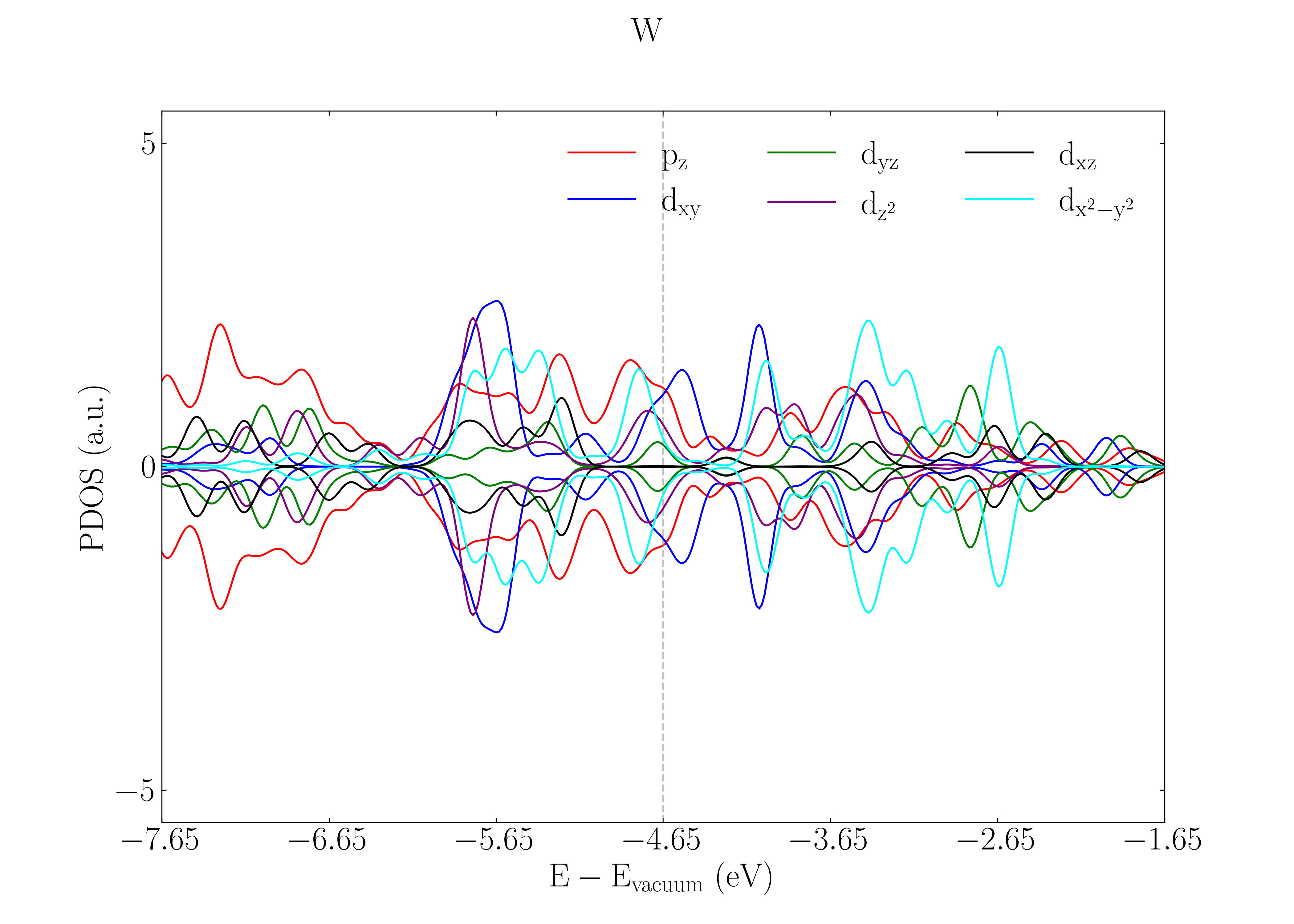}
  \end{subfigure}

  \caption{The PDOS of $p_z$ and $d$-orbitals for B/M/B for M = Cu, Nb, Ru and W, where the lines blue is d$_{xy}$-orbital, green is d$_{yz}$-orbital, purple is d$_{z^2}$-orbital, black is d$_{xy}$-orbital, cyan is d$_{x^2-y^2}$-orbital, and the red line is $p_z$-orbital from C$^{468}$ atoms. The dashed line indicates the Fermi energy level.}\label{fig:dos-all3}
\end{figure}

\begin{figure}[H]
  \centering
  \begin{subfigure}[b]{0.5\textwidth}
    \centering
    \includegraphics[width=\linewidth,height=0.5\textheight,keepaspectratio]{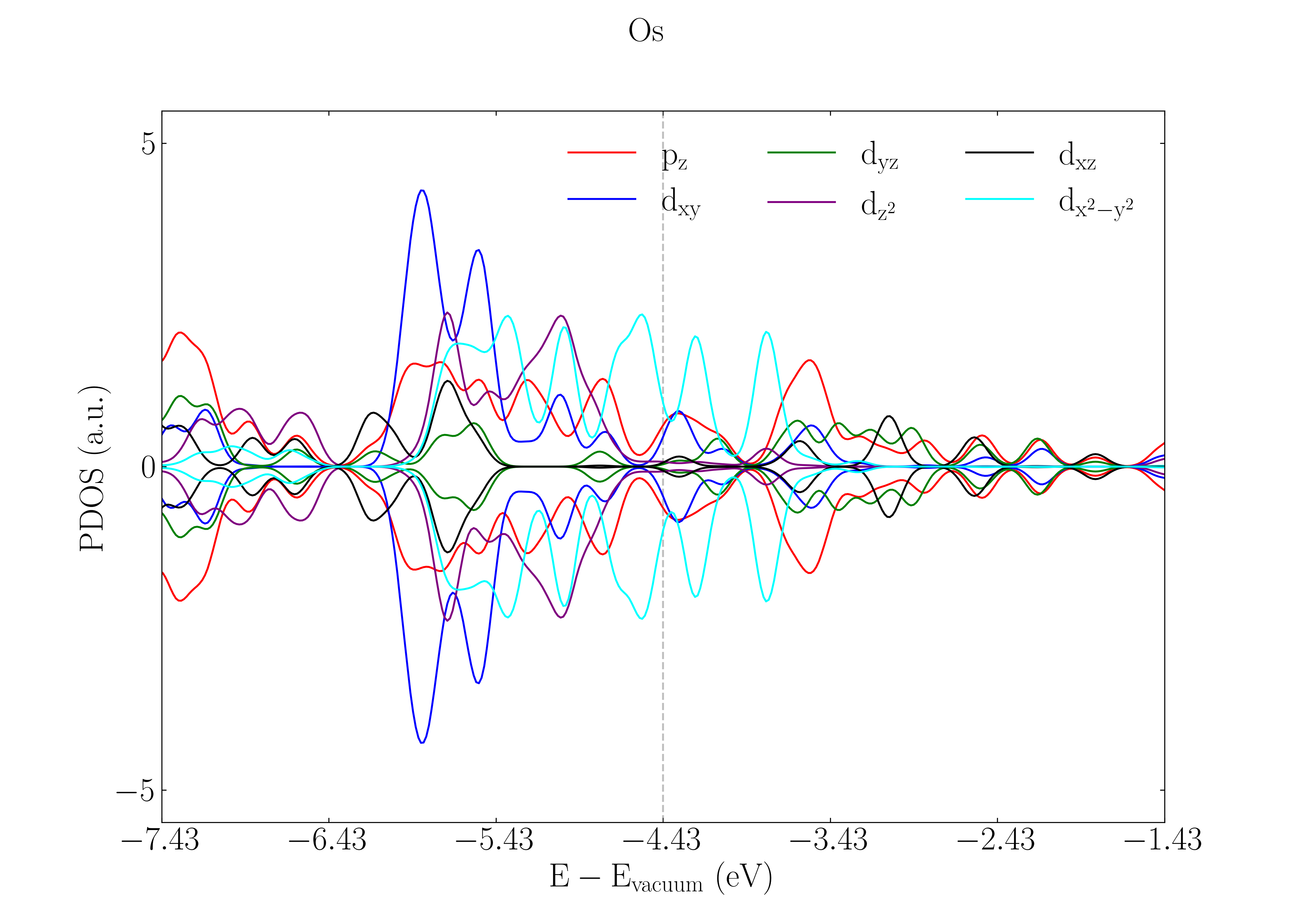}
  \end{subfigure}\hfill
  \begin{subfigure}[b]{0.5\textwidth}
    \centering
    \includegraphics[width=\linewidth,height=0.5\textheight,keepaspectratio]{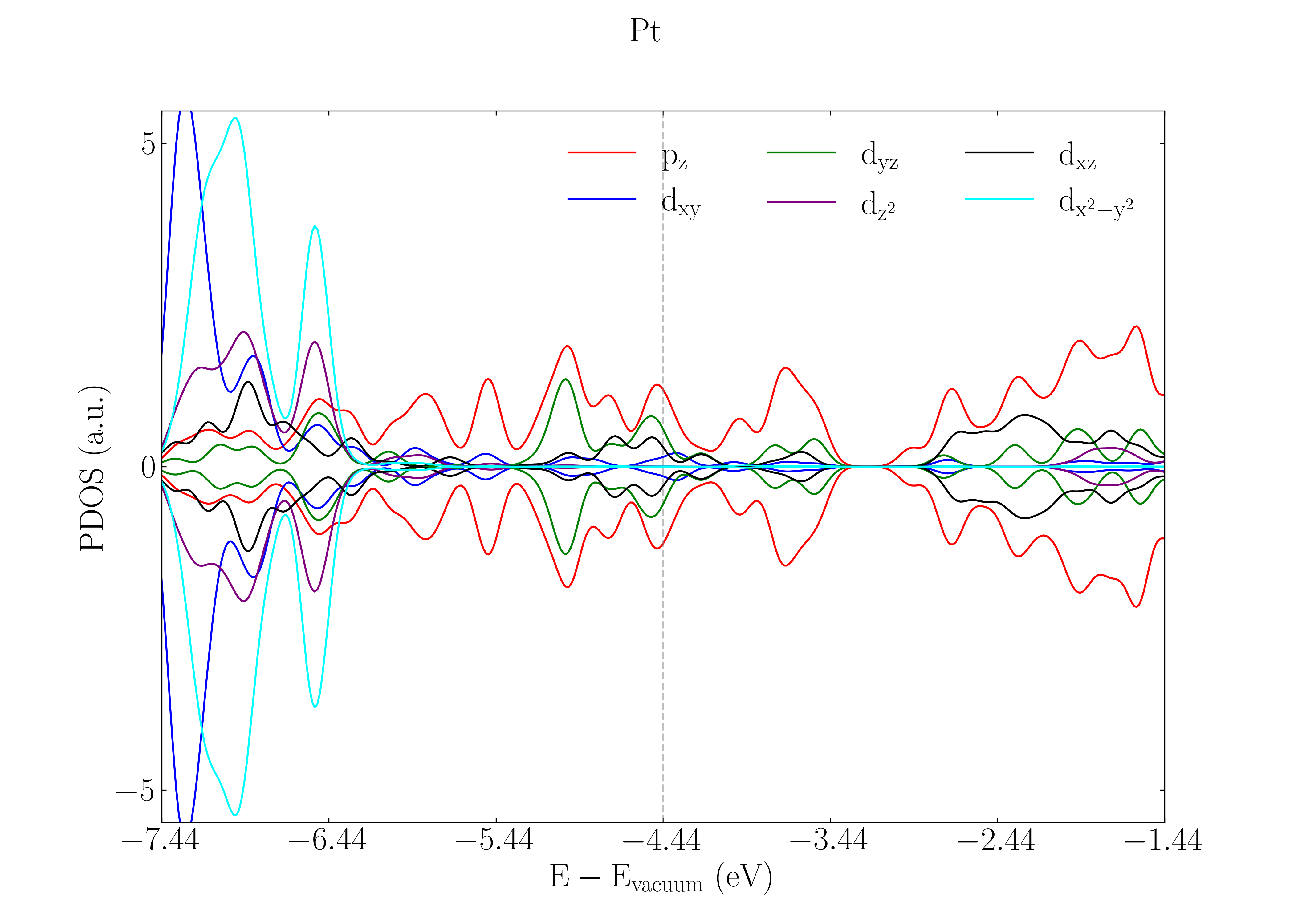}
  \end{subfigure}

  \caption{The PDOS of $p_z$ and $d$-orbitals for B/M/B for M = Os and Pt, where the lines blue is d$_{xy}$-orbital, green is d$_{yz}$-orbital, purple is d$_{z^2}$-orbital, black is d$_{xy}$-orbital, cyan is d$_{x^2-y^2}$-orbital, and the red line is $p_z$-orbital from C$^{468}$ atoms. The dashed line indicates the Fermi energy level.}\label{fig:dos-all4}
\end{figure}

In Figs.\,\ref{fig:dos-all1}, \ref{fig:dos-all2}, \ref{fig:dos-all3} and \ref{fig:dos-all4}, we present the PDOS of the d$-$orbitals and the p$_z-$orbital of the C$^{468}$ atoms for each B/M/B system in the vicinity of the Fermi level.

\bibliography{apssamp}